\documentclass[structabstract]{aa} 

\usepackage{epstopdf}
\usepackage{graphicx}
\usepackage[table]{xcolor}
\usepackage{bigstrut}
\usepackage{booktabs}

\usepackage{array}
\newcolumntype{+}{>{\global\let\currentrowstyle\relax}}
\newcolumntype{^}{>{\currentrowstyle}}

\usepackage[a4paper,colorlinks=true,pdffitwindow=true,pdfpagelayout=SinglePage,pdfpagemode=UseOutlines,bookmarks=true,bookmarksopen=true,bookmarksnumbered=true]{hyperref}

\usepackage{savesym}

\usepackage{amsmath}

\savesymbol{iint} 
\usepackage{txfonts} 
\restoresymbol{TXF}{iint} 

\DeclareSymbolFont{TXFontLetterFix}{OML}{cmm}{m}{it}
\DeclareMathSymbol{\Substitutev}{\mathalpha}{TXFontLetterFix}{118}

\usepackage{natbib} 
\bibpunct{(}{)}{;}{a}{}{,} 

\usepackage[printonlyused,withpage]{acronym} 

\usepackage{pdflscape}
\usepackage{longtable}
\usepackage{multirow}

\usepackage[normalem]{ulem}
\usepackage{pifont}

\usepackage{xspace}

\usepackage{mhchem}

\savesymbol{arcmin}
\savesymbol{arcsec}
\savesymbol{fs}
\usepackage{siunitx}
\restoresymbol{SIUX}{arcmin}
\restoresymbol{SIUX}{arcsec}
\restoresymbol{SIUX}{fs}
\newunit{\Year}{yr}
\newunit{\dex}{dex}
\newunit{\magn}{mag}
\newunit{\parsec}{pc}

\newcommand{\abratio}[2]{[\mathrm{#1}/\mathrm{#2}]\xspace}
\newcommand{\ElementDegree}[2]{#1\,#2\xspace}
\newcommand{\turbospectrum}[0]{\emph{turbospectrum}\xspace}
\newcommand{\daospec}[0]{DAOSPEC\xspace}
\newcommand{\rms}[0]{\mathrm{r.m.s}\xspace}
\newcommand{\ExcerptWarning}[0]{Excerpt of the full table shown here for guidance regarding its form and content. Full table available online at CDS.\xspace}

\begin{document}
\acrodef{LMC}[LMC]{Large Magellanic Cloud}
\acrodef{MC}[MC]{Magellanic Clouds}
\acrodef{SMC}[SMC]{Small Magellanic Cloud}
\acrodef{MW}[MW]{Milky Way}
\acrodef{EW}[EW]{equivalent width}
\acrodef{EW'}[EW']{normalised equivalent width}
\acrodef{CaT}[CaT]{\ion{Ca}{II} triplet}
\acrodef{RGB}[RGB]{red giant branch}
\acrodef{AGB}[AGB]{asymptotic giant branch}
\acrodef{CMD}[CMD]{colour-magnitude diagram}
\acrodef{SNR}[$\mathrm{S}/\mathrm{N}$ ratio]{signal-to-noise ratio}
\acrodef{SNR2}[$\mathrm{S}/\mathrm{N}$]{$\mathrm{S}/\mathrm{N}$}
\acrodef{RC}{Red Clump}
\acrodef{LTE}{Local Thermodynamic Equilibrium}
\acrodef{SS}[SS]{spectrum synthesis}
\acrodef{hfs}{hyperfine structure}
\acrodef{GC}{globular cluster}
\acrodef{SFH}{star formation history}
\acrodef{dSph}{dwarf spheroidal}
\acrodef{SNII}[SNII]{type II supernova}
\acrodef{SNIa}[SNIa]{type Ia supernova}
\acrodef{SNeII}[SNII]{type II supernovae}
\acrodef{SNeIa}[SNIa]{type Ia supernovae}
\acrodef{ISM}[ISM]{intestellar medium}
\acrodef{IMF}[IMF]{initial mass function}

\title{Chemical abundances in LMC stellar populations.}
\subtitle{II. The bar sample\thanks{Proposals 072.B-0293(B) and 078.B-0323(A), P.I. Vanessa Hill},\thanks{Tables $3$, $5$, $7$, $9$, and $11$ are only available in electronic form at the CDS via anonymous ftp to cdsarc.u-strasbg.fr or via \url{http://cdsweb.u-strasbg.fr}}}

\author{M. Van der Swaelmen\inst{\ref{institute1},\ref{institute2}}
  \and V. Hill\inst{\ref{institute1}}
  \and F. Primas\inst{\ref{institute2}}
  \and A. A. Cole\inst{\ref{institute3}}
}

\institute{Observatoire de la C\^ote d'Azur, CNRS UMR\,7293, BP4229, 06304, Nice Cedex 4, France\\
  \email{Mathieu.Van-der-Swaelmen@oca.eu}\\
  \email{Vanessa.Hill@oca.eu}\label{institute1}
  \and European Southern Observatory, Karl Schwarzschild Str. 2, 85748 Garching b. M\"unchen, Germany\\
  \email{fprimas@eso.org}\label{institute2}
  \and School of Mathematics \& Physics, University of Tasmania, Private Bag 37, Hobart, Tasmania 7001, Australia\\
  \email{andrew.cole@utas.edu.au}\label{institute3}
}

\date{Received 15/01/2013; accepted 27/03/2013}

\abstract{}{ 
  This paper compares the chemical evolution of the \ac{LMC} to that of the \ac{MW} and investigates the relation between the bar and the inner disc of the \acs{LMC} in the context of the formation of the bar.}{ 
  We obtained high-resolution and mid signal-to-noise ratio spectra with FLAMES/GIRAFFE at ESO/VLT and performed a detailed chemical analysis of \num{106} and \num{58} \acs{LMC} field red giant stars (mostly older than \SI{1}{\giga\Year}), located in the bar and the disc of the \acs{LMC} respectively. To validate our stellar parameter determinations and abundance measurement procedures, we performed thorough tests using the well-known mildly metal-poor Milky-Way thick disc giant Arcturus (HD \num{124897}, $\alpha$ Boo). 
 We measured elemental abundances for O, Mg, Si, Ca, Ti ($\alpha$-elements), Na (light odd element), Sc, V, Cr, Co, Ni, Cu (iron-peak elements), Y, Zr, Ba, La and Eu (\emph{s}- and \emph{r}-elements).}{ 
 We find that the $\alpha$-element ratios $\abratio{Mg}{Fe}$ and $\abratio{O}{Fe}$ are lower in the \acs{LMC} than in the \acs{MW} while the \acs{LMC} has similar $\abratio{Si}{Fe}$, $\abratio{Ca}{Fe}$, and $\abratio{Ti}{Fe}$ to the \acs{MW}. As for the heavy elements, $\abratio{Ba, La}{Eu}$ exhibit a strong increase with increasing metallicity starting from $\abratio{Fe}{H} \approx \SI{-0.8}{\dex}$, and the \acs{LMC} has lower $\abratio{Y+Zr}{Ba+La}$ ratios than the \acs{MW}. Cu is almost constant over all metallicities and about \SI{0.5}{\dex} lower in the \acs{LMC} than in the \acs{MW}. The \acs{LMC} bar and inner disc exhibit differences in their $\abratio{\alpha}{Fe}$ (slightly larger scatter for the bar in the metallicity range $[-1, -0.5]$), their Eu (the bar trend is above the disc trend for $\abratio{Fe}{H} \ge \SI{-0.5}{\dex}$, their Y and Zr, their Na and their V (offset between the bar and the disc distributions).}{ 
  Our results show that the chemical history of the \acs{LMC} experienced a strong contribution from type Ia supernovae as well as a strong \emph{s}-process enrichment from metal-poor AGB winds. Massive stars made a smaller contribution to the chemical enrichment compared to the \acs{MW}. The observed differences between the bar and the disc speak in favour of an episode of enhanced star formation a few \SI{}{\giga\Year} ago, occurring in the central parts of the \acs{LMC} and leading to the formation of the bar. This is in agreement with recently derived star formation histories.} 


\maketitle

\section{Introduction}
Despite decades of intensive observational and theoretical work, we are still far from a complete and clear understanding of the nearby universe, the \ac{MW} and its neighbours. Among the satellites of the \ac{MW}, the \ac{SMC} and the \ac{LMC} are of particular interest since they form the closest example of galaxies in gravitational and hydrodynamical interaction, and therefore constitute a unique laboratory to study the effect of tides and matter exchange on the chemical evolution and star formation history of a galaxy.

The \acs{LMC} is an almost face-on, gas-rich galaxy with regions of active stellar formation located at a distance of \SI{50}{\kilo\parsec} \citep{2004NewAR..48..659A}. It has a mass of $10^{10}\rm M_{\odot}$ \citep{2002AJ....124.2639V}, intermediate between massive spirals and dwarf galaxies. Because of its bar-like feature embedded in a disc and its single spiral arm, the \acs{LMC} is classified as a Barred Magellanic Spiral (SBm) \citep{1972VA.....14..163D}. The young population exhibits an irregular morphology, likely the stigmata of a very recent interaction with the \acs{SMC}, while the old and intermediate-age populations are located within a regular disc and a prominent and luminous off-centre bar. However, the morphology of the \acs{LMC} is not well understood. For instance, the \acs{GC} population of the \acs{LMC} is intriguing since no object of age between \num{3} and \SI{10}{\giga\Year} is found \citep[e.g.][]{1991IAUS..148..183D, 2001AJ....122..842R}; this age gap is not observed in the \acs{SMC} \acs{GC} population. We still do not know the origin and the true nature of the asymmetric bar-like structure: is it a dynamical bar driven by disc instabilities like the one found at the centre of the \acs{MW} or is it a new stellar population? In addition, distance measurements based on Red Clump stars or RR Lyrae variables located in the \acs{LMC} bar suggest that the bar is about \SI{5}{\kilo\parsec} above the disc plane \citep{2012AJ....144..106H}. This feature is also puzzling and difficult to understand: is it a deformation of the \acs{LMC} disc due to gravitational interaction with the \acs{SMC}? Another interesting feature is that the bar is off-centre: the centroids of the bar and the disc differ \citep{2001AJ....122.1827V}. \cite{2004ApJ...614L..37Z} showed that these features can be explained by a triaxial stellar bulge embedded in a highly obscuring thin disc: unfortunately, this solution is not completely satisfactory since it requires a strong reddening (or a very inclined disc which has equivalent effect). If this were the case, one would then have to understand the origin of such a stellar bulge (driven by a dynamical instability in the past or similar to early-type bulges?). The Magellanic Bridge, made of gas and stars, connects the \acs{LMC} and the \acs{SMC} and is the site of matter exchange between the two Clouds.

Numerous authors \citep[e.g][]{2007ApJ...668..949B,2005MNRAS.356..680B,2009MNRAS.393L..60B} have tried to self-consistently reproduce the large- and small-scale structure of the \acs{LMC} (asymmetric off-centre bar, \acs{GC} age gap, Magellanic Bridge...) in their dynamical models, taking into account the interaction with the \acs{SMC} and/or with the \acs{MW}. However, because of uncertainties on the proper motions \citep{2006ApJ...638..772K,2006ApJ...652.1213K}, we still do not know whether the \{\acs{LMC}+\acs{SMC}\} system is performing its first passage about the \acs{MW}, or whether the two Magellanic Clouds formed as separate entities and became gravitationally bound later on; a variety of dynamical models have therefore been proposed. For instance, \cite{2012MNRAS.421.2109B} tested two first infall models: at the beginning, the \acs{MC} are a binary pair, evolving in isolation until their first passage close to the \acs{MW}. Their model 2 reproduces most of the morphological and dynamical features. In particular, \cite{2012MNRAS.421.2109B} explains the asymmetric off-centre bar: as the \acs{LMC} disc is bar unstable, the bar is present from the beginning; it becomes asymmetric off-centre due to a close encounter of the \acs{LMC} and \acs{SMC} a few \SI{}{\mega\Year} ago. On the other hand, \cite{2002ApJ...566..239S} derived the star formation histories for field stars located in the LMC bar and the inner part of the disc from deep \ac{CMD}. They show that the \ac{SFH} of the bar and the inner disc were similar at old epochs (between \num{7} and \SI{14}{\giga\Year}); but while the SFR of the inner disc has remained rather constant, the bar experienced a dramatic increase SFR between \num{4} to \SI{6} {\giga\Year} ago. Thus, the \acs{SFH} supports the scenario of a new burst of stellar formation at the centre of the \acs{LMC}, which could lead to the appearance of the bar-like structure.

Kinematic and chemical tagging of stellar populations is a powerful tool to help reconstruct the past history of a given galactic environment. \cite{2008A&A...480..379P} provided for the first time a detailed chemical analysis of a large sample of \acs{LMC} \ac{RGB} stars located in the \acs{LMC} disc, $\sim$\SI{2}{degrees} South of the \acs{LMC} bar, hereafter the inner disc. \cite{2012ApJ...761...33L} measured the [$\alpha$/Fe] of 89 stars in a field close to the \acs{LMC} globular cluster NGC\,1786, some \SI{3}{\degree} North-West of the bar. The present work aims at bringing new light on the nature of the bar: to this end, we provide a detailed chemical tagging of a large sample of \acs{LMC} bar \ac{RGB} stars and compare the elemental trends to the reanalysed trends of the inner disc. Section~\ref{data} describes the sample selection and the data reduction. Section~\ref{Stellar_parameters} and \ref{Abundance_analysis} explain the stellar parameter and abundance measurement methods. Section~\ref{Results_discussion} provides the results and their interpretation. Section~\ref{Conclusion} summarises the main results of this work.

\section{Observations and data reduction}
\label{data}

\subsection{Sample selection}
\label{Sample_selection}
\begin{figure}
  \begin{centering}
    \newlength{\HeightMDFCMD}
    \setlength{\HeightMDFCMD}{4cm}
    \includegraphics[height=\HeightMDFCMD]{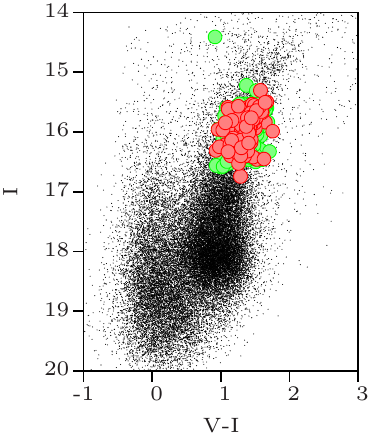}
    \includegraphics[height=\HeightMDFCMD]{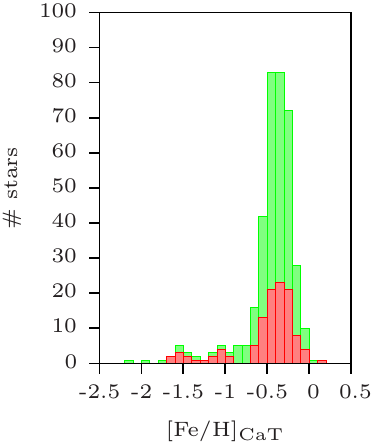}
    \caption{\label{vdsetal2012_bar_mdf}\label{vdsetal2012_bar_cmd} Colour-Magnitude diagram (left panel) and metallicity distribution (right panel) of \cite{2005AJ....129.1465C} \acs{RGB} sample (green) and our selected \acs{RGB} stars (red). V and I magnitude are from the OGLE catalogue \cite{1997AcA....47..319U, 2000AcA....50..307U, 2005AcA....55...43S} for the samples of \cite{2005AJ....129.1465C} and the present paper, while black dots in the \acs{CMD} are photometric data from \cite{2004AJ....128.1606Z} catalogue. Metallicities in the right panel are those derived from the infrared \ion{Ca}{II} triplet index by \cite{2005AJ....129.1465C}.}
  \end{centering}
\end{figure}

\cite{2005AJ....129.1465C} observed \num{373} \ac{RGB} stars in the field of the \ac{LMC} bar and derived radial velocities and metallicities from low-resolution infrared \ion{Ca}{II} triplet spectra. We used their metallicity distribution to select \num{113} \ac{RGB} stars (the maximum number we could accommodate in a single multi-object fibre configuration of FLAMES) belonging to the \ac{LMC} bar, taking care to sample as evenly as possible the whole metallicity range from $\abratio{Fe}{H}_{\mathrm{CaT}} = \SI{-1.69}{\dex}$ to $\abratio{Fe}{H}_{\mathrm{CaT}}=\SI{0.14}{\dex}$. Because metal-poor stars ($\abratio{Fe}{H}\le\SI{-1.}{\dex}$) are rare, a random selection would not provide many metal-poor stars; hence this metallicity pre-selection is a necessary precaution to trace the early epochs of the \ac{LMC} history (represented in the low-metallicity tail of the distribution). Figure~\ref{vdsetal2012_bar_cmd} shows the location of the \num{373} stars from \cite{2005AJ....129.1465C} and our \num{113} targets on a I,(V-I) \ac{CMD} and the metallicity distribution function of these two samples. We obtained high resolution spectra of our \num{113} stars at VLT/ESO with the FLAMES/GIRAFFE multifibre spectrograph \citep{2002Msngr.110....1P}. In order to measure numerous elemental abundances, we used three setups HR11 ($\lambda_{\mathrm{central}}=\SI{572.8}{\nano\meter}$, $R_{\lambda_{\mathrm{central}}}\simeq \num{24200} $), HR13 ($\lambda_{\mathrm{central}}=\SI{627.3}{\nano\meter}$, $R_{\lambda_{\mathrm{central}}}\simeq \num{22500} $) and HR14 ($\lambda_{\mathrm{central}}=\SI{651.5}{\nano\meter}$, $R_{\lambda_{\mathrm{central}}}\simeq \num{17740} $)\footnote{for technical details see \url{http://www.eso.org/sci/facilities/paranal/instruments/flames/doc/VLT-MAN-ESO-13700-2994_v86.0.pdf}}, covering a total of $\approx \SI{1000}{\angstrom}$. The spectra thus cover lines belonging to the $\alpha$ (Ca, O, Mg, Ti, Si), iron-peak (Sc, V, Cr, Co, Ni, Cu), \emph{s}-process and \emph{r}-process elements (Ba, La, Zr, Y, Eu). Thanks to the MEDUSA mode of the GIRAFFE spectrograph up to 135 objects can be observed simultaneously in a single exposure. For our purposes, around 10 to 20 fibres were allocated to sky positions and the other remaining fibres were devoted to the observation of LMC bar stars. In addition, three hot (O-B type) stars in the LMC were allocated to fibres, to allow an accurate correction for telluric absorption lines.

\subsection{Data reduction}

We carried out the data reduction with the help of the ESO GIRAFFE pipeline (built upon the Geneva Giraffe pipeline described in \citealp{2000SPIE.4008..467B}), part of the \emph{esorex} framework\footnote{pipeline available at \url{http://www.eso.org/projects/dfs/dfs-shared/web/vlt/vlt-instrument-pipelines.html}}. The reduction steps include the bias and dark current correction, wavelength calibration (using a Th-Ar lamp), spectrum extraction and flat fielding. As the pipeline does not support sky subtraction nor radial velocity correction, we carried out those operations separately. 

\begin{figure}
  \begin{centering}
    \includegraphics{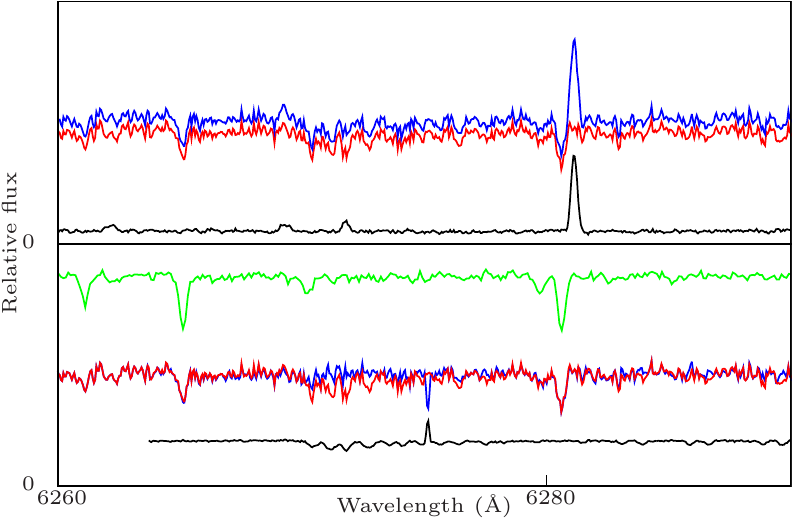}
    \caption{\label{vdsetal2012_data_reduction} Top panel: an example of raw spectrum (blue) for the star 05244301-6943412 and the corresponding master-sky (black) we used to obtain the sky-subtracted spectrum (red). In the red spectrum sky emission lines have been correctly removed. A cosmic ray remains at $\sim\SI{6290}{\angstrom}$. Bottom panel: the sky-subtracted spectrum of 05244301-6943412 before (red) and after (blue) the correction for the telluric lines. The spectrum of the fast rotator is plotted in black to show the position of telluric features. The green spectrum is the co-added spectrum: the $\sigma$-clipping has removed the remaining cosmic rays and the signal-to-noise ratio is clearly improved compared to the single exposure. All spectra are plotted in the same scale, except for the fast rotator spectrum (it has been scaled for legibility).}
  \end{centering}
\end{figure}

\paragraph{Sky subtraction}
We visually ranked the sky spectra according to their quality and discarded those showing the lowest \ac{SNR} or spectral contamination like jumps in flux due to scattered light (stellar light, CCD glow, simultaneous calibration lamp) or a CCD defect. After this quality selection, we ended up with a handful of sky spectra (at least five to eight) in most cases. The selected sky spectra were averaged with k-$\sigma$ clipping rejection and the resulting master-sky was subtracted from each stellar spectrum (see Fig.~\ref{vdsetal2012_data_reduction}). This procedure was repeated for each observation of the \num{113} bar stars and for each setup.

\paragraph{Correction of telluric absorption band around $\SI{6300}{\angstrom}$}
Among the two oxygen lines (at \SI{6300}{\angstrom} and \SI{6363}{\angstrom}) that are available in the optical wavelength range, the first is the strongest line and is more appropriate for abundance measurements. Unfortunately, it is in a region contaminated by atmospheric spectral features (from \SI{6270}{\angstrom} to \SI{6330}{\angstrom}). To measure abundance reliably, it is mandatory to correct for the telluric lines. Three hot stars were observed simultaneously to our science targets for that purpose: 05235121-6934233, 05235885-6952357, 05242945-6937236. We inspected the spectra of the three stars. As the star 05235885-6952357 showed the broadest stellar spectral features (highest rotation) and the highest \ac{SNR}, we used its spectrum for a telluric correction: in the wavelength region $[\SI{6270}{\angstrom}, \SI{6330}{\angstrom}]$, we divided our individual stellar spectra by the hot star spectrum. We checked that no discontinuities were introduced (see Fig.~\ref{vdsetal2012_data_reduction}).

\begin{figure}
  \begin{centering}
    \includegraphics{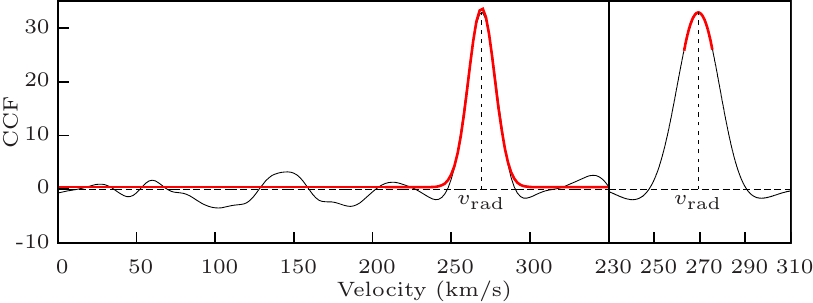}
    \caption{\label{vdsetal_ccf_radvel} Left panel: example of cross-correlation function (black line) and its Gaussian fit (red thick line). Right panel: Zoom in on the maximum of the cross-correlation function and its parabolic fit (red thick line). We first used a Gaussian fit to locate the position of the maximum and then defined a smaller velocity range (Gaussian fit FWHM) and compute the cross-correlation function over a finer grid to improve the determination of the radial velocity.}
  \end{centering}
\end{figure}

\paragraph{Radial velocity measurements and correction}
We obtained multiple observations of the same star in a given setup: \num{10} exposures with HR11, \num{5} exposures with HR13 and \num{4} exposures with HR14, which represents a total of more than \num{2000} spectra. Table~\ref{vdsetal2012_exposures} lists the observations, the dates and the total exposure times. We wrote a cross-correlation routine using our own Gaussian masks (continuum with discrete Gaussian absorption profiles centred at the position of particular stellar lines) to perform the radial velocity measurement. In order to build Gaussian masks resembling our spectra in terms of stellar parameters (temperature, gravity, metallicity, microturbulent velocity) and spectral resolution, we used a set of our \ac{LMC} spectra: in each spectrum, we selected a high number ($\ge \num{30}$) of strong spectral features (iron, calcium... lines), fitted them with a Gaussian profile, computed an average absorption line profile, and then built a mask for each setup. The cross-correlation routine returned the radial velocity in the Earth frame; to correct it for the Earth motion and obtain the barycentric velocity $\Substitutev_{\mathrm{rad}}$, we used the MIDAS task \emph{barycor}. Using a k-$\sigma$ clipping rejection (over the radial velocity) allowed us to point out suspicious spectra requiring a special investigation, and we discarded them if justified (\emph{e.g.} low \ac{SNR} leading to a poor determination of the radial velocity). For instance, in the setup HR14, for the star 05231321-6946382, we measured four radial velocities: \SI{270.2}{\kilo\meter\per\second}, \SI{270.3}{\kilo\meter\per\second}, \SI{270.5}{\kilo\meter\per\second} and \SI{275.4}{\kilo\meter\per\second}; based on the above procedure, we flagged the observation leading to a velocity of \SI{275.4}{\kilo\meter\per\second}. As the poor \ac{SNR} ($\sim \num{2}$) explains the disagreement, we discarded this observation. We only excluded a few spectra with this test. Figure~\ref{vdsetal_ccf_radvel} shows an example of a cross-correlation function and the parabolic fit used to determine the radial velocity.

\begin{table}
  \begin{center}
    \caption{\label{vdsetal2012_exposures} For each setup, the exposures, the total exposure times and the observation dates are given.}
    \begin{tabular}{cccc}
      \hline
      \hline
      Setups & \# & Total exp. time & Dates\\
\hline
\begin{tabular}{l}HR11\\ \\\end{tabular} & \begin{tabular}{l}10\\ \\\end{tabular} & \begin{tabular}{l}7 h 42 min\\ \\\end{tabular} & \begin{tabular}{ll} 2006-10-[6, 7, 10, 26]\phantom{17}\\ 2006-11-[8, 22]\\\end{tabular}\\
\begin{tabular}{l}HR13\\ \\ \\\end{tabular} & \begin{tabular}{l}5\\ \\ \\\end{tabular}  & \begin{tabular}{l}5 h 50 min\\ \\ \\\end{tabular} & \begin{tabular}{l} 2004-01-15\phantom{[, 17, 18, 20]}\\ 2004-02-[16, 21]\\ 2004-03-26\\\end{tabular}\\
\begin{tabular}{l}HR14\end{tabular} & \begin{tabular}{l}4\end{tabular}  & \begin{tabular}{l}4 h 08 min\end{tabular} &\begin{tabular}{l} 2004-02-[16, 17, 18, 20]\end{tabular}\\
\hline

    \end{tabular}
  \end{center}
\end{table}

\begin{figure}
  \begin{centering}
    \includegraphics{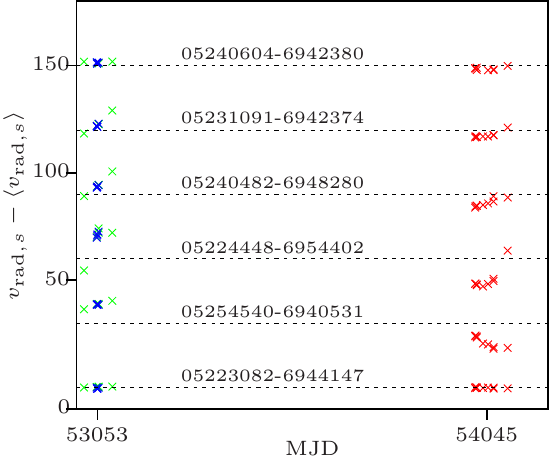}
    \caption{\label{vdsetal_radvel_curves} Radial velocity curves for six stars. The relative radial velocities $\Substitutev_{\mathrm{rad, }s} - \langle \Substitutev_{\mathrm{rad, }s} \rangle$ determined for each exposure are plotted as a function of MJD for five stars where the T-test failed and we suspect a variability in the radial velocity. The curves were shifted for legibility (the dashed line represents the offset). The bottom curve (star 05223082-6944147) is a star with no radial velocity variability and shown here for reference. In red: HR11, in green: HR13, in blue: HR14 (for HR13 data and HR14 data obtained at the same epoch, the green crosses are below the blue ones). Error bars are smaller than the symbols ($\le \SI{1}{\kilo\meter\per\second}$). %
    }
  \end{centering}
\end{figure}

In order to detect any systematic effect (from one setup to the other) or possible variations of the radial velocity, we computed a mean barycentric radial velocity for each star in each setup. For a given setup $s$, we used the $\mathrm{N}_{s}$ estimates to compute the mean radial velocity $\Substitutev_{\mathrm{rad, }s}$. As the setup HR11 provides the highest number of exposures, the standard deviation of the radial velocity distribution is better defined in this setup; we therefore employed it to estimate the precision of a single velocity measurement by computing the mean of the standard deviations over the sample, finding a value of  \SI{0.6}{\kilo\meter\per\second}. %
For each setup $s$, we compute the standard error of the mean radial velocity per setup $\Substitutev_{\mathrm{rad, }s}$ as $e_{\mathrm{rad, }s}=\max(\sqrt{\mathrm{var}(\Substitutev_{\mathrm{rad, }s})/\mathrm{N}_{s}}, 0.6/\sqrt{\mathrm{N}_{s}})$. We performed a T-test to compare $\Substitutev_{\mathrm{rad, }\mathrm{HR\num{11}}}$, $\Substitutev_{\mathrm{rad, }\mathrm{HR\num{13}}}$, and $\Substitutev_{\mathrm{rad, }\mathrm{HR\num{14}}}$, taken two by two:
\begin{itemize}
\item[$\bullet$] $\mathrm{H}_0$: the two mean radial velocities are equal 
\item[$\bullet$] $\mathrm{H}_1$: they are different; significance level: 1\%
\item[$\bullet$]  hypothesis of equal variance 
\end{itemize}
For 103 stars (91\% of the sample), we conclude that the three mean radial velocities are equal at the significance level of 1\%. %
For those stars, the radial velocity measurements in the various setups are remarkably similar ($\rms$ in parenthesis): $\langle \Substitutev_{\mathrm{rad, }\mathrm{HR\num{11}}} - \Substitutev_{\mathrm{rad, }\mathrm{HR\num{13}}} \rangle = \SI{-0.1}{\kilo\meter\per\second}$ (\SI{0.6}{\kilo\meter\per\second}), $\langle \Substitutev_{\mathrm{rad, }\mathrm{HR\num{11}}} - \Substitutev_{\mathrm{rad, }\mathrm{HR\num{14}}} \rangle = \SI{0.3}{\kilo\meter\per\second}$ (\SI{0.7}{\kilo\meter\per\second}) and $\langle \Substitutev_{\mathrm{rad, }\mathrm{HR\num{13}}} - \Substitutev_{\mathrm{rad, }\mathrm{HR\num{14}}} \rangle = \SI{0.4}{\kilo\meter\per\second}$ (\SI{0.3}{\kilo\meter\per\second}). %
For the 10 stars reported in Table~\ref{vdsetal2012_radvel_failing_ttest} however, at least one of the three T-tests failed. We remark that for all reported cases, the radial velocities measured for HR13 and HR14 agree rather well, while the radial velocity measured for HR11 is discrepant with the two others. The stars 05240482-6948280, 05254540-6940531, and 05224448-6954402 show the most dramatic disagreement with differences of about \num{8}, \num{17}, \SI{18}{\kilo\meter\per\second} (respectively) between HR11 and HR13 or between HR11 and HR14. The mean epoch and time span of the observations for each setup are given in the last columns of the Table~\ref{vdsetal2012_radvel_failing_ttest}: the observations in HR13 and HR14 were run at similar epochs (for HR13, more than two months separate the first and the last observation, which explains the large values of standard deviations observed for this setup for stars with a variable $\Substitutev_{\mathrm{rad}}$) while those in HR11 were carried out two years later. The discrepancies between setups for those stars therefore reveal a true radial velocity variation, most probably due to an internal stellar variability or a binary system. Figure~\ref{vdsetal_radvel_curves} displays the radial velocity curves for the five stars with the most extreme variations. The period of variation seems to be large, which is expected for giant stars.

\begin{table*}
  \begin{center}
    \caption{\label{vdsetal2012_radvel_failing_ttest} Stars showing a disagreement in their mean radial velocities from one setup to another. %
    }
    \begin{tabular}{cccccccccc}
      \hline
      \hline

\multirow{2}{*}{2MASS ID} & \multicolumn{3}{c}{HR11} &  \multicolumn{3}{c}{HR13} &  \multicolumn{3}{c}{HR14}\\
\cline{2-10}
                       & $\Substitutev_{\mathrm{rad}}$ & $\sigma(\Substitutev_{\mathrm{rad}})$ & \# & $\Substitutev_{\mathrm{rad}}$ & $\sigma(\Substitutev_{\mathrm{rad}})$ & \# & $\Substitutev_{\mathrm{rad}}$ & $\sigma(\Substitutev_{\mathrm{rad}})$ & \#\\
                       & \SI{}{\kilo\meter\per\second} & \SI{}{\kilo\meter\per\second} &  & \SI{}{\kilo\meter\per\second} & \SI{}{\kilo\meter\per\second} &  & \SI{}{\kilo\meter\per\second} & \SI{}{\kilo\meter\per\second} & \\
\hline
05223316-6951389 & 220.9 & 0.9 & 8 & 219.3 & 0.8 & 5 & 219.1 & 0.6 & 4\\
05224448-6954402 & 243.8 & 5.2 & 9 & 262.2 & 8.0 & 5 & 264.7 & 1.2 & 4\\
05230867-6956329 & 266.6 & 0.6 & 7 & 262.7 & 1.2 & 5 & 262.5 & 0.6 & 4\\
05231074-6939184 & 201.2 & 0.6 & 9 & 204.1 & 0.6 & 5 & 203.7 & 0.6 & 4\\
05231091-6942374 & 248.9 & 1.4 & 9 & 254.4 & 3.9 & 5 & 253.4 & 0.7 & 4\\
05240482-6948280 & 222.1 & 1.9 & 9 & 230.7 & 4.1 & 5 & 229.9 & 0.6 & 4\\
05240604-6942380 & 255.3 & 0.7 & 8 & 258.2 & 0.6 & 5 & 258.1 & 0.6 & 4\\
05240613-6953529 & 217.7 & 1.2 & 9 & 215.8 & 0.6 & 5 & 216.1 & 0.7 & 4\\
05254540-6940531 & 269.3 & 2.6 & 9 & 286.8 & 1.4 & 5 & 286.7 & 0.6 & 4\\
05255801-6937309 & 257.5 & 0.6 & 9 & 256.9 & 0.6 & 5 & 256.4 & 0.6 & 4\\ 
\hline
\hline
$\langle \mathrm{MJD} \rangle$ (\SI{}{\dday})& \multicolumn{3}{c}{54045.347} & \multicolumn{3}{c}{53054.700} & \multicolumn{3}{c}{53053.109}\\
$\mathrm{MJD}_{\mathrm{max}} - \mathrm{MJD}_{\mathrm{min}}$ (\SI{}{\dday})& \multicolumn{3}{c}{82.94} & \multicolumn{3}{c}{71.84} & \multicolumn{3}{c}{4.86}\\
\hline

    \end{tabular}
  \end{center}
\end{table*}

\begin{figure}
  \begin{centering}
    \includegraphics{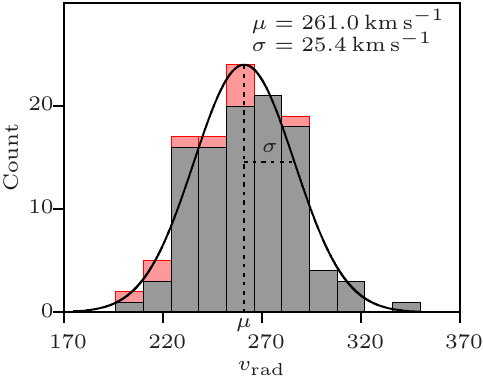}
    \caption{\label{vdsetal2012_bar_radvel_histogram} Distribution of the radial velocities of the \ac{LMC} bar stars. The histogram overplotted in red denotes the stars for which we suspect a radial velocity variability. The mean and the standard deviation of the distribution are respectively \SI{261}{\kilo\meter\per\second} and \SI{25}{\kilo\meter\per\second}. A Gaussian curve $(\mu;\sigma)$ is plotted over the histogram.}
  \end{centering}
\end{figure}

Table~\ref{vdsetal2012_bar_kinematics2_excerpt} provides the weighted mean radial velocity defined by $\langle \Substitutev_{\mathrm{rad}} \rangle = \sum_s \omega_{s} \Substitutev_{\mathrm{rad, }s}$, where the sum is over the three setups $s$ and $\omega_{s} = ({e_{\mathrm{rad, }s}}^{2})^{-1} / \sum_s ({e_{\mathrm{rad, }s}}^{2})^{-1}$, together with the associated error defined by $e_{\mathrm{rad}} = \sqrt{\sum_s \omega_{s}^{2} {e_{\mathrm{rad, }s}}^{2}}$. The typical error on the final radial velocity is $\sim \SI{0.16}{\kilo\meter\per\second}$. Figure~\ref{vdsetal2012_bar_radvel_histogram} shows the distribution of the radial velocities in the \ac{LMC} bar; the mean of the $\langle \Substitutev_{\mathrm{rad}} \rangle$ distribution is \SI{261}{\kilo\meter\per\second} and the standard deviation of the distribution is \SI{25}{\kilo\meter\per\second}, in good agreement with values reported in \cite{2005AJ....129.1465C} (\SI{257}{\kilo\meter\per\second}, $\sigma = $\SI{24.7}{\kilo\meter\per\second}).

\begin{table*}
  \begin{center}
    \caption{\label{vdsetal2012_bar_kinematics2_excerpt} Radial velocities of LMC bar stars. 2MASS identifiers, $\Substitutev_{\mathrm{rad}}$, $\sigma(\Substitutev_{\mathrm{rad}})$, number of independent measurements and \acs{SNR} for each setup, final mean $\Substitutev_{\mathrm{rad}}$ and its error. Be aware that the table provides $\sigma(\Substitutev_{\mathrm{rad}})$ and not $e_{\mathrm{rad}, s}$. The exponents l, m and h of the \acs{SNR} indicate respectively whether the spectrum was classified as low, median or high \acs{SNR2}.%
}
    \begin{tabular}{ccccccccccccccc}
      \hline
      \hline
      \multirow{2}{*}{2MASS ID} & \multicolumn{4}{c}{HR11} &  \multicolumn{4}{c}{HR13} &  \multicolumn{4}{c}{HR14} & \multicolumn{2}{c}{Average}\\
\cline{2-15}
                          & $\Substitutev_{\mathrm{rad}}$ & $\sigma(\Substitutev_{\mathrm{rad}})$ & \# & $\mathrm{S} / \mathrm{N}$ & $\Substitutev_{\mathrm{rad}}$ & $\sigma(\Substitutev_{\mathrm{rad}})$ & \# & $\mathrm{S} / \mathrm{N}$ & $\Substitutev_{\mathrm{rad}}$ & $\sigma(\Substitutev_{\mathrm{rad}})$ & \# & $\mathrm{S} / \mathrm{N}$ & $\Substitutev_{\mathrm{rad}}$ & $e_{\mathrm{rad}}$\\
                          & \SI{}{\kilo\meter\per\second} & \SI{}{\kilo\meter\per\second} & & & \SI{}{\kilo\meter\per\second} & \SI{}{\kilo\meter\per\second} & & & \SI{}{\kilo\meter\per\second} & \SI{}{\kilo\meter\per\second} & & & \SI{}{\kilo\meter\per\second} & \SI{}{\kilo\meter\per\second}\\
\hline
05223082-6944147 & 250.0 & 0.2 & 9 & $28^{\mathrm{m}}$ & 250.4 & 0.2 & 5 & $38^{\mathrm{m}}$ & 249.7 & 0.3 & 4 & $50^{\mathrm{m}}$ & 250.1 & 0.2\\
05223112-6945292 & 263.1 & 0.2 & 9 & $32^{\mathrm{m}}$ & 262.9 & 0.4 & 5 & $48^{\mathrm{m}}$ & 262.5 & 0.3 & 4 & $59^{\mathrm{m}}$ & 263.0 & 0.2\\
05223186-6947159 & 271.2 & 0.4 & 9 & $25^{\mathrm{m}}$ & 271.7 & 0.4 & 5 & $36^{\mathrm{m}}$ & 271.2 & 0.2 & 4 & $43^{\mathrm{m}}$ & 271.3 & 0.2\\
05223309-6946595 & 258.3 & 0.4 & 9 & $35^{\mathrm{h}}$ & 258.2 & 1.1 & 5 & $45^{\mathrm{m}}$ & 258.1 & 0.2 & 4 & $51^{\mathrm{m}}$ & 258.3 & 0.2\\
\dots	&	\ldots	&	\ldots	&	\ldots	&	\ldots	&	\ldots	&	\ldots	&	\ldots	&	\ldots  &	\ldots  &	\ldots	&	\ldots &	\ldots &	\ldots	&	\ldots\\
\hline

    \end{tabular}
    \tablefoot{\ExcerptWarning}
  \end{center}
\end{table*}

\paragraph{Co-addition and \acs{SNR}}
\label{Coaddition_and_snr}
To compute the \ac{SNR}, we used the error spectrum produced by the ESO GIRAFFE pipeline, so that the \ac{SNR} at pixel $i$ is given by $\mathrm{SNR}_i = \mathrm{flux}_i / \mathrm{error}_i$ (actually, this estimator of the \ac{SNR} underestimates the \ac{SNR} because the errors are overestimated in the GIRAFFE pipeline; but it is still a good index to compare the quality of a spectrum to another). Before co-adding multiple exposures, we selected the spectra according to their median \ac{SNR}, requiring it to be higher than $\sim \num{3}$. Some observations were taken with the old FLAMES CCD, affected by the so-called glow (polluting light in one corner of the CCD); when necessary, we removed the part of the spectrum altered by this extra source of light. Some observations were obtained with the simultaneous calibration (simcal) lamp turned on: the light of a Th-Ar lamp feeds 5 MEDUSA fibres and allows for small corrections to the wavelength calibration. %
However, some well-known strong emission lines of the Th-Ar gas leak and contaminate the stellar light of the $\sim 5$ science fibres adjacent to a given simcal fibre; we removed these wavelength regions when needed. Once all exposures of the same star were in the same velocity frame, we averaged them with k-$\sigma$ clipping rejection (over the fluxes at a given wavelength) to clean for cosmic rays and increase \ac{SNR}. We ended up with a typical final \ac{SNR} of around \num{25} for HR11, \num{40} for HR13 and \num{48} for HR14. Table~\ref{vdsetal2012_rosetta_stone_snr_hypothesis} lists the typical lowest, median and highest values of \ac{SNR} as well as an empirically corrected \ac{SNR} (see Sec.~\ref{Arcturus_benchmark}).

\subsection{Arcturus as a benchmark star}
\label{Arcturus_benchmark}
To control any systematic effect that could hamper the comparison of our derived abundances to literature measurements, we have tested and applied our methods to the well-known mildly metal-poor Milky-Way thick disc giant Arcturus (HD \num{124897}, $\alpha$ Boo). Indeed, well-known stars such as the Sun, or the giant stars $\mu$Leo or Arcturus are often chosen \citep[\emph{e.g.,}][]{2009A&A...508L..17R, 2009ApJ...704L..66M, 2007A&A...465..799L, 2009MNRAS.400.1039W, 2010A&A...513A..35A} as reference stars for differential analysis, since the literature is broad and provides a good knowledge of their stellar parameters and atmospheric chemical composition (from independent and less model-dependent methods). Arcturus, with $\mathrm{T}_{\mathrm{eff}} = \SI{4286}{\kelvin}$, $\log g = \num{1.66}$ and $\abratio{Fe}{H} = \SI{-0.52}{\dex}$ \citep[][and see references therein]{2011ApJ...743..135R}, is very similar to the stars of our \acs{LMC} sample, hence the choice of this star as a benchmark for our sample.

We sliced the \cite{2000vnia.book.....H} spectral atlas of Arcturus (high resolution $R\sim \num{150000}$, high \ac{SNR} $\sim \num{1000}$) into three pieces to simulate an HR11 (\SI{550}{\nano\meter} to \SI{589}{\nano\meter}), an HR13 (\SI{609}{\nano\meter} to \SI{641}{\nano\meter}) and an HR14 (\SI{629}{\nano\meter} to \SI{671}{\nano\meter}) spectrum. We then degraded the resolution (according to the setup, see Sec.~\ref{Sample_selection}) and sampling of these spectra to reach a best quality spectrum for each setup (referred to as \{low-resolution, low sampling, $\infty$ \acs{SNR2}\} in the following). We finally added Gaussian noise according to the typical noise encountered in our \acs{LMC} sample for each setup, to match four assumptions of \acs{SNR}: an $\infty$ \acs{SNR2}, which is the original quality of the \cite{2000vnia.book.....H} atlas; a high \acs{SNR2}, which corresponds to the median of the ninth decile of the \ac{SNR} distribution (the best 10\% of the sample); a median \acs{SNR2}, which corresponds to the $\sim$ median of the \ac{SNR} distribution; a low \acs{SNR2}, which corresponds to the median of the first decile of the \ac{SNR} distribution (the worst 10\% of the sample).

\begin{figure}
  \begin{centering}
    \includegraphics{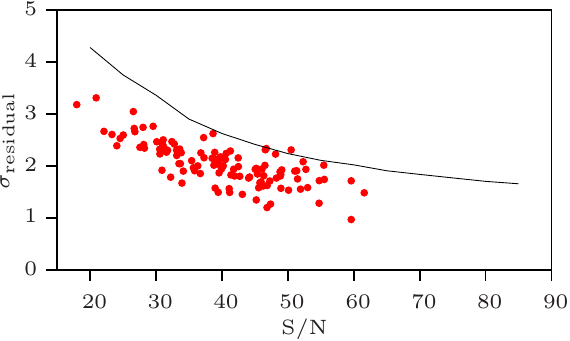}
    \caption{\label{vdsetal2012_residual_snr} $\sigma_{\mathrm{residual}}$ vs. $\mathrm{S}/\mathrm{N}$. Black solid line: we measured $\sigma_{\mathrm{residual}}$ of the Arcturus HR13 spectrum for different values of artificially added noise ${\sigma_{\mathrm{noise}}}^2 = (\mathrm{S}/\mathrm{N})^{-2}$. Red dots: \ac{LMC} HR13 spectra. For Arcturus with artificial noise, $\sigma_{\mathrm{residual}} \simeq 2$ when $\mathrm{S}/\mathrm{N} \simeq 55$ while for our \ac{LMC} spectra, $\sigma_{\mathrm{residual}} \simeq 2$ when $\mathrm{S}/\mathrm{N} \simeq 40$.}
  \end{centering}
\end{figure}

As mentioned in Section~\ref{Coaddition_and_snr}, the \ac{SNR} computed from the GIRAFFE pipeline products (as the ratio of the flux over its propagated error) is not accurate and likely underestimated.
In order to empirically find a correspondence between the measured $\mathrm{S}/\mathrm{N}$ and the genuine $\mathrm{S}/\mathrm{N}$, we employed the automated tool \daospec \citep{2008PASP..120.1332S}, designed to measure \ac{EW}: when it performs this task, the software splits the input spectrum $S$ into a fitted continuum component $C_{f}$ and a fitted line component $L_{f}$ and returns a number $\sigma_{\mathrm{residual}}$ called ``relative flux dispersion in residual spectrum'', which is the dispersion (expressed in percentage) of $\lvert {S}_{i} - ({C_{f}}_{i} + {L_{f}}_{i}) \rvert / {C_{f}}_{i}$ (where $i$ is the pixel index). Therefore, $\sigma_{\mathrm{residual}}$ depends on the \ac{SNR} with an observed dependence as shown in Figure~\ref{vdsetal2012_residual_snr}. 
For each setup and \acs{SNR} regime (low, median or high) observed in our \acs{LMC} spectra, we investigated various values of ${\sigma_{\mathrm{noise}}}^2$ until the $\sigma_{\mathrm{residual}}$ matched the targeted \acs{SNR}. Table~\ref{vdsetal2012_rosetta_stone_snr_hypothesis} gives for each setup, the values of \ac{SNR} (measured and corrected values) corresponding to the qualifiers high, median, and low.

\begin{table}
  \begin{center}
    \caption{\label{vdsetal2012_rosetta_stone_snr_hypothesis} Values of \ac{SNR} corresponding to the qualifiers high (best 10\% of the sample), median, and low (worst 10\% of the sample). $\mathrm{S}/\mathrm{N}$ is the \ac{SNR} measured with the pipeline products and $\mathrm{S}/\mathrm{N}^{*}$ is the empirically corrected \ac{SNR}. %
    }
    \begin{tabular}{lcccccc}
      \hline
      \hline
      \multirow{2}{*}{Qualifier} & \multicolumn{2}{c}{HR11} & \multicolumn{2}{c}{HR13} & \multicolumn{2}{c}{HR14} \\
\cline{2-7}
                           & $\mathrm{S}/\mathrm{N}$ & $\mathrm{S}/\mathrm{N}^{*}$ & $\mathrm{S}/\mathrm{N}$ & $\mathrm{S}/\mathrm{N}^{*}$ & $\mathrm{S}/\mathrm{N}$ & $\mathrm{S}/\mathrm{N}^{*}$ \\
\hline
low       & 16 & 25 & 24 & 35 & 31 & 35 \\
median    & 25 & 40 & 40 & 55 & 48 & 60 \\
high      & 36 & 55 & 55 & 75 & 64 & 75 \\
\hline

    \end{tabular}
  \end{center}
\end{table}

Thus we added a Gaussian noise (with zero-mean and variance ${\sigma_{\mathrm{noise}}}^2 = (\mathrm{S}/\mathrm{N}^{*})^{-2}$) to the three \{low-resolution, low sampling, $\infty$ \acs{SNR2}\} spectra according to \ac{SNR} values listed in Table~\ref{vdsetal2012_rosetta_stone_snr_hypothesis}. We drew \num{101} realisations for each high, median and low \ac{SNR} version of the Arcturus spectra. In the following, we will employ the single $\infty$ \acs{SNR2}, the \num{101} high \acs{SNR2}, the \num{101} median \acs{SNR2}, and the \num{101} low \acs{SNR2} spectra when we determine the stellar parameters (Sec.~\ref{Stellar_parameters}) and when we measure the chemical abundances (Sec.~\ref{Abundance_analysis}).

\section{Stellar parameters}
\label{Stellar_parameters}
To derive the stellar parameters of our \ac{LMC} stars (the temperature $\mathrm{T}_{\mathrm{phot}}$, the gravity $\log g$, the overall metallicity $\abratio{M}{H}$ and the microturbulent velocity $\xi_{\mathrm{micro}}$), we used a combination of photometric and spectroscopic methods. We used our set of Arcturus spectra to assess our iterative procedure and estimate the errors on $\abratio{M}{H}$ and $\xi_{\mathrm{micro}}$.

\subsection{Photometric temperature $\mathrm{T}_{\mathrm{phot}}$}
For our stars, visible (V and I magnitude, from the OGLE catalogue \citealp{1997AcA....47..319U,2000AcA....50..307U,2005AcA....55...43S}) and infrared (J, H and K magnitude, from the 2MASS catalogue \citealp{2006AJ....131.1163S}) photometry is available. Table~\ref{vdsetal2012_bar_photometry_cat_excerpt} gives the V magnitude, the four colour indices we used and the \acs{CaT} metallicity index for our \acs{LMC} bar stars. We used the \citet{2005ApJ...626..446R,2005ApJ...626..465R} photometric calibrations for giants (the calibrations are functions of the colour index and the $\abratio{Fe}{H}$; they are available for our photometric systems, so no conversion of magnitude from one photometric system to another is needed) to compute four scales of photometric temperatures, using the de-reddened $\left(\mathrm{V}-\mathrm{I}\right)_{0}$, $\left(\mathrm{V}-\mathrm{J}\right)_{0}$, $\left(\mathrm{V}-\mathrm{H}\right)_{0}$ and $\left(\mathrm{V}-\mathrm{K}\right)_{0}$ colour indices. Table~\ref{vdsetal2012_comparison_photometric_scales} shows that the agreement between the four photometric temperature scales is very good with a mean difference always smaller than \SI{100}{\kelvin} (in absolute value), therefore we simply averaged the four estimates to derive our final $\mathrm{T}_{\mathrm{phot}}$.

\begin{table*}
  \begin{center}
    \caption{\label{vdsetal2012_bar_photometry_cat_excerpt} Photometry and Cat metallicity of LMC bar stars. 2MASS identifiers, V, $\left(\mathrm{V}-\mathrm{I}\right)$, $\left(\mathrm{V}-\mathrm{J}\right)$, $\left(\mathrm{V}-\mathrm{H}\right)$ and $\left(\mathrm{V}-\mathrm{K}\right)$ \citep{1997AcA....47..319U,2000AcA....50..307U,2005AcA....55...43S} and $\abratio{Fe}{H}_{\mathrm{CaT}}$ \citep{2005AJ....129.1465C}. Errors are provided for each quantity.}
    \resizebox{\textwidth}{!}{%
      \begin{tabular}{ccccccccccccc}
        \hline
        \hline
        2MASS ID                & $\mathrm{V}$ & $e(\mathrm{V})$ & $\mathrm{V} - \mathrm{I}$ & $e(\mathrm{V} - \mathrm{I})$ & $\mathrm{V} - \mathrm{J}$ & $e(\mathrm{V} - \mathrm{J})$ & $\mathrm{V} - \mathrm{H}$ & $e(\mathrm{V} - \mathrm{H})$ & $\mathrm{V} - \mathrm{K}$ & $e(\mathrm{V} - \mathrm{K})$ & $\abratio{Fe}{H}_{\mathrm{CaT}}$ & $e(\abratio{Fe}{H}_{\mathrm{CaT}})$ \\
                        & \SI{}{\magn}  & \SI{}{\magn}  & \SI{}{\magn}  & \SI{}{\magn}  & \SI{}{\magn}  & \SI{}{\magn}  & \SI{}{\magn}  & \SI{}{\magn}  & \SI{}{\magn}  & \SI{}{\magn}  & \SI{}{\dex}   & \SI{}{\dex} \\
\hline
05223082-6944147	&	17.228	&	0.031	&	 1.554	&	 0.038	&	2.702	&	 0.053	&	3.519	&	 0.053	&	3.663	&	0.052	&	-0.14	&	0.14 \\
05223112-6945292	&	17.163	&	0.021	&	 1.584	&	 0.025	&	2.729	&	 0.051	&	3.524	&	 0.059	&	3.676	&	0.050	&	-0.41	&	0.14 \\
05223186-6947159	&	17.450	&	0.030	&	 1.385	&	 0.037	&	2.349	&	 0.064	&	3.129	&	 0.066	&	3.323	&	0.085	&	-0.35	&	0.14 \\
05223309-6946595	&	17.106	&	0.037	&	 1.342	&	 0.041	&	2.268	&	 0.061	&	3.027	&	 0.070	&	2.945	&	0.094	&	-0.40	&	0.14 \\
\ldots                  & \ldots        & \ldots        & \ldots        & \ldots        & \ldots        & \ldots        & \ldots        & \ldots        & \ldots        & \ldots        & \ldots        & \ldots     \\
\hline

      \end{tabular}
    }
    \tablefoot{\ExcerptWarning}
  \end{center}
\end{table*}

\begin{table*}
  \begin{center}
    \caption{\label{vdsetal2012_comparison_photometric_scales} Mean difference between the four photometric scales and standard deviation.}
    \begin{tabular}{lccc}
      \hline
      \hline
      
                                                      & $\mathrm{T}((\mathrm{V}_{\mathrm{Johnson}}-\mathrm{J}_{\mathrm{2MASS}})_{0})$ & $\mathrm{T}((\mathrm{V}_{\mathrm{Johnson}}-\mathrm{H}_{\mathrm{2MASS}})_{0})$ & $\mathrm{T}((\mathrm{V}_{\mathrm{Johnson}}-\mathrm{K}_{\mathrm{2MASS}})_{0})$\\
                                                      & \SI{}{\kelvin} & \SI{}{\kelvin} & \SI{}{\kelvin}\\
\hline
$\mathrm{T}(\mathrm{V}_{\mathrm{Johnson}, 0}-\mathrm{I}_{\mathrm{Cousins}, 0})$ & $-40\pm110$ & $-95\pm\phantom{0}80$ & $-90\pm100$\\
$\mathrm{T}(\mathrm{V}_{\mathrm{Johnson}, 0}-\mathrm{J}_{\mathrm{2MASS}, 0})$  & -- & $-50\pm110$ & $-50\pm120$\\
$\mathrm{T}(\mathrm{V}_{\mathrm{Johnson}, 0}-\mathrm{H}_{\mathrm{2MASS}, 0})$  & -- & -- & $\phantom{+}10\pm100$\\
\hline

    \end{tabular}
  \end{center}
\end{table*}

The photometric calibrations are subject to, at least, four sources of uncertainty: the dispersion $\sigma_{\mathrm{calib}}$ of the calibration relation itself, the uncertainty $\sigma_{\mathrm{colour}}$ of the two magnitudes combined to form the colour index, the uncertainty $\sigma \left( \mathrm{E}(\mathrm{B} - \mathrm{V}) \right)$ of the reddening $\mathrm{E}(\mathrm{B} - \mathrm{V})$, and the uncertainty $\sigma \left( \abratio{Fe}{H} \right)$ of the $\abratio{Fe}{H}$ ratio. The dispersion of the calibration relations can be taken from \cite{2005ApJ...626..465R} (their Table~3); they are smaller than \SI{50}{\kelvin} and account for less than \SI{20}{\kelvin} in the error on the final temperature (for the colour indices we used in this study). The errors on the magnitudes were taken from the OGLE and 2MASS catalogues. The typical error is of the order of \SI{0.05}{\magn}, and the error on the colour index translates in a typical error of $\sim \SI{35}{\kelvin}$ on the final mean $\mathrm{T}_{\mathrm{phot}}$. We used the \ac{CaT} metallicity from \citet{2005AJ....129.1465C} as the initial estimator of the $\abratio{Fe}{H}$ ratio. Although the \ac{CaT} metallicity is not a very precise estimator of $\abratio{Fe}{H}$ (mean error of $\sim \SI{0.20}{\dex}$), the calibrations are not very sensitive to this parameter (typical error of $ < \SI{5}{\kelvin}$ on the final mean $\mathrm{T}_{\mathrm{phot}}$). For the reddening, we used $\mathrm{E}(\mathrm{B} - \mathrm{V}) = \SI{0.14}{\magn}$ and a conservative error $\sigma \left( \mathrm{E}(\mathrm{B} - \mathrm{V}) \right) = \SI{0.07}{\magn}$ (see Sec.~\ref{Choice_reddening}). It results in a typical error on the final mean $\mathrm{T}_{\mathrm{phot}}$ of the order of \SI{130}{\kelvin}. Among the four sources of uncertainty denoted above, the reddening is the least constrained quantity and accounts for most of the final error on the final mean temperature. After propagating all the errors, we end up with a typical error on the mean photometric temperature $\mathrm{T}_{\mathrm{phot}}$ of about \SI{150}{\kelvin}.

\subsection{Surface gravity $\log g$}
The surface gravities $\log g$ were derived using the Bayesian estimation algorithm of stellar parameters of \cite{2006A&A...458..609D}, based on evolutionary tracks\footnote{web interface at \url{http://stev.oapd.inaf.it/cgi-bin/param}}. The required input parameters (and their associated errors) are: the effective temperature, $\abratio{Fe}{H}$, the dereddened V magnitude $\mathrm{V}_{0}$, and the parallax $\pi_{LMC}$. We assumed a constant star formation rate and an initial mass function from \cite{2001ApJ...554.1274C}. We used the photometric temperature as the effective temperature, and the \ac{CaT} metallicity index as an initial guess of $\abratio{Fe}{H}$. The V magnitude was taken from the OGLE catalogue, and was dereddened using the reddening value defined above. The parallax of the \ac{LMC} was set to $(20\pm 1)\times 10^{-6}\mathrm{~arcsec}$ which corresponds to a distance modulus of $(18.5\pm0.1)\SI{}{\magn}$ \citep{2004NewAR..48..659A}. The typical error on $\log g$ returned by the method is of the order of \SI{0.16}{\dex}.

\subsection{Overall metallicity $\abratio{M}{H}$ and microturbulent velocity $\xi_{\mathrm{micro}}$}
\label{Metallicity_microturbulent_velocity}
The overall metallicity and the microturbulent velocity were derived simultaneously by requiring that different \ion{Fe}{I} lines of different \ac{EW} give the same iron abundance $\abratio{\ion{Fe}{I}}{H}$. We used the automated tool \daospec \citep{2008PASP..120.1332S} to measure the \ac{EW} and their associated error, and we used the grid of OSMARCS model atmospheres\footnote{models available at \url{http://marcs.astro.uu.se/}} \citep{2008A&A...486..951G} together with the  spectrum synthesis code \turbospectrum (\turbospectrum is described in \citealp{1998A&A...330.1109A} and improved along the years by B. Plez) to convert the \ac{EW} into abundances. Since our stars are giants, atmosphere models and radiative transfer were both in spherical geometry. We built the atmosphere model for a given set of stellar parameters by interpolation onto the OSMARCS grid with the interpolation routine written by Masseron (2006, PhD thesis).

The iterative procedure is as follows: 
\begin{enumerate}
\item for a given set of stellar parameters $\{ T_{\mathrm{phot}},\log g,\xi_{\mathrm{micro}},\abratio{M}{H} \}$, abundances of around \num{45} \ion{Fe}{I} lines are derived from their \acs{EW}.
\item the mean $\abratio{\ion{Fe}{I}}{H}$ is computed and compared to the input metallicity; if $\left|\langle \abratio{\ion{Fe}{I}}{H} \rangle- \abratio{M}{H}\right| > \SI{0.01}{dex}$, then the global metallicity is updated ($\abratio{M}{H}\leftarrow \langle \abratio{\ion{Fe}{I}}{H}\rangle$) and we go back to step \num{1}. If the convergence is not reached after \num{10} iterations, we release the previous criterion and increase the threshold by \SI{0.01}{dex}.
\item the linear regression of $\abratio{\ion{Fe}{I}}{H}$ vs. $\mathrm{EW'}$ is made, where $\mathrm{EW'}$ is the reduced equivalent width ($\log \mathrm{EW}/\lambda$). As previously noted, the errors on the \ac{EW} are given by \daospec and are turned into errors on $\abratio{\ion{Fe}{I}}{H}$ abundances by \turbospectrum. There is no analytical solution to the problem of linear regression with errors on both coordinates, so various recipes exist to answer this question. In our case, the errors of the \ac{EW} (explanatory variable) and the abundances (dependent variable) are correlated because we used the former to derive the latter. In order to handle the errors on both coordinates as properly as possible, we used a linear regression algorithm based on bootstrapping, as it turns out that the low statistics (number of \ion{Fe}{I} lines) dominate the uncertainty on the slope of the regression.
\end{enumerate}

This procedure is repeated for each value of $\xi_{\mathrm{micro}}$ in the range $\{ {1.0, 1.1, ... , 2.5} \}$ (\SI{}{\kilo\meter\per\second}). We then selected the set of parameters which gives a minimum slope, smaller than its error (in absolute value). The estimate of the error on the metallicity $\abratio{M}{H}$ and the microturbulent velocity $\xi_{\mathrm{micro}}$ is not a straightforward task and a method is proposed in Section~\ref{Arcturus_stellar_parameters}.

\begin{figure}
  \begin{centering}
    \includegraphics{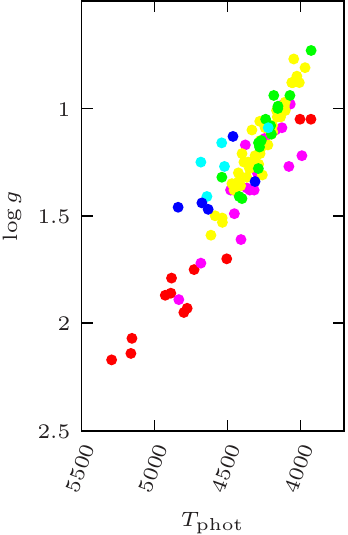}
    \includegraphics{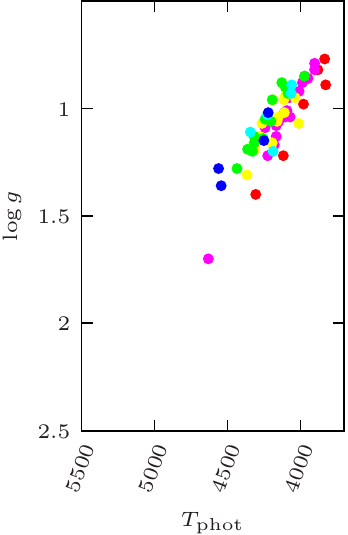}
    \caption{\label{vdsetal2012_hr_diagram_Bar} \label{vdsetal2012_hr_diagram_Disc01} Location of the LMC bar (left panel) and disc (right panel) stars in the Hertzsprung-Russel diagram. Legend: blue: $\SI{-2}{\dex} \le \abratio{\ion{Fe}{I}}{H} \le \SI{-1.3}{\dex}$, cyan: $\SI{-1.3}{\dex} \le \abratio{\ion{Fe}{I}}{H} \le \SI{-1.}{\dex}$, green: $\SI{-1.}{\dex} \le \abratio{\ion{Fe}{I}}{H} \le \SI{-0.8}{\dex}$, yellow: $\SI{-0.8}{\dex} \le \abratio{\ion{Fe}{I}}{H} \le \SI{-0.6}{\dex}$, magenta: $\SI{-0.6}{\dex} \le \abratio{\ion{Fe}{I}}{H} \le \SI{-0.4}{\dex}$, red: $\SI{-0.4}{\dex} \le \abratio{\ion{Fe}{I}}{H} \le \SI{0.}{\dex}$.}
  \end{centering}
\end{figure}

Table~\ref{vdsetal2012_bar_stellar_parameters_excerpt} gives the final stellar parameters for our \ac{LMC} bar stars. For the stars 05223316-6951389, 05225632-6942269, 05225980-6954368, 05224240-6940567, 05232554-6943388, 05244301-6943412, 05253235-6943137, the procedure did not converge onto a satisfactory solution. Figure~\ref{vdsetal2012_hr_diagram_Bar} (left panel) shows the location of the \ac{LMC} bar stars in the Hertzsprung-Russel diagram.

\begin{table*}
  \begin{center}
    \caption{\label{vdsetal2012_bar_stellar_parameters_excerpt} Stellar parameters of LMC bar stars. 2MASS identifiers, $T_{\mathrm{phot}}$, $\log g$, $\abratio{M}{H}$, $\xi_{\mathrm{micro}}$, $\abratio{\ion{Fe}{I}}{H}$, $\abratio{\ion{Fe}{II}}{H}$. Errors are given for each quantity.}
    \resizebox{\textwidth}{!}{%
      \begin{tabular}{ccccccccccccc}
        \hline
        \hline
        2MASS ID & $T_{\mathrm{phot}}$ & $\sigma(T_{\mathrm{phot}})$ & $\log g$ & $\sigma(\log g)$ & $\abratio{M}{H}$ & $\sigma(\abratio{M}{H})$ & $\xi_{\mathrm{micro}}$ & $\sigma(\xi_{\mathrm{micro}})$ & $\abratio{\ion{Fe}{I}}{H}$ & $\sigma(\abratio{\ion{Fe}{I}}{H})$ & $\abratio{\ion{Fe}{II}}{H}$ & $\sigma(\abratio{\ion{Fe}{II}}{H})$\\
      & \SI{}{\kelvin} & \SI{}{\kelvin} & & & \SI{}{\dex} & \SI{}{\dex} & \SI{}{\kilo\meter\per\second} & \SI{}{\kilo\meter\per\second} & \SI{}{\dex} & \SI{}{\dex} & \SI{}{\dex} & \SI{}{\dex}\\
\hline
05223082-6944147	&	4070	&	102	&	0.98	&	0.15	&	-0.49	&	0.10	&	1.80	&	0.15	&	-0.48	&	0.04	&	-0.49	&	0.15\\
05223112-6945292	&	4025	&	98	&	0.85	&	0.14	&	-0.71	&	0.10	&	1.80	&	0.15	&	-0.71	&	0.03	&	-0.65	&	0.11\\
05223186-6947159	&	4277	&	134	&	1.21	&	0.16	&	-0.70	&	0.10	&	1.90	&	0.15	&	-0.71	&	0.03	&	-0.78	&	0.13\\
05223309-6946595	&	4401	&	151	&	1.21	&	0.19	&	-0.68	&	0.10	&	2.00	&	0.15	&	-0.68	&	0.03	&	-0.90	&	0.10\\
\ldots	& \ldots & \ldots & \ldots & \ldots & \ldots & \ldots & \ldots & \ldots	& \ldots & \ldots & \ldots & \ldots\\
\hline

      \end{tabular}
    }
    \tablefoot{\ExcerptWarning}
  \end{center}
\end{table*}

\subsection{Arcturus}
\label{Arcturus_stellar_parameters}
Our set of Arcturus spectra served as a test of our iterative procedure giving the overall metallicity $\abratio{M}{H}$ and the microturbulent velocity $\xi_{\mathrm{micro}}$. To this end, we used the effective temperature and the gravity published by \cite{2011ApJ...743..135R}: $\mathrm{T}_{\mathrm{eff}} = \SI{4286}{\kelvin}$ (spectral energy distribution fitting) and $\log g = \num{1.66}$ (isochrone fitting.  These were kept constant while we applied the iterative procedure described in Section~\ref{Metallicity_microturbulent_velocity} on the \num{101} realisations of high, median and low \acs{SNR2} version of the Arcturus spectra. Table~\ref{vdsetal2012_arcturus_snr_metallicity_microturbulent_velocity} gives the mean $\abratio{M}{H}$, $\abratio{\ion{Fe}{I}}{H}$, $\abratio{\ion{Fe}{II}}{H}$ and $\xi_{\mathrm{micro}}$ and the standard deviation around the mean; results for the $\infty$ \acs{SNR2} are also given for reference. As could be anticipated, results differ according to the \acs{SNR2}, but the differences are mild: the lower the \ac{SNR}, the higher the metallicity (or $\abratio{\ion{Fe}{I}}{H}$, or $\abratio{\ion{Fe}{II}}{H}$), the higher the difference $\Delta = \abratio{\ion{Fe}{I}}{H}$ - $\abratio{\ion{Fe}{II}}{H}$, and the lower the microturbulent velocity. We note that the standard deviation around the mean value increases when the \ac{SNR} decreases, which is again an expected behaviour. Our procedure tends to lead to lower metallicities and higher microturbulent velocity than the reference values in the literature (in fact, in our procedure, the bias in $\abratio{\ion{Fe}{I}}{H}$ varies linearly with the bias in $\xi_{\mathrm{micro}}$), although this effect is paradoxically alleviated at the median and low \acs{SNR2} of our \ac{LMC} sample: \cite{2009MNRAS.400.1039W} found $\abratio{\ion{Fe}{I}}{H} = \SI{-0.6}{\dex}$ and $\xi_{\mathrm{micro}} = \SI{1.5}{\kilo\meter\per\second}$; \cite{2011ApJ...743..135R} found $\abratio{\ion{Fe}{I}}{H} = \SI{-0.52}{\dex}$ and $\xi_{\mathrm{micro}} = \SI{1.74}{\kilo\meter\per\second}$.

\begin{table}
  \begin{center}
    \caption{\label{vdsetal2012_arcturus_snr_metallicity_microturbulent_velocity} Mean and standard deviation of the distribution of $\abratio{M}{H}$, $\abratio{\ion{Fe}{I}}{H}$, $\abratio{\ion{Fe}{II}}{H}$ and $\xi_{\mathrm{micro}}$ for the \num{101} realisations of high, median and low \acs{SNR2} version of the Arcturus spectra. The bottom line give the results for the $\infty$ \acs{SNR2} spectrum.}
    \resizebox{\columnwidth}{!}{%
      \begin{tabular}{lllll}
        \hline
        \hline
        \acs{SNR2} & $\langle \abratio{M}{H} \rangle$ & $\langle \abratio{FeI}{H} \rangle$ & $\langle \abratio{FeII}{H} \rangle$ & $\langle \xi_{\mathrm{micro}} \rangle$ \\
 & \SI{}{\dex} & \SI{}{\dex} & \SI{}{\dex} & \SI{}{\kilo\meter\per\second} \\
\hline
low       & $-0.58 \pm 0.11$                               & $-0.58 \pm 0.11$                               & $-0.49 \pm 0.11$                               & $1.82 \pm 0.15$ \\
median    & $-0.65 \pm 0.06$                               & $-0.65 \pm 0.06$                               & $-0.59 \pm 0.07$                               & $1.83 \pm 0.09$ \\
high      & $-0.69 \pm 0.05$                               & $-0.69 \pm 0.05$                               & $-0.63 \pm 0.05$                               & $1.87 \pm 0.07$ \\
$\infty$  & $-0.71 \phantom{\pm 0.00}$                     & $-0.72 \phantom{\pm 0.00}$                     & $-0.70 \phantom{\pm 0.00}$                     & $1.9 \phantom{\pm 0.00}$ \\
\hline

      \end{tabular}
    }
  \end{center}
\end{table}

Standard deviations reported in Table~\ref{vdsetal2012_arcturus_snr_metallicity_microturbulent_velocity} can also be used as an estimator of the (random) error on the determined metallicity and microturbulent velocity in the \acs{LMC} sample, due to the (random) error on the \acs{EW} measurements (itself originating in the noise present in the stellar spectra). In the following, we will keep the conservative estimates: $\sigma(\abratio{M}{H}) = \SI{0.1}{\dex}$ and $\sigma(\xi_{\mathrm{micro}}) = \SI{0.15}{\kilo\meter\per\second}$.

\subsection{Choice of the reddening}
\label{Choice_reddening}
The mapping of the reddening in the \ac{LMC} has been a longstanding issue, and depending on the targeted stars and the technique used, different reddenings are derived. \cite{2004AJ....128.1606Z} published a reddening map of the \acs{LMC} based on a colour decomposition. We estimated the reddening for our \acs{LMC} bar field from their catalogue\footnote{online \acs{LMC} reddening estimator at \url{http://djuma.as.arizona.edu/~dennis/lmcext.html}}. As all of our stars were not studied by \cite{2004AJ....128.1606Z} and as individual reddening values are reported to be too uncertain, we extracted all of the \cite{2004AJ....128.1606Z} stars located in our field of view (\num{4287} extracted stars for a search radius of \SI{12}{\arcmin}), and computed a median value of the extinction $A_{V}$: ${A_{V}} = \SI{0.44}{\magn}$, hence a median reddening $\mathrm{E}(\mathrm{B}-\mathrm{V}) = {A_{V}}/3.24 \approx \SI{0.14}{\magn}$ \citep{1989ApJ...345..245C}. This value of reddening is similar to what was found by \cite{1999AcA....49..223U} from \ac{RC} stars (comparison of the observed and the theoretical \acs{RC} colour): $\mathrm{E}(\mathrm{B}-\mathrm{V}) \simeq \SI{0.13}{\magn}$ in the bar region. \cite{2012AJ....143...48H} have derived optical reddening maps using two different techniques: \ac{RC} stars and RR Lyrae (comparison of the apparent and intrinsic colour, the latter being computed from the period and  metallicity). They found similar results with the two techniques and their reddening map gives $\mathrm{E}(\mathrm{B}-\mathrm{V}) \simeq \SI{0.06}{\magn}$ and $\sigma \left( \mathrm{E}(\mathrm{B} - \mathrm{V}) \right) = \SI{0.05}{\magn}$ for our bar field. They found good agreement with other works from \citet[][\acs{RC} stars]{2005A&A...430..421S} and \citet[][RR Lyrae]{2009ApJ...704.1730P}. Based on these variations for the reddening in our region, we decided to use a conservative error of \SI{50}{\percent}, $\sigma \left( \mathrm{E}(\mathrm{B} - \mathrm{V}) \right) = \SI{0.07}{\magn}$ to propagate the errors on our stellar parameters (see Sec.~\ref{Stellar_parameters}), which covers the range of possible reddenings in this field.

\begin{figure*}
  \begin{centering}
    \includegraphics{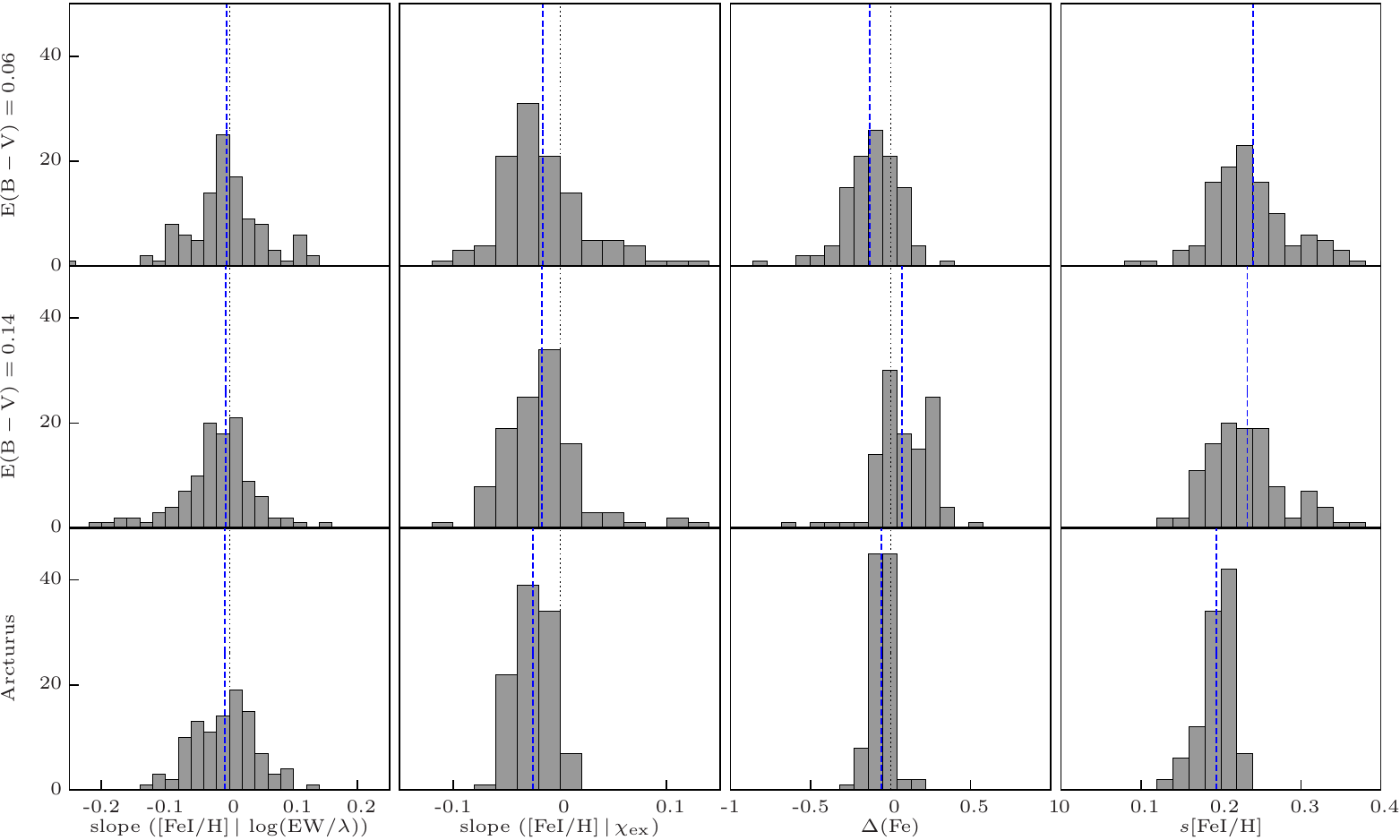}
    \caption{\label{vdsetal2012_test_reddening_spectroscopic_criteria} First and second row: distribution of a given quantity for our \ac{LMC} stars assuming $\mathrm{E}(\mathrm{B}-\mathrm{V}) = \SI{0.06}{\magn}$ and $\mathrm{E}(\mathrm{B}-\mathrm{V}) = \SI{0.14}{\magn}$ respectively. Third row: distribution of a given quantity for the \num{101} median \acs{SNR2} Arcturus spectra. First column: distributions of the slopes ($\abratio{\ion{Fe}{I}}{H}$ $\vert$ $\log (\mathrm{EW}/\lambda)$) (spectroscopic criterion used to derive $\xi_{\mathrm{micro}}$). Second column: distributions of the slopes ($\abratio{\ion{Fe}{I}}{H}$ $\vert$ $\chi_{\mathrm{ex}}$) (excitation equilibrium). Third column: distribution of the difference $\Delta(\mathrm{Fe}) = \abratio{\ion{Fe}{I}}{H} - \abratio{\ion{Fe}{II}}{H}$ (ionisation equilibrium). Fourth column: distribution of the sample standard deviation of $\abratio{\ion{Fe}{I}}{H}$.}
  \end{centering}
\end{figure*}

To further investigate the most probable reddening for our field, we tested two possible values: $\mathrm{E}(\mathrm{B}-\mathrm{V}) = \SI{0.06}{\magn}$ and $\mathrm{E}(\mathrm{B}-\mathrm{V}) = \SI{0.14}{\magn}$. The choice of reddening has a strong effect on the photometric temperature scale, a moderate effect on the microturbulent velocity (comparable to the typical error on the parameter), and a small effect on the gravity and overall metallicity (lower than the typical error): $\langle \mathrm{T}_{\mathrm{phot}} [\mathrm{E}(\mathrm{B}-\mathrm{V}) = 0.14] - \mathrm{T}_{\mathrm{phot}} [\mathrm{E}(\mathrm{B}-\mathrm{V}) = 0.06] \rangle = \SI{140}{\kelvin}$, with a r.m.s of \SI{40}{\kelvin}; $\langle \log g [\mathrm{E}(\mathrm{B}-\mathrm{V}) = 0.14] - \log g [\mathrm{E}(\mathrm{B}-\mathrm{V}) = 0.06] \rangle = \SI{0.03}{}$, with a r.m.s of \SI{0.07}{}; $\langle \abratio{M}{H} [\mathrm{E}(\mathrm{B}-\mathrm{V}) = 0.14] - \abratio{M}{H} [\mathrm{E}(\mathrm{B}-\mathrm{V}) = 0.06] \rangle = \SI{-0.02}{\dex}$, with a r.m.s of \SI{0.12}{\dex}; $\langle \xi_{\mathrm{micro}} [\mathrm{E}(\mathrm{B}-\mathrm{V}) = 0.14] - \xi_{\mathrm{micro}} [\mathrm{E}(\mathrm{B}-\mathrm{V}) = 0.06] \rangle = \SI{0.17}{\kilo\meter\per\second}$, with a r.m.s of \SI{0.13}{\kilo\meter\per\second}. Figure~\ref{vdsetal2012_test_reddening_spectroscopic_criteria} shows the results of the determination of the stellar parameters for the \acs{LMC} bar stars (first row: $\mathrm{E}(\mathrm{B}-\mathrm{V}) = \SI{0.06}{\magn}$; second row: $\mathrm{E}(\mathrm{B}-\mathrm{V}) = \SI{0.14}{\magn}$) and for the median \acs{SNR2} Arcturus spectra (third row). The choice of reddening has a small effect on the distribution of the slopes ($\abratio{\ion{Fe}{I}}{H}$ $\vert$ $\log (\mathrm{EW}/\lambda)$) (first column): for $\mathrm{E}(\mathrm{B}-\mathrm{V}) = \SI{0.06}{\magn}$ and \SI{0.14}{\magn} respectively, the medians are $\SI{-0.008}{\dex}$ and \SI{-0.017}{\dex}, the semi-interquartile ranges are \SI{0.037}{\dex} and \SI{0.030}{\dex} respectively. We note however that the distribution is narrower when the reddening is higher. Similarly, the effect on the distribution of the slopes ($\abratio{\ion{Fe}{I}}{H}$ $\vert$ $\chi_{\mathrm{ex}}$) (excitation equilibrium, second column) is also small: the medians are similar in both cases and close to zero ($\simeq \SI{-0.023}{\dex\per\electronvolt}$), but the distribution is narrower when the reddening is higher: the semi-interquartile ranges are \SI{0.023}{\dex\per\electronvolt} and \SI{0.020}{\dex\per\electronvolt} for $\mathrm{E}(\mathrm{B}-\mathrm{V}) = \SI{0.06}{\magn}$ and \SI{0.14}{\magn} respectively. Whatever the assumed reddening, we see that our photometric scales do not deviate dramatically from excitation equilibrium and that the higher reddening value seems to slightly improve the general trends. The largest effect is observed for the ionisation equilibrium: changing the reddening will shift the distribution of the difference $\Delta(\mathrm{Fe}) = \abratio{\ion{Fe}{I}}{H} - \abratio{\ion{Fe}{II}}{H}$ (ionisation equilibrium, third column). Indeed, the medians are \SI{-0.12}{\dex} (over-ionisation) and \SI{0.06}{\dex} (under-ionisation) for $\mathrm{E}(\mathrm{B}-\mathrm{V}) = \SI{0.06}{\magn}$ and \SI{0.14}{\magn} respectively, with similar semi-interquartile ranges of \SI{0.12}{\dex} and \SI{0.11}{\dex} respectively. The last column of Figure~\ref{vdsetal2012_test_reddening_spectroscopic_criteria} shows that the reddening (thus the temperature) has a small effect on the distribution of the standard deviations of \ion{Fe}{I} abundances (though the situation improves slightly for $\mathrm{E}(\mathrm{B}-\mathrm{V}) = \SI{0.14}{\magn}$: smaller median, less prominent tail in the distribution).  Therefore, the change in reddening has negligible effect on the agreement of \ion{Fe}{I} lines (when the pipeline has converged). As the high reddening tends to slightly improve the determination of parameters (distribution of slopes are narrower, the departure from the ionisation equilibrium is reduced), we decided to adopt this value. 

We checked that results obtained for the \num{101} median \acs{SNR2} Arcturus spectra share the same properties: the distribution of the slopes ($\abratio{\ion{Fe}{I}}{H}$ $\vert$ $\log (\mathrm{EW}/\lambda)$) is centred around zero; the excitation equilibrium is not exactly fulfilled (the median of the distribution is \SI{-0.024}{\dex\per\electronvolt}, similar to the median of our \ac{LMC} sample), and we found a small over-ionisation (the median of $\Delta(\mathrm{Fe})$ is \SI{-0.05}{\dex}).

\subsection{Re-analysis of \cite{2008A&A...480..379P} sample}
In the following, we will compare our results for the \ac{LMC} bar to the results for the \ac{LMC} inner disc published by \cite{2008A&A...480..379P}. \cite{2008A&A...480..379P} chemically analysed \num{59} \acs{LMC} \acs{RGB} field stars located in the \acs{LMC} inner disc, \SI{2}{\degree} South of the bar. Similar to ours, their spectroscopic data were obtained with FLAMES/GIRAFFE in three setups HR$11$, HR$13$ and HR$14$ (they used an older version of the setup definition, where the resolution was slightly higher than what we used for the bar, at the cost of a slightly smaller wavelength coverage). Therefore, the wavelength coverage, the resolution and the \acs{SNR} of the \cite{2008A&A...480..379P} sample and ours are nearly identical. In order to remove systematic effects due to differences in the analysis procedures, we re-analysed the \num{59} stars of the inner disc field and derived a new set of stellar parameters, assuming a reddening $\mathrm{E}(\mathrm{B}-\mathrm{V}) = \SI{0.12}{\magn}$ (computed from \cite{2004AJ....128.1606Z} catalogue as for the bar). We found a good agreement, within the error bars, between our newly derived stellar parameters and those of \cite{2008A&A...480..379P}: $\langle \mathrm{T}_{\mathrm{spec, Pompeia}} - \mathrm{T}_{\mathrm{phot}} \rangle = \SI{-25}{\kelvin}$ ($\mathrm{r.m.s} = \SI{65}{\kelvin}$); $\langle {\log g}_{\mathrm{spec, Pompeia}} - \log g \rangle = \SI{-0.13}{}$ ($\mathrm{r.m.s} = \SI{0.14}{}$); $\langle {\abratio{\ion{Fe}{I}}{H}}_{\mathrm{Pompeia}} - \abratio{\ion{Fe}{I}}{H} \rangle = \SI{-0.06}{\dex}$ ($\mathrm{r.m.s} = \SI{0.15}{\dex}$); $\langle {\abratio{\ion{Fe}{II}}{H}}_{\mathrm{Pompeia}} - \abratio{\ion{Fe}{II}}{H} \rangle = \SI{-0.11}{\dex}$ ($\mathrm{r.m.s} = \SI{0.17}{\dex}$); $\langle \xi_{\mathrm{micro, Pompeia}} -\xi_{\mathrm{micro}} \rangle = \SI{0.05}{\kilo\meter\per\second}$ ($\mathrm{r.m.s} = \SI{0.22}{\kilo\meter\per\second}$). It is remarkable to find such a good agreement between physical quantities (temperature, gravity) derived by different methods (photometry/spectroscopy, isochrone/spectroscopy respectively).

Table~\ref{vdsetal2012_disc01_stellar_parameters_excerpt} gives the final stellar parameters for the \ac{LMC} disc stars. Compared to \cite{2008A&A...480..379P}, our procedure did not converge onto a satisfactory solution for the star \num{0758}. Figure~\ref{vdsetal2012_hr_diagram_Disc01} (right panel) shows the location of the \ac{LMC} disc stars in the Hertzsprung-Russel diagram. In Section~\ref{Pompeia_abundances}, we will use our new set of stellar parameters to re-derive the abundances for the inner disc stars.

\begin{table*}
  \begin{center}
    \caption{\label{vdsetal2012_disc01_stellar_parameters_excerpt} Stellar parameters of LMC disc stars. Star identifiers, $T_{\mathrm{phot}}$, $\log g$, $\abratio{M}{H}$, $\xi_{\mathrm{micro}}$, $\abratio{\ion{Fe}{I}}{H}$, $\abratio{\ion{Fe}{II}}{H}$. Errors are given for each quantity.}
    \resizebox{\textwidth}{!}{%
      \begin{tabular}{ccccccccccccc}
        \hline
        \hline
        2MASS ID & $T_{\mathrm{phot}}$ & $\sigma(T_{\mathrm{phot}})$ & $\log g$ & $\sigma(\log g)$ & $\abratio{M}{H}$ & $\sigma(\abratio{M}{H})$ & $\xi_{\mathrm{micro}}$ & $\sigma(\xi_{\mathrm{micro}})$ & $\abratio{\ion{Fe}{I}}{H}$ & $\sigma(\abratio{\ion{Fe}{I}}{H})$ & $\abratio{\ion{Fe}{II}}{H}$ & $\sigma(\abratio{\ion{Fe}{II}}{H})$\\
      & \SI{}{\kelvin} & \SI{}{\kelvin} & & & \SI{}{\dex} & \SI{}{\dex} & \SI{}{\kilo\meter\per\second} & \SI{}{\kilo\meter\per\second} & \SI{}{\dex} & \SI{}{\dex} & \SI{}{\dex} & \SI{}{\dex}\\
\hline
0499	&	4264	&	117	&	1.07	&	0.15	&	-0.69	&	0.10	&	1.90	&	0.15	&	-0.71	&	0.03	&	-0.78	&	0.07\\
0512	&	4128	&	99	&	0.88	&	0.13	&	-0.91	&	0.10	&	1.80	&	0.15	&	-0.91	&	0.03	&	-0.78	&	0.04\\
0522	&	4101	&	97	&	0.91	&	0.15	&	-0.66	&	0.10	&	1.90	&	0.15	&	-0.67	&	0.03	&	-0.73	&	0.09\\
0533	&	4188	&	107	&	0.96	&	0.15	&	-0.78	&	0.10	&	2.10	&	0.15	&	-0.77	&	0.04	&	-0.81	&	0.09\\
\ldots	& \ldots & \ldots & \ldots & \ldots & \ldots & \ldots & \ldots & \ldots & \ldots & \ldots & \ldots & \ldots\\
\hline

      \end{tabular}
    }
    \tablefoot{\ExcerptWarning}
  \end{center}
\end{table*}

\subsection{$\abratio{Fe}{H}_{\mathrm{CaT}}$ vs. $\abratio{Fe}{H}_{\mathrm{spectro}}$}

\begin{figure}
  \begin{centering}
    \includegraphics{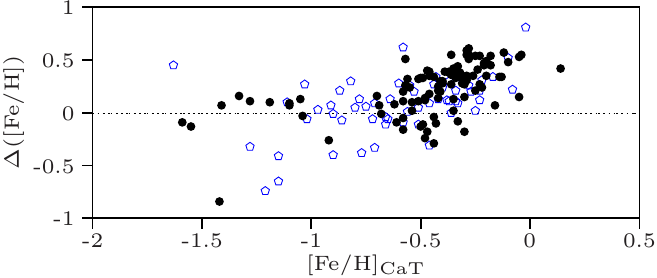}
    \caption{\label{vdsetal2012_comparison_cat_spectroscopy} Comparison of $\abratio{Fe}{H}_{\mathrm{CaT}}$ and $\abratio{Fe}{H}_{\mathrm{spectro}}$. $\Delta(\abratio{Fe}{H}) = \abratio{Fe}{H}_{\mathrm{CaT}} - \abratio{Fe}{H}_{\mathrm{spectro}}$ vs. $\abratio{Fe}{H}_{\mathrm{CaT}}$. Legend: black filled circles: LMC bar; blue open pentagons: LMC disc.%
}
  \end{centering}
\end{figure}

Figure~\ref{vdsetal2012_comparison_cat_spectroscopy} compares for both bar and inner disc fields the $\abratio{\ion{Fe}{I}}{H}$ ratio derived from high-resolution spectroscopy to the metallicity derived from the infrared \acf{CaT} index. 
The typical error bar on $\abratio{Fe}{H}_{\mathrm{CaT}}$ is \num{0.1}-\SI{0.2}{\dex} \citep{2005AJ....129.1465C}, and the typical error bar on $\abratio{Fe}{H}_{\mathrm{spectro}}$ is \SI{0.11}{\dex} (quadratic sum of the typical random and systematic errors on the mean Fe abundance). We see a rather good agreement, within the errors, between the two indices up to $\abratio{Fe}{H}_{\mathrm{CaT}} \approx \SI{-0.5}{\dex}$; then, for higher $\abratio{Fe}{H}_{\mathrm{CaT}}$, we have $\abratio{Fe}{H}_{\mathrm{CaT}} \ge \abratio{Fe}{H}_{\mathrm{spectro}}$. A possible explanation is that for metal-rich stars the continuum placement in the \acs{CaT} region becomes difficult and leads to poor abundance determinations. A possible contribution to the discrepancy could also be due to the presence of stars in the \num{0.8}-\SI {1.2}{\giga\Year} age range in the field samples, where the Red Clump magnitude is changing very quickly and few calibrators of the \acs{CaT} method are available. Based on the trends in the Padova stellar isochrones and with reference to the empirical data in \cite{2004MNRAS.347..367C} there might be a bias of order \SI{+0.1}{\dex} in the \acs{CaT} abundances for stars aged $\sim \SI{1}{\giga\Year}$. Good agreement between \acs{CaT} metallicities \citep[\emph{e.g.},][]{2006AJ....132.1630G, 1991AJ....101..515O} and spectroscopic abundances \citep[\emph{e.g.},][]{2008AJ....136..375M} has been seen for \acs{LMC} \acs{GC} with $\abratio{Fe}{H} = -0.4 \pm 0.1$ and ages around $\sim$\SI{2}{Gyr}. Very few to no \acs{LMC} \acs{GC} with abundances $\abratio{Fe}{H} > -0.3$ are known, so we have no direct tests of the correspondence between the two methods for \acs{LMC} stars. For the remainder of this paper we take the spectroscopic $\abratio{\ion{Fe}{I}}{H}$ to be the true metallicity. 

In the metal-poor range, one \acs{LMC} bar star (05232680-6953109) and four \acs{LMC} disc stars (0606, 0633, 0699, 1105) have very discrepant $\abratio{Fe}{H}_{\mathrm{CaT}}$ and $\abratio{Fe}{H}_{\mathrm{spectro}}$ ($\vert \Delta \vert \ge \SI{0.4}{\dex}$). Except for the disc star 0606, we could not find any anomaly in the stellar parameters determination or the abundance measurements. The star 0606 with $\abratio{Fe}{H} = \SI{-2.07}{\dex}$ has normal $\alpha$-ratios ($\abratio{Ca}{Fe} = \SI{0.39}{\dex}$) but overabundant \emph{s}- and \emph{r}-ratios ($\abratio{Ba}{Fe} = \SI{0.57}{\dex}$, $\abratio{La}{Fe} = \SI{0.51}{\dex}$). This is in agreement with \cite{2008A&A...480..379P} who found $\abratio{Fe}{H} = \SI{-1.74}{\dex}$, $\abratio{Ca}{Fe} = \SI{0.13}{\dex}$ (our \acs{LMC} disc Ca ratios are \SI{+0.1}{\dex} higher in the mean, see Sec.~\ref{Abundance_analysis}), $\abratio{Ba}{Fe} = \SI{0.80}{\dex}$, $\abratio{La}{Fe} = \SI{0.30}{\dex}$. The high fraction of \emph{s}-process elements in this star could be the sign that it is part of a binary system (the \emph{s}-process elements would have been transfer from a former AGB companion).

\section{Abundance analysis}
\label{Abundance_analysis}

\subsection{Abundance measurements}
\paragraph{Methods}
We used both equivalent widths and fitting of absorption profiles to measure elemental abundances. As mentioned in Section~\ref{Metallicity_microturbulent_velocity}, we used \daospec to measure the \acs{EW}. We converted \acs{EW} into abundances, and computed synthetic spectra with \turbospectrum (in spherical geometry, with LTE spherical radiative transfer) together with the grid of OSMARCS spherical model atmospheres.

The fitting of absorption profiles consists of computing a grid of theoretical spectra by varying the abundance of an element, and searching the grid for the best fit to an observed absorption line of the element.  We set up the following procedure:
\begin{enumerate}
\item for a given absorption line $\mathcal{L}$ of an element $\mathrm{X}$, with a central wavelength $\lambda_0$, we define a small wavelength interval $I$ in which the profile fitting is performed. The interval is defined by the compromise between three contradictory requirements: covering as many pixels as possible, avoiding neighbouring lines and including continuum on both sides of the line. The typical width of the wavelength interval considered ranges from \num{3} to \SI{5}{\angstrom}.
\item we compute a grid of theoretical spectra by varying the abundance ratio $\abratio{X}{Fe}$\footnote{$\abratio{X}{Fe} = \abratio{X}{H} - \abratio{Fe}{H}$} with \turbospectrum, from \SI{-1}{\dex} up to \SI{1}{\dex}, by increments of \SI{0.1}{\dex}. We compute the spectra over a wavelength range centred on $\lambda_0$ and convolve them with a Gaussian profile to take into account the combined effects of rotation, macroturbulence and instrumental response.
\item we normalise the theoretical spectra and the observed spectrum in the same way and then compute the quantity: $${T}^{2}(\abratio{X}{Fe}) = \frac{1}{\sum_{i=1}^{n} \mathcal{\hat S}_{i}} \sum_{i=1}^{n} \mathcal{\hat S}_{i} (\mathcal{S(\abratio{X}{Fe})}_{i} - \mathcal{O}_{i})^{2}$$ where $i$ is the pixel index, $n$ is the number of pixel in the interval $I$, $\mathcal{\hat S}$ is the (normalised) theoretical spectrum \emph{without} the element $\mathrm{X}$, $\mathcal{S(\abratio{X}{Fe})}$ is the (normalised) theoretical spectrum for a given value of $\abratio{X}{Fe}$, and $\mathcal{O}$ is the (normalised) observed spectrum. $\mathcal{\hat S}$ allows the weighting of each pixel by its contamination: if the flux at pixel $i$ is only due to the absorption by the element $\mathrm{X}$, then $\mathcal{\hat S}_{i} = 1$; if the flux at pixel $i$ is the result of the absorption by the element $\mathrm{X}$ and by one or more other chemical entities, then $\mathcal{\hat S}_{i} < 1$. Therefore, the more contaminated, the less it counts in ${T}^{2}(\abratio{X}{Fe})$.
\item ${T}^{2}$ is (generally) a convex function of $\abratio{X}{Fe}$; the position of the minimum ${T}^{2}_{\mathrm{nominal}} = {T}^{2}(\abratio{X}{Fe}_{\mathrm{nominal}})$ gives us the best-fit abundance $\abratio{X}{Fe}_{\mathrm{nominal}}$. ${T}^{2}(\abratio{X}{Fe})$ is not a genuine $\chi^{2}$ since we do not divide the quadratic difference $(\mathcal{S(\abratio{X}{Fe})}_{i} - \mathcal{O}_{i})^{2}$ by the error at pixel $i$ (the errors returned by the pipeline do not take into account the correlation) but we can still use it to find the best fit. 
\item the last step aims at accepting or rejecting the solution. Once again, as ${T}^{2}(\abratio{X}{Fe})$ is not a genuine $\chi^{2}$, we cannot apply usual statistics theorems, and for instance, we cannot check the goodness of fit. Therefore, to decide whether the solution has to be rejected, we checked the shape of the ${T}^{2}$ curve according to simple geometric criteria. Indeed, the shape of the ${T}^{2}$ curve is not accidental and reveals the curve of growth of the measured line. For instance, a saturated line is likely to produce a very open curve; a weak line is likely to produce a curve with a well defined minimum, but with a left branch that becomes flat for the smallest abundance; the mismatch between the synthetic and observed spectrum will influence the value of ${T}^{2}_{\mathrm{nominal}}$. So it is easier to work with the normalised ${T}^{2}$ given by $${\hat{T}}^{2}(\abratio{X}{Fe}) = \frac{{T}^{2}(\abratio{X}{Fe}) - {T}^{2}_{\mathrm{nominal}}}{{T}^{2}_{\mathrm{nominal}}}$$

  \emph{Non-detection:} as we cover a broad abundance range ($\abratio{X}{Fe}$ varies from \num{-1} to \SI{+1}{\dex}), we expect (in general) a strong variation of ${T}^{2}$ over this interval. A flat ${T}^{2}$ curve (or at least, if the curve has a completely flat left branch) is the symptom of a non-detection of the line. If the line is very weak, then the profile of the absorption line will slightly change from one abundance point of the grid to the next, at least as long as $\abratio{X}{Fe}$ is small (say $\lessapprox \SI{-0.3}{\dex}$). Therefore, $(\mathcal{S(\abratio{X}{Fe})}_{i} - \mathcal{O}_{i})^{2} \approx \mathrm{constant}$, thus ${T}^{2}(\abratio{X}{Fe}) \approx \mathrm{cst}$. For larger abundances, the line appears in the synthesis and ${T}^{2}$ (or ${\hat{T}}^{2}$) steeply increases. We can detect a flat left (right) branch with this criterion: ${\hat{T}}^{2}(- 1.0) < \epsilon_{1}$ (${\hat{T}}^{2}(+ 1.0) < \epsilon_{1}$, respectively). We empirically fixed $\epsilon_{1}$ to \num{4.0}. In other words, we require ${T}^{2}$ to be five times as high as ${T}^{2}_{\mathrm{nominal}}$ at the grid border for the solution to be meaningful. It may happen that the nominal abundance is close to the grid edge; thus the left (right) branch will not be complete and the solution will be mistakenly rejected. In such cases, we checked the local symmetry of the  ${\hat{T}}^{2}$ curve around the nominal abundance. If the curve is non-symmetric, the solution is rejected. In the mean, the rejection rate is of about ten lines/star; the rejection is minimum for stars with a metallicity between \num{-1} and \SI{-0.5}{\dex} and tends to be higher for metallicities lower than \SI{-1.}{\dex} or larger than \SI{-0.5}{\dex}.
\end{enumerate}

We used the \ac{EW} or the fitting of absorption profile depending on the line: if the number of lines was high ($\ge 5$), we preferred the \acs{EW}; if only few lines were available or if a blend was present or if the element has a hyperfine structure (\acs{hfs}), we preferred the fitting of absorption profile.

\paragraph{List of elements} Because of our broad wavelength coverage, we are in position to measure elemental abundances for \num{17} elements (the method used, \acf{EW} or \acf{SS}, and the number of available lines are given in parenthesis): \ElementDegree{O}{I} (\ac{SS}, \num{1}), \ElementDegree{Mg}{I} (\ac{SS}, \num{3}), \ElementDegree{Si}{I} (\ac{EW}, \num{3}), \ElementDegree{Ca}{I} (\ac{EW}, \num{13}), \ElementDegree{Ti}{I} (\ac{EW}, \num{8}), \ElementDegree{Ti}{II} (\ac{EW}, \num{3}), \ElementDegree{Na}{I} (\ac{SS}, \num{4}), \ElementDegree{Sc}{II} (\ac{SS}, \num{6}), \ElementDegree{V}{I} (\ac{SS}, \num{12}), \ElementDegree{Cr}{I} (\ac{SS}, \num{4}), \ElementDegree{Co}{I} (\ac{SS}, \num{3}), \ElementDegree{Ni}{I} (\ac{EW}, \num{7}), \ElementDegree{Cu}{I} (\ac{SS}, \num{1}), \ElementDegree{Y}{I} (\ac{SS}, \num{1}), \ElementDegree{Zr}{I} (\ac{SS}, \num{4}), \ElementDegree{Ba}{II} (\ac{SS}, \num{2}), \ElementDegree{La}{II} (\ac{SS}, \num{3}), \ElementDegree{Eu}{II} (\ac{SS}, \num{2}). We compiled the atomic line lists from the line database VALD\footnote{\url{http://www.astro.uu.se/~vald/php/vald.php}} \citep{1999A&AS..138..119K, 2000BaltA...9..590K}; for the measured lines, we used the $\log gf$ quoted in \cite{2008A&A...480..379P}. We took into account the \ac{hfs} for \ElementDegree{Sc}{II} (\citealp{1966atp..book.....W}: $\SI{5640}{\angstrom}$, $\SI{5667}{\angstrom}$, $\SI{5669}{\angstrom}$, $\SI{6245}{\angstrom}$; \citealp{1988atps.book.....M}: $\SI{5657}{\angstrom}$, $\SI{6604}{\angstrom}$), \ElementDegree{V}{I} (\citealp{1988atps.book.....M}: \SI{6119}{\angstrom}, \SI{6135}{\angstrom}, \SI{6150}{\angstrom}, \SI{6199}{\angstrom}, \SI{6224}{\angstrom}, \SI{6274}{\angstrom}, \SI{6285}{\angstrom}, \SI{6292}{\angstrom}, \SI{6357}{\angstrom}, \SI{6452}{\angstrom}, \SI{6531}{\angstrom}; \citealp{1988TIAU..XXB.....K}: \SI{6224}{\angstrom}), \ElementDegree{Co}{I} (\citealp{1988JPCRD..17S....F}: $\SI{5647.240}{\angstrom}$, $\SI{6117.000}{\angstrom}$, $\SI{6282.600}{\angstrom}$), \ElementDegree{Cu}{I} (\citealp{1975JQSRT..15..463B}: $\SI{5782.127}{\angstrom}$),  \ElementDegree{Ba}{II} (\citealp{1978SoPh...56..237R}: $\SI{6496.912}{\angstrom}$; no \ac{hfs} data for $\SI{6141.713}{\angstrom}$), \ElementDegree{La}{II} (\citealp{2001ApJ...556..452L}: $\SI{6262.287}{\angstrom}$, $\SI{6390.477}{\angstrom}$; no \ac{hfs} data for $\SI{6320.430}{\angstrom}$), and \ElementDegree{Eu}{II} (\citealp{2001ApJ...563.1075L}: $\SI{6437.640}{\angstrom}$, $\SI{6645.064}{\angstrom}$). We extracted the \ac{hfs} data from the Kurucz database\footnote{\url{http://kurucz.harvard.edu}} \citep{1995ASPC...78..205K} for Sc, V, Co, and Cu; we computed the hyperfine splitting for Ba, La and Eu using the published hyperfine constants. As our stars are cool ($T_{\mathrm{phot}} \sim \SI{4500}{\kelvin}$), molecules in the stellar atmospheres produce absorption bands in the stellar spectra: we included the molecular line lists of ${}^{12}\mathrm{C}{}^{14}\mathrm{N}$, ${}^{13}\mathrm{C}{}^{14}\mathrm{N}$ (Plez, private communication) and TiO \citep{1998A&A...337..495P} in the spectrum synthesis. We consider the solar composition from \cite{1998SSRv...85..161G}. For all our \acs{LMC} stars, we fixed the carbon and nitrogen abundances: $\abratio{C}{Fe} = \SI{-0.65}{\dex}$ and $\abratio{N}{Fe} = \SI{0.3}{\dex}$ (values derived from \citealp{2002AJ....124.3241S}). Knowing $\abratio{C}{Fe}$ and $\abratio{N}{Fe}$ is necessary for the \ce{CN} and \ce{CO} equilibria. The \ce{CO} equilibrium has an effect on the derived \ce{O} abundance. If we assume $\abratio{C}{Fe} = 0. $ and $\abratio{N}{Fe} = 0.$ instead of $\abratio{C}{Fe} = -0.65$ and $\abratio{N}{Fe} = +0.3$, then in the mean, $\abratio{O}{Fe}$ is increased by \SI{0.09}{\dex} (r.m.s. of the difference = \SI{0.04}{\dex}). However, we estimated $\abratio{C}{Fe}$ and $\abratio{N}{Fe}$ ratios from measurements in \acs{LMC} \acs{RGB} stars \citep{2002AJ....124.3241S}, \emph{i.e.} stars in the same evolutionary stage as ours. Therefore, we do not expect an error of \num{0.3} or \SI{0.6}{\dex} on $\abratio{C}{Fe}$ and $\abratio{N}{Fe}$ respectively, and the corresponding systematic error on $\abratio{O}{Fe}$ should be much lower than \SI{0.1}{\dex}. The \ce{CN} equilibrium plays a role in modelling \ce{CN} lines and some \ce{CN} lines contribute to blends. For instance, the Ba line at \SI{6496}{\angstrom} is blended with \ce{CN} lines. However, the assumption on $\abratio{C}{Fe}$ and $\abratio{N}{Fe}$ has a marginal effect on Ba measurements.

\paragraph{Calibration of the line lists}
When a line of interest is blended with one arising from another chemical species (atom or molecule), the abundance measurement becomes more difficult. This is most problematic if the absorption profile of the contaminant is poorly predicted (lack of accurate experimental quantum data or reliable theoretical predictions), as is the case for the CN lines. We therefore calibrated a number of CN lines that contaminate crucial lines of Eu, La, Y, Ba, and Zr using Arcturus.

\subsection{Arcturus}
\label{Arcturus_abudances}
In the following, we will derive the abundances for Arcturus so that it will provide the zeropoint of our abundance scale. In order to have a unique Arcturus atmosphere model for any \acs{SNR2} hypothesis, we chose as stellar parameters $\mathrm{T}_{\mathrm{eff}} = \SI{4286}{\kelvin}$, $\log g = \num{1.66}$, $\abratio{M}{H} = \SI{-0.65}{\dex}$ and $\xi_{\mathrm{micro}} = \SI{1.8}{\kilo\meter\per\second}$: the former two are from \cite{2011ApJ...743..135R} while we determined the latter two in Section~\ref{Arcturus_stellar_parameters} (median \ac{SNR} hypothesis).

We followed the same procedure described above to derive the abundances from our Arcturus spectra. For the high, median, and low \acs{SNR2} hypotheses, we computed a mean abundance and dispersion (over the \num{101} realisations) for each individual line of a given element, and then we computed the final mean abundance ratio (over the $N_\mathrm{lines}$) following the procedure described in Section~\ref{Final_elemental_abundances}. The error on the individual line abundance (dispersion over the \num{101} realisations) was propagated when we computed the final mean abundance. We did the same for the $\infty$ \acs{SNR2} hypothesis (except for the averaging over the realisations); as we have only one realisation for this \acs{SNR2} hypothesis, we used the standard error of the mean as an error estimator (hence the lack of error bar when only one line was used). Table~\ref{vdsetal2012_arcturus_abundances} gives the results for the $\infty$, high, median and low \acs{SNR2} version of the Arcturus spectra as well as the abundance ratios (and their errors) published by \cite{2011ApJ...743..135R} and \cite{2009MNRAS.400.1039W}. 

\begin{table*}
  \begin{center}
    \caption{\label{vdsetal2012_arcturus_abundances} Elemental abundances and errors for our $\infty$, high, median and low \acs{SNR2} version of the Arcturus spectra as well as abundance ratios (and their errors) published by \cite{2011ApJ...743..135R} and \cite{2009MNRAS.400.1039W}. The number of lines used and the method to derive the abundances are recalled.}
    \begin{tabular}{crrrrrcrr}
      \hline
      \hline
      $\abratio{X}{Fe}$ & $\infty$ \acs{SNR2} & high \acs{SNR2} & median \acs{SNR2} & low \acs{SNR2} & \# & Method & Ramirez et al. & Worley et al. \\
 & \SI{}{\dex} & \SI{}{\dex} & \SI{}{\dex} & \SI{}{\dex} &  &  & \SI{}{\dex} & \SI{}{\dex} \\
\hline

\ion{O}{I} & $0.45 \phantom{\pm 0.00}$ & $0.45 \pm 0.03$ & $0.43 \pm 0.04$ & $0.41 \pm 0.06$ & $1$ & SS & $0.50 \pm 0.03$ & $0.57 \pm 0.02$ \\
\ion{Mg}{I} & $0.33 \pm 0.06$ & $0.31 \pm 0.03$ & $0.32 \pm 0.04$ & $0.30 \pm 0.07$ & $3$ & SS & $0.37 \pm 0.03$ & $0.34 \pm 0.15$ \\
\ion{Si}{I} & $0.31 \pm 0.04$ & $0.30 \pm 0.05$ & $0.32 \pm 0.06$ & $0.33 \pm 0.11$ & $2$ & EW & $0.33 \pm 0.04$ & $0.24 \pm 0.14$ \\
\ion{Ca}{I} & $0.03 \pm 0.04$ & $0.04 \pm 0.04$ & $0.04 \pm 0.04$ & $0.03 \pm 0.04$ & $10$ & EW & $0.11 \pm 0.04$ & $0.19 \pm 0.06$ \\
\ion{Ti}{I} & $0.35 \pm 0.07$ & $0.36 \pm 0.02$ & $0.35 \pm 0.03$ & $0.34 \pm 0.04$ & $8$ & EW & $0.27 \pm 0.05$ & $0.35 \pm 0.12$ \\
\ion{Ti}{II} & $0.31 \pm 0.07$ & $0.31 \pm 0.04$ & $0.30 \pm 0.04$ & $0.32 \pm 0.08$ & $3$ & EW & $0.21 \pm 0.04$ & $0.33 \pm 0.10$ \\
\ion{Na}{I} & $0.10 \pm 0.04$ & $0.09 \pm 0.03$ & $0.08 \pm 0.04$ & $0.06 \pm 0.07$ & $3$ & SS & $0.11 \pm 0.03$ & $0.15 \pm 0.04$ \\
\ion{Sc}{II} & $0.25 \pm 0.04$ & $0.23 \pm 0.03$ & $0.23 \pm 0.04$ & $0.21 \pm 0.07$ & $4$ & SS & $0.23 \pm 0.04$ & $0.24 \pm 0.01$ \\
\ion{V}{I} & $0.01 \pm 0.02$ & $0.00 \pm 0.02$ & $-0.01 \pm 0.02$ & $-0.04 \pm 0.03$ & $8$ & SS & $0.20 \pm 0.05$ & -\phantom{ $0.00$} \\
\ion{Cr}{I} & $-0.06 \pm 0.06$ & $-0.07 \pm 0.04$ & $-0.08 \pm 0.05$ & $-0.09 \pm 0.08$ & $3$ & SS & $-0.05 \pm 0.04$ & -\phantom{ $0.00$} \\
\ion{Co}{I} & $0.20 \pm 0.11$ & $0.18 \pm 0.03$ & $0.18 \pm 0.04$ & $0.15 \pm 0.07$ & $2$ & SS & $0.09 \pm 0.04$ & -\phantom{ $0.00$} \\
\ion{Ni}{I} & $0.07 \pm 0.04$ & $0.07 \pm 0.03$ & $0.07 \pm 0.03$ & $0.07 \pm 0.05$ & $6$ & EW & $0.06 \pm 0.03$ & -\phantom{ $0.00$} \\
\ion{Cu}{I} & $-0.15 \phantom{\pm 0.00}$ & $-0.18 \pm 0.06$ & $-0.16 \pm 0.08$ & $-0.18 \pm 0.15$ & $1$ & SS & -\phantom{ $0.00$} & -\phantom{ $0.00$} \\
\ion{Y}{I} & $0.01 \phantom{\pm 0.00}$ & $-0.01 \pm 0.07$ & $-0.02 \pm 0.08$ & $-0.02 \pm 0.13$ & $1$ & SS & -\phantom{ $0.00$} & $0.07 \pm 0.24$ \\
\ion{Zr}{I} & $-0.07 \pm 0.03$ & $-0.09 \pm 0.03$ & $-0.10 \pm 0.04$ & $-0.13 \pm 0.05$ & $3$ & SS & -\phantom{ $0.00$} & $0.01 \pm 0.07$ \\
\ion{Ba}{II} & $-0.19 \pm 0.03$ & $-0.18 \pm 0.06$ & $-0.20 \pm 0.07$ & $-0.21 \pm 0.11$ & $2$ & SS & -\phantom{ $0.00$} & $-0.19 \pm 0.08$ \\
\ion{La}{II} & $-0.04 \pm 0.05$ & $-0.04 \pm 0.04$ & $-0.05 \pm 0.06$ & $-0.06 \pm 0.09$ & $3$ & SS & -\phantom{ $0.00$} & $0.04 \pm 0.08$ \\
\ion{Eu}{II} & $0.40 \pm 0.02$ & $0.41 \pm 0.07$ & $0.40 \pm 0.07$ & $0.39 \pm 0.18$ & $2$ & SS & -\phantom{ $0.00$} & $0.36 \pm 0.04$ \\
\hline

    \end{tabular}
  \end{center}
\end{table*}

Chemical differences appeared for two elements: calcium and vanadium. When we first computed $\abratio{Ca}{Fe}$, we obtained $\approx \SI{-0.12}{\dex}$ which is not the expected ratio for a disc star and is very different from the \cite{2011ApJ...743..135R} and \cite{2009MNRAS.400.1039W} ratios. The $\log gf$ we used had been taken from the NIST\footnote{\url{http://physics.nist.gov/PhysRefData/ASD/index.html}} database and used in \cite{2008A&A...480..379P}. We tested two other sets of $\log gf$: the Kurucz $\log gf$ gave also $\approx \SI{-0.12}{\dex}$ (the Kurucz and the NIST $\log gf$ of our \ion{Ca}{I} lines are almost equal); the VALD $\log gf$ gave $\approx \SI{0.05}{\dex}$, which is closer to the quoted $\abratio{Ca}{Fe}$. We decided to keep the VALD $\log gf$ \citep{1977ZPhys...41..125D, 1981JPhB...14.4015S, 1981A&A...103..351S, 1988JPhB...21.2827S} in order to alleviate the disagreement. Vanadium has a hyperfine structure: when we take into account the \ac{hfs}, $\abratio{V}{Fe} \approx \SI{0.01}{\dex}$, while without the \ac{hfs}, $\abratio{V}{Fe} \approx \SI{0.23}{\dex}$. The latter value is closer to the value that \mbox{\cite{2011ApJ...743..135R}} seemingly derived without taking into account the hyperfine splitting for V (we have five \ion{V}{I} lines in common). So the \ac{hfs} seems to explain the disagreement. In the following, we will derive the V ratios \emph{with} the \ac{hfs}.

Except for V, our derived elemental ratios are in good agreement within the errors with \cite{2011ApJ...743..135R} or \cite{2009MNRAS.400.1039W} and are perfectly understandable if we consider all the possible differences between our study and theirs (stellar parameters, atomic data, method to derive the abundances). We refer the reader to \cite{2012arXiv1209.2656L} who offer a broad analysis of the effects of models, input data and procedures on the derived stellar parameters and chemical composition; for instance, differences of up to $\sim \SI{0.3}{\dex}$ are observed for $\abratio{Ca}{Fe}$ between the different works. 

In our determinations, we note that, in general, when \acs{SNR2} decreases, $\abratio{X}{Fe}$ slightly decreases ($\lessapprox \SI{0.05}{\dex}$) and the error increases. The generally good agreement for Arcturus between our results and the literature makes us confident of the detailed chemical analysis of our \ac{LMC} sample. There is no strong bias and we are able to compare directly the abundance trends of the \ac{LMC} to those of the \ac{MW}, at all \acs{SNR}s.

\subsection{Final elemental abundances}
\label{Final_elemental_abundances}
\paragraph{Computation of the mean abundance}
As shown in the previous section, for a number of elements, two lines or more are available in the full spectral coverage, and we measured all of them whenever possible.

To combine the abundances from multiple lines, we distinguished three different cases to compute the quantity $\langle \abratio{X}{Fe} \rangle$. If $N_{\mathrm{lines}} = 1$, then the final elemental abundance is simply equal to the single measurement. If $2 \le N_{\mathrm{lines}} < 5$, then we computed the simple mean of the $N_{\mathrm{lines}}$ measurements. If  $N_{\mathrm{lines}} \ge 5$, then we applied a $3\sigma$-clipping to remove discrepant measurement, and computed the simple mean of the remaining measurements. The complete abundance table for our \acs{LMC} bar stars is available online at CDS and provides the reader with all abundance ratios and their corresponding random and systematic errors.

\paragraph{Cleaning of the line lists}

We used Arcturus and our \acs{LMC} stars to study the behaviour of each absorption line. It helped us to identify discrepant lines which were then removed from the abundance analysis. In the end, we discarded a few lines for \ion{Ca}{I} (\SI{5601}{\angstrom}, \SI{6162}{\angstrom}, \SI{6572}{\angstrom}), \ion{Cr}{I} (\SI{6362}{\angstrom}), \ion{Co}{I} (\SI{6117}{\angstrom}), \ion{Na}{I} (\SI{5682}{\angstrom}), \ion{Ni}{I} (\SI{6314}{\angstrom}), \ion{Sc}{II} (\SI{5657}{\angstrom}, \SI{6245}{\angstrom}), \ion{Si}{I} (\SI{5665}{\angstrom}), \ion{V}{I} (\SI{6119}{\angstrom}, \SI{6199}{\angstrom}, \SI{6357}{\angstrom}, \SI{6452}{\angstrom}) and \ion{Zr}{I} (\SI{6140}{\angstrom}) and updated the computation of the mean abundances accordingly. We decided to keep in our abundance analysis the Ba line at \SI{6141.713}{\angstrom} (resp. the La line at \SI{6320.430}{\angstrom}) for which no \ac{hfs} data is available since we noted a good agreement with the other Ba (resp. La) line, with a difference of \SI{0.2}{\dex} for Ba (resp. \SI{0.1}{\dex} for La) in the mean (over the whole sample) between the line with and without \acs{hfs}. Table~\ref{vdsetal2012_final_line_list_excerpt} gives the final line list.

\begin{table*}
  \begin{center}
    \caption{\label{vdsetal2012_final_line_list_excerpt} Line list. For each line, the wavelength $\lambda$ (column 4), excitation potential $\chi_{\mathrm{exc}}$ (column 5), oscillator strength $\log gf$ (column 6) and literature reference (column 9) are given. The abundance measurement method is recalled (column 7). If a line has hyperfine structure, the label \emph{equivalent} appear across the column 2 and 3: we provide first the wavelength and oscillator strength of the equivalent line, and below the detailed hyperfine structure for the different isotopes (isotope in column 2, isotopic fraction $f$ in column 3, isotope-scaled $\log gf$ in column 6). The column before the last indicates lines identical to \cite{2008A&A...480..379P}.}
    \begin{tabular}{ccccccccl}
      \hline
      \hline
      Element                       & Isotope                      & $f$           & $\lambda$ & $\chi_{\mathrm{exc}}$ & $\log gf$      & Method &   & Source \\
                              &                              &               & \SI{}{\angstrom} & \SI{}{\electronvolt} &                &        &   &         \\
\hline

\ion{\ce{^{}_{8}O}}{I}        &                              &               &  6300.304 &                 0.000 &         -9.819 & SS     &   & VALD                                 \bigstrut\\
\hline\noalign{\smallskip}\hline

\ion{\ce{^{}_{12}Mg}}{I}      &                              &               &  5711.088 &                 4.346 &         -1.833 & SS     &   & VALD                                 \bigstrut\\

\ion{\ce{^{}_{12}Mg}}{I}      &                              &               &  6318.717 &                 5.108 &         -1.730 & SS     &   & VALD                                 \bigstrut\\

\ion{\ce{^{}_{12}Mg}}{I}      &                              &               &  6319.237 &                 5.108 &         -1.950 & SS     &   & VALD                                 \bigstrut\\
\hline\noalign{\smallskip}\hline

\ion{\ce{^{}_{14}Si}}{I}      &                              &               &  5690.425 &                 4.930 &         -1.870 & EW     & x & \cite{2000AJ....119.1239S}           \bigstrut\\

\ldots                        & \ldots                       & \ldots        &  \ldots   &                \ldots &         \ldots & \ldots & \ldots & \ldots           \bigstrut\\
\hline

    \end{tabular}
    \tablefoot{\ExcerptWarning}
  \end{center}
\end{table*}

\subsection{Re-analysis of \cite{2008A&A...480..379P} sample}
\label{Pompeia_abundances}
To derive the abundances for the \ac{LMC} disc stars, we used the same \ac{EW} and the same reduced spectra that were used by \cite{2008A&A...480..379P}. The differences between their work and ours lie in the stellar parameters and the methods to derive and compute the final abundances. Table~\ref{vdsetal2012_comparison_with_pompeia_abundances} gives a comparison of our new abundances for the \ac{LMC} disc stars and those published in \cite{2008A&A...480..379P}. For most of the elements, the agreement between our abundance ratios and those from \cite{2008A&A...480..379P} is good, with a mean difference less than $\approx \SI{0.15}{\dex}$, \emph{i.e.} of the order of the error. Thus, it is reasonable to attribute the observed differences to the differences in the stellar parameters, and in the measurement of the individual abundances and their combination. For six elements, \ion{Mg}{I}, \ion{Na}{I}, \ion{Sc}{II}, \ion{V}{I}, \ion{Y}{I}, and \ion{Zr}{I}, the differences are larger. Those elements, as well as \ion{Ca}{I}, are discussed below:
\begin{itemize}
\item \ion{Mg}{I}: \cite{2008A&A...480..379P} used the \ion{Mg}{I} line at \SI{5711}{\angstrom} while we used in addition two other lines (\SI{6318}{\angstrom} and \SI{6319}{\angstrom}). If we had used only the line at \SI{5711}{\angstrom}, then $\langle \abratio{Mg}{Fe}_{\mathrm{us}} - \abratio{Mg}{Fe}_{\mathrm{P08}} \rangle = \SI{-0.09}{\dex}$ ($\mathrm{r.m.s} = \SI{0.12}{\dex}$), instead of \SI{-0.22}{\dex}.
\item \ion{Ca}{I}: we recall that we changed the $\log gf$ of the \ion{Ca}{I} lines (see Section~\ref{Arcturus_abudances}). Consequently, all the abundances are shifted by about \SI{0.2}{\dex}. With the old $\log gf$, $\langle \abratio{Ca}{Fe}_{\mathrm{us}} - \abratio{Ca}{Fe}_{\mathrm{P08}} \rangle = \SI{-0.08}{\dex}$; with the new $\log gf$, $\langle \abratio{Ca}{Fe}_{\mathrm{us}} - \abratio{Ca}{Fe}_{\mathrm{P08}} \rangle = \SI{+0.09}{\dex}$.
\item \ion{Na}{I}: \cite{2008A&A...480..379P} used four lines and derived the individual abundances from \ac{EW} while we used only three lines after having discarded the \ion{Na}{I} line at \SI{5862}{\angstrom} (that we found systematically discrepant) and derived the individual abundances from \ac{SS}. If we had used all four lines, then $\langle \abratio{Na}{Fe}_{\mathrm{us}} - \abratio{Na}{Fe}_{\mathrm{P08}} \rangle = \SI{+0.03}{\dex}$ ($\mathrm{r.m.s} = \SI{0.18}{\dex}$).
\item \ion{Sc}{II}: \cite{2008A&A...480..379P} used only the \ion{Sc}{II} line at \SI{5657}{\angstrom} instead of four lines and took into account the \ac{hfs} when deriving the abundance. If we limit ourselves to the line at \SI{5657}{\angstrom}, then $\langle \abratio{Sc}{Fe}_{\mathrm{us}} - \abratio{Sc}{Fe}_{\mathrm{P08}} \rangle = \SI{+0.05}{\dex}$ ($\mathrm{r.m.s} = \SI{0.11}{\dex}$).
\item \ion{V}{I}: as explained in Section~\ref{Arcturus_abudances}, we took into account the \ac{hfs} in the abundance measurement, while \cite{2008A&A...480..379P} did not. This explains the disagreement.
\item \ion{Y}{I}, \ion{Zr}{I}: for \ion{Y}{I} we used the same line and the same method (fitting of line profile) to derive the abundance as \cite{2008A&A...480..379P} did. For \ion{Zr}{I}, \cite{2008A&A...480..379P} used the \ion{Zr}{I} line at \SI{6134}{\angstrom} while we used three lines. But if we restrict the analysis to the same line, we still have $\langle \abratio{Zr}{Fe}_{\mathrm{us}} - \abratio{Zr}{Fe}_{\mathrm{P08}} \rangle = \SI{+0.46}{\dex}$ ($\mathrm{r.m.s} = \SI{0.21}{\dex}$). For those two elements, the lines are weak and difficult to measure. Therefore, the abundance measurement is likely less robust and more sensitive to the method (\emph{e.g.} the wavelength range where the synthesis is compared to the data, the continuum placement).
\end{itemize}

In addition, we derived the Eu abundances for the \ac{LMC} disc stars. The wavelength coverage of \cite{2008A&A...480..379P}'s spectra is not exactly the same as ours since the setup HR14 they used was different. Consequently, the \ion{Eu}{II} line at \SI{6645}{\angstrom} is not available; but the \ion{Eu}{II} line at \SI{6437}{\angstrom} is present. Although this line is weaker than the other, we could use it successfully for most of the \ac{LMC} disc stars. The complete abundance table for our \acs{LMC} disc stars is available online at CDS and provides the reader with all abundance ratios and their corresponding random and systematic errors.

\begin{table}
  \begin{center}
    \caption{\label{vdsetal2012_comparison_with_pompeia_abundances} Comparison of our new abundances for the \ac{LMC} disc stars and those published in \cite{2008A&A...480..379P}: mean $m$ and r.m.s $s$ of the distribution of $\abratio{X}{Fe}_{\mathrm{us}} - \abratio{X}{Fe}_{\mathrm{P08}}$.}
    \begin{tabular}{ccc}
      \hline
      \hline
      Element & $m$ & $s$ \\
 & \SI{}{\dex} & \SI{}{\dex} \\
\hline
\ion{O}{I}	& -0.12	& 0.13\\
\ion{Mg}{I}	& -0.22	& 0.14\\
\ion{Si}{I}	& +0.11	& 0.12\\
\ion{Ca}{I}	& +0.09	& 0.12\\
\ion{Ti}{I}	& +0.12	& 0.14\\
\ion{Ti}{II}	& +0.15	& 0.11\\
\ion{Na}{I}	& -0.15	& 0.17\\
\ion{Sc}{II}	& +0.12	& 0.14\\
\ion{V}{I}	& -0.25	& 0.17\\
\ion{Cr}{I}	& +0.07	& 0.11\\
\ion{Co}{I}	& -0.01	& 0.13\\
\ion{Ni}{I}	& +0.06	& 0.09\\
\ion{Cu}{I}	& -0.12	& 0.15\\
\ion{Y}{I}	& +0.24	& 0.21\\
\ion{Zr}{I}	& +0.43	& 0.18\\
\ion{Ba}{II}	& +0.04	& 0.17\\
\ion{La}{II}	& +0.02	& 0.11\\

\hline

    \end{tabular}
  \end{center}
\end{table}

\subsection{Error budget}
Four main sources of uncertainty exist: uncertainties on the atomic data describing the measured lines, uncertainties due to the modelling of the absorption line, uncertainties on the abundance measurement (for both \acs{EW} or \acs{SS}, due to the noise in the fluxes, the continuum placement, the profile integration or profile fitting, and if the line is blended, the hypothesis on the contaminant abundance), and uncertainties on the stellar parameters.

\begin{figure}
  \begin{centering}
    \includegraphics{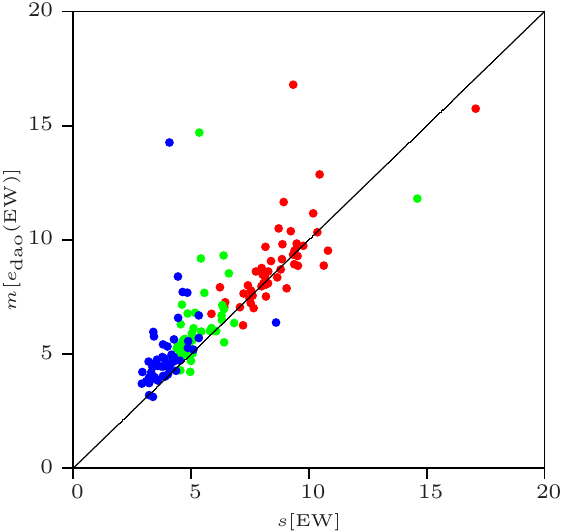}

    \includegraphics{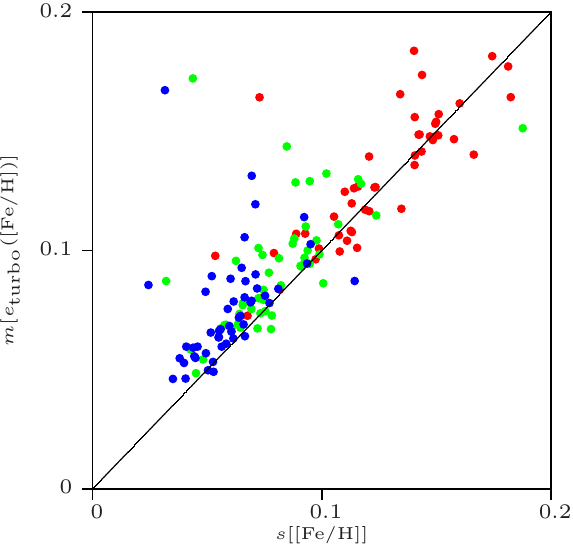}
    \caption{\label{vdsetal2012_comparison_error_daospec_monte_carlo} Left panel: $m[e_{\mathrm{dao}}(\mathrm{EW})]$ vs. $s[\mathrm{EW}]$. Righ panel: $m[e_{\mathrm{turbo}}(\abratio{Fe}{H})]$ vs. $s[\abratio{Fe}{H}]$. Red dots: low \acs{SNR2}; green dots: median \acs{SNR2}; Blue dots: high \acs{SNR2}.}
  \end{centering}
\end{figure}

\paragraph{Abundance measurement}
\daospec provides us with an error on the \ac{EW}, which is obtained during the least-square fit of the line. As mentioned in \cite{2008PASP..120.1332S}, this error is not a genuine $1\sigma$ confidence interval (\emph{e.g.}, the correlation between the pixels is not taken into account). We checked it using our Arcturus spectra and the set of \ion{Fe}{I} lines (51 lines measured which cover a broad range of line strengths and wavelengths). For each \acs{SNR2} hypothesis and for each \ion{Fe}{I} line, we computed the sample standard deviation $s[\mathrm{EW}]$ of the \ac{EW} distribution, as well as the mean $m[e_{\mathrm{dao}}(\mathrm{EW})]$ of the error returned by \daospec. $s[\mathrm{EW}]$ is a good estimator of the error on the \ac{EW} since it encompasses the effect of the noise in the fluxes and the continuum placement. Figure~\ref{vdsetal2012_comparison_error_daospec_monte_carlo} shows the comparison of $m[e_{\mathrm{dao}}(\mathrm{EW})]$ and $s[\mathrm{EW}]$. There is a fairly good agreement between the two: the mean of $(s[\mathrm{EW}] - 
m[e_{\mathrm{dao}}(\mathrm{EW})])$ is \SI{-0.18}{\milli\angstrom}, \SI{-0.54}{\milli\angstrom}, \SI{-0.76}{\milli\angstrom} for the low, median and high \acs{SNR2} respectively; when the Monte-Carlo simulation predicts large errors, \daospec does also; the error decreases when the \ac{SNR} increases. In the mean, \daospec tends to mildly overestimate the error bar, especially when the \ac{SNR} gets better. So it is reasonable to use the error computed by \daospec.

Another pitfall is the conversion of the error on the \ac{EW} into an error on the abundance. Indeed, when we feed \turbospectrum with the pair $(\mathrm{EW}, e_{\mathrm{dao}}(\mathrm{EW}))$, it computes the abundances corresponding to $\mathrm{EW}$, and $\mathrm{EW} \pm e_{\mathrm{dao}}(\mathrm{EW})$ and often provides asymmetric (right and left) errors. This is not \emph{a priori} a proper way to find the error on the abundance since the relationship between $\abratio{X}{Fe} \pm e(\abratio{X}{Fe})$ and $\mathrm{EW} \pm e_{\mathrm{dao}}(\mathrm{EW})$ is not known. We performed similar tests for the abundances as we did for \ac{EW} in the previous paragraph. For each \acs{SNR2} hypothesis and for each \ion{Fe}{I} line, we computed the sample standard deviation $s[\abratio{Fe}{H}]$ of the $\abratio{Fe}{H}$ distribution, as well as the mean $m[e_{\mathrm{turbo}}(\abratio{Fe}{H})]$ of the error returned by \turbospectrum. Figure~\ref{vdsetal2012_comparison_error_daospec_monte_carlo} shows the comparison of $m[e_{\mathrm{turbo}}(\abratio{Fe}{H})]$ and $s[\abratio{Fe}{H}]$. We obtain a similar pattern for the abundances as for the \ac{EW}: the agreement is fairly good but the errors tend to be mildly overestimated when the \ac{SNR} increases (though the effect is $< \SI{0.05}{\dex}$ at high \acs{SNR2}). Here again, we consider it safe to keep the error returned by \turbospectrum (\emph{i.e.}, the mean of the right and left errors).

Unfortunately, for the lines measured by absorption line fitting, we cannot use classical theorems to derive an error on the abundance measurement. Indeed, ${T}^{2}$ is not a random variable which follows a $\chi^{2}$ distribution since we do not divide each term of the quadratic sum by the error on the flux at pixel $i$ (the GIRAFFE pipeline certainly provides an error for each pixel but it is overestimated and correlated, see Section~\ref{Coaddition_and_snr}) and the $\mathcal{O}_{i}$ are correlated due to the interpolation or rebinning performed during the data reduction. One way to get an estimator of the the $1\sigma$ error is to do Monte-Carlo simulations. We used the Arcturus spectra to estimate the error $e_{\mathrm{Arcturus}}(\abratio{X}{Fe})$ on each single line (by computing the standard deviation of the abundance distribution) and to assign the error to the \ac{LMC} stars depending on the \ac{SNR} category in which they fall.

For a given element $\mathrm{X}$, we propagated the errors on the individual lines $e_{\mathrm{turbo}}(\abratio{X}{Fe})$ or $e_{\mathrm{Arcturus}}(\abratio{X}{Fe})$, which gave us $e_{\mathrm{prop}}(\langle \abratio{X}{Fe} \rangle)$.

\paragraph{Atomic data and line modelling}
Our capacity to model an absorption line correctly, and thus to measure the abundances accurately, depends on the quality of the atomic data describing the radiative transitions but also on our understanding of the underlying physics. Line lists are often a compilation of various sources aiming at giving the best parameters for a given line, and therefore, the precision of these parameters (among which $\log gf$ holds the main role) varies from line to line. The resulting synthetic spectrum is model-dependent (systematic error due to the choice of the grid of model atmospheres, the assumptions on the thermodynamic equilibrium, the atom models) and data-dependent (random error due to the $\log gf$ provided by the line lists). The sample dispersion $s[\abratio{X}{Fe}]$ of the individual abundances about the mean can be used to estimate the combination of these effects (if enough lines are available to estimate it). We derived conservative errors as follows:
\begin{itemize}
\item if $N_{\mathrm{lines}} < 5$: $$e_{\mathrm{data}}(\langle \abratio{X}{Fe} \rangle) = e_{\mathrm{prop}}(\langle \abratio{X}{Fe} \rangle)$$
\item if $N_{\mathrm{lines}} \ge 5$: $$e_{\mathrm{data}}(\langle \abratio{X}{Fe} \rangle) = \max \left( \frac{s[\abratio{X}{Fe}]}{\sqrt{N_{\mathrm{lines}}}}, e_{\mathrm{prop}}(\langle \abratio{X}{Fe} \rangle) \right)$$
\end{itemize}

To assess our method of error estimation, we compared the standard error of the mean to the propagated error for Ca, Ni, Sc and V. We recall that for these elements, we have enough lines to compute a meaningful variance, and that we derived Ca and Ni abundances from \ac{EW} and Sc and V abundances from \ac{SS}. We found a median difference of $\sim \SI{0.02}{\dex}$: thus, this check validates the use of $e_{\mathrm{turbo}}(\abratio{X}{Fe})$ or $e_{\mathrm{Arcturus}}(\abratio{X}{Fe})$ for the individual measurements.

\paragraph{Stellar parameters}
The error $e_{\mathrm{params}}$ on chemical abundances due to the adopted stellar parameters is a thorny question. The four stellar parameters are mutually dependent and changing one of them will imply a change of the others (see \citealp{1995AJ....109.2757M,2006ApJ...640..801J} for a discussion on covariance terms). When the propagation of error is not straightforward, a usual practise is to perturb the explanatory variable (input) by $\pm$~its error and to look at the corresponding shift of the dependent variable (output). For the abundances, it would come down to repetition of this procedure for each parameter, keeping the other three constant. The pitfall is to work with a set of parameters that do not satisfactorily describe the atmosphere of the star under study. For instance, when the temperature is changed by, say, $\SI{150}{\kelvin}$, and $\{\log g, \abratio{M}{H}, \xi_{\mathrm{micro}}\}_{\mathrm{nominal}}$ (which were found for the nominal temperature) are kept, it is likely that the spectroscopic criterion used to find $\xi_{\mathrm{micro}}$ does not hold anymore and therefore the determination of abundances from strong lines will not be correct. We followed the prescription from \cite{2004A&A...416.1117C}: as $\mathrm{T}_{\mathrm{phot}}$ has the major effect on the abundance determination, we change it by $\pm$~its error and determine the three other stellar parameters corresponding to this new temperature $\{\log g, \abratio{M}{H}, \xi_{\mathrm{micro}}\}_{\pm\sigma(\mathrm{T}_{\mathrm{phot}})}$; we derive the chemical abundances corresponding to this perturbed solution and compare them to those given by the nominal solution. The final systematic error on $\abratio{A}{B}$ due to errors on effective temperature is then given by:
\begin{align*}
  e_{\mathrm{params}} = \max & \left( \lvert \abratio{A}{B}_{+\sigma(\mathrm{T}_{\mathrm{phot}})} - \abratio{A}{B}_{\mathrm{nominal}} \rvert, \right.\\
  & \left. \lvert \abratio{A}{B}_{-\sigma(\mathrm{T}_{\mathrm{phot}})} - \abratio{A}{B}_{\mathrm{nominal}} \rvert \right)
\end{align*}
and the total error by:
\begin{align*}
  e_{\mathrm{total}} = \sqrt{e_{\mathrm{data}}^{2} + e_{\mathrm{params}}^{2}}
\end{align*}

Table~\ref{vdsetal2012_typical_errors2} gives the typical (i.e. the mean over the sample) $e_{\mathrm{data}}$ and $e_{\mathrm{params}}$ (given as $\abratio{A}{B}_{+\sigma(\mathrm{T}_{\mathrm{phot}})} - \abratio{A}{B}_{\mathrm{nominal}}$ and $\abratio{A}{B}_{-\sigma(\mathrm{T}_{\mathrm{phot}})} - \abratio{A}{B}_{\mathrm{nominal}}$) for different elemental ratios for our \ac{LMC} bar stars. In the vast majority of cases, the errors due to stellar parameters dominate over the random measurement errors. Both these sources of error are plotted in Fig.~\ref{vdsetal2012_OvsFe_MgvsFe_OMgvsFe}-\ref{vdsetal2012_CuvsFe}.

\begin{table}
  \begin{center}
    \caption{\label{vdsetal2012_typical_errors2} Typical $e_{\mathrm{data}}$ and $e_{\mathrm{params}}$, given as $\abratio{A}{B}_{-\sigma(\mathrm{T}_{\mathrm{phot}})} - \abratio{A}{B}_{\mathrm{nominal}}$ and $\abratio{A}{B}_{+\sigma(\mathrm{T}_{\mathrm{phot}})} - \abratio{A}{B}_{\mathrm{nominal}}$, for different elemental ratios for our \ac{LMC} bar stars.}
    \resizebox{\columnwidth}{!}{%
      \begin{tabular}{lccc}
        \hline
        \hline
        Elemental ratio & $e_{\mathrm{data}}$ & $e_{\mathrm{params}}(-\sigma(\mathrm{T}_{\mathrm{phot}}))$ & $e_{\mathrm{params}}(+\sigma(\mathrm{T}_{\mathrm{phot}}))$\\
                & dex              & dex & dex\\
\hline

$\abratio{\ion{Fe}{I}}{\ion{H}{}}$		& 0.03	& 0.04	& 0.04\\
$\abratio{\ion{Fe}{II}}{\ion{H}{}}$		& 0.07	& 0.23	&-0.12\\
$\abratio{\ion{O}{I}}{\ion{Fe}{I}}$		& 0.04	&-0.10	& 0.06\\
$\abratio{\ion{Mg}{I}}{\ion{Fe}{I}}$		& 0.04	&-0.03	& 0.01\\
$\abratio{\ion{OI+MgI}{}}{\ion{Fe}{I}}$		& 0.03	&-0.07	& 0.04\\
$\abratio{\ion{Si}{I}}{\ion{Fe}{I}}$		& 0.08	& 0.03	&-0.05\\
$\abratio{\ion{Ca}{I}}{\ion{Fe}{I}}$		& 0.04	&-0.09	& 0.05\\
$\abratio{\ion{Ti}{I}}{\ion{Fe}{I}}$		& 0.07	&-0.20	& 0.17\\
$\abratio{\ion{Ti}{II}}{\ion{Fe}{I}}$		& 0.04	& 0.02	&-0.04\\
$\abratio{\ion{Na}{I}}{\ion{Fe}{I}}$		& 0.04	&-0.14	& 0.05\\
$\abratio{\ion{Sc}{II}}{\ion{Fe}{I}}$		& 0.04	&-0.03	& 0.00\\
$\abratio{\ion{V}{I}}{\ion{Fe}{I}}$		& 0.03	&-0.27	& 0.24\\
$\abratio{\ion{Cr}{I}}{\ion{Fe}{I}}$		& 0.05	&-0.13	& 0.11\\
$\abratio{\ion{Co}{I}}{\ion{Fe}{I}}$		& 0.04	&-0.12	& 0.10\\
$\abratio{\ion{Ni}{I}}{\ion{Fe}{I}}$		& 0.05	&-0.04	& 0.03\\
$\abratio{\ion{Cu}{I}}{\ion{Fe}{I}}$		& 0.08	&-0.10	& 0.08\\
$\abratio{\ion{Y}{I}}{\ion{Fe}{I}}$		& 0.08	&-0.40	& 0.31\\
$\abratio{\ion{Zr}{I}}{\ion{Fe}{I}}$		& 0.04	&-0.28	& 0.26\\
$\abratio{\ion{Ba}{II}}{\ion{Fe}{I}}$		& 0.07	& 0.06	&-0.06\\
$\abratio{\ion{La}{II}}{\ion{Fe}{I}}$		& 0.06	&-0.04	& 0.04\\
$\abratio{\ion{YI+ZrI}{}}{\ion{BaII+LaII}{}}$		& 0.06	&-0.33	& 0.29\\
$\abratio{\ion{Eu}{II}}{\ion{Fe}{I}}$		& 0.07	&-0.05	& 0.02\\
$\abratio{\ion{Ba}{II}}{\ion{Eu}{II}}$		& 0.11	& 0.08	&-0.08\\
$\abratio{\ion{La}{II}}{\ion{Eu}{II}}$		& 0.09	& 0.00	& 0.02\\
\hline

      \end{tabular}
    }
  \end{center}
\end{table}

\section{Results and discussion}
\label{Results_discussion}
In this section, we present the results for the key elements: O, Mg, Si, Ca, Ti ($\alpha$-elements), Na (light odd element), Sc, V, Cr, Co, Ni, Cu (iron-peak elements), Y, Zr, Ba, La and Eu (\emph{s}- and \emph{r}-elements). We compare our results for the \ac{LMC} field stars (bar and inner disc) to \ac{LMC} \ac{GC} stars \citep{2006ApJ...640..801J, 2008AJ....136..375M, 2010ApJ...717..277M}, and to the \ac{MW} stellar populations (thin and thick disc \citealp{2005A&A...433..185B, 2003MNRAS.340..304R, 2006MNRAS.367.1329R}; halo \citealp{2000AJ....120.1841F, 2002AJ....123.1647S, 2006MNRAS.367.1329R}; Eu and La ratios: \citealp{2004ApJ...617.1091S, 2006AJ....131..431B}; O ratios of halo stars: \citealp{2000A&A...356..238C}). Our results for Arcturus are plotted as well to check our abundance scale (Arcturus) versus the literature abundance scales (the \acs{MW} thick disc compilation).

\subsection{$\alpha$-elements}
\label{Results_alpha}
\begin{figure}
  \begin{centering}
    \includegraphics{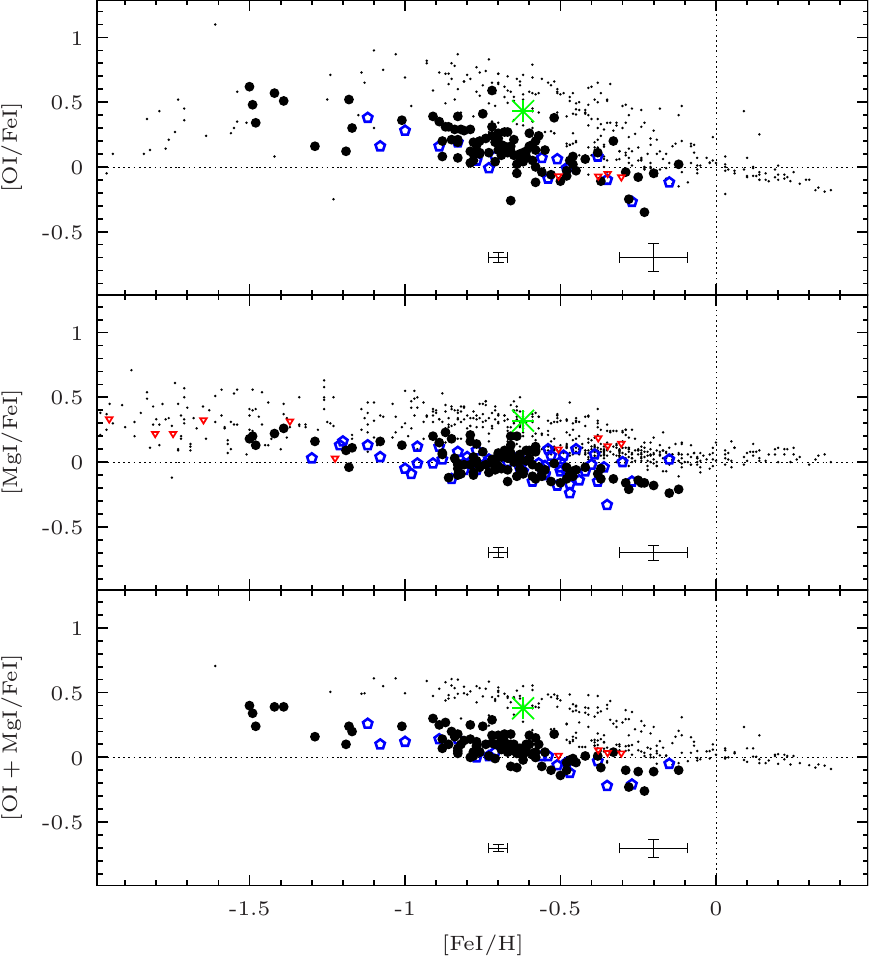}
    \caption{\label{vdsetal2012_OvsFe_MgvsFe_OMgvsFe} First row: $\abratio{\ion{O}{I}}{\ion{Fe}{I}}$ vs. $\abratio{\ion{Fe}{I}}{H}$. Second row: $\abratio{\ion{Mg}{I}}{\ion{Fe}{I}}$ vs. $\abratio{\ion{Fe}{I}}{H}$. Third row: $\abratio{\ion{O}{I} + \ion{Mg}{I}}{2\ion{Fe}{I}}$ vs. $\abratio{\ion{Fe}{I}}{H}$. Legend: black filled circles: \acs{LMC} bar (this work); blue open pentagons: \acs{LMC} inner disc (this work); green asterisk: Arcturus (this work, data for median \acs{SNR}); red downward triangle: \acs{LMC} \acs{GC} \citep{2006ApJ...640..801J, 2008AJ....136..375M, 2010ApJ...717..277M}; black tiny dots: \acs{MW} thin and thick disc \citep{2005A&A...433..185B, 2003MNRAS.340..304R, 2006MNRAS.367.1329R}, halo \citep{2000AJ....120.1841F, 2002AJ....123.1647S, 2006MNRAS.367.1329R}, and additional \acs{MW} data for O from \cite{2000A&A...356..238C}. Typical random (left) and systematic (right) error bars on both coordinates are provided for our \acs{LMC} samples.}
  \end{centering}
\end{figure}

\begin{figure}
  \begin{centering}
    \includegraphics{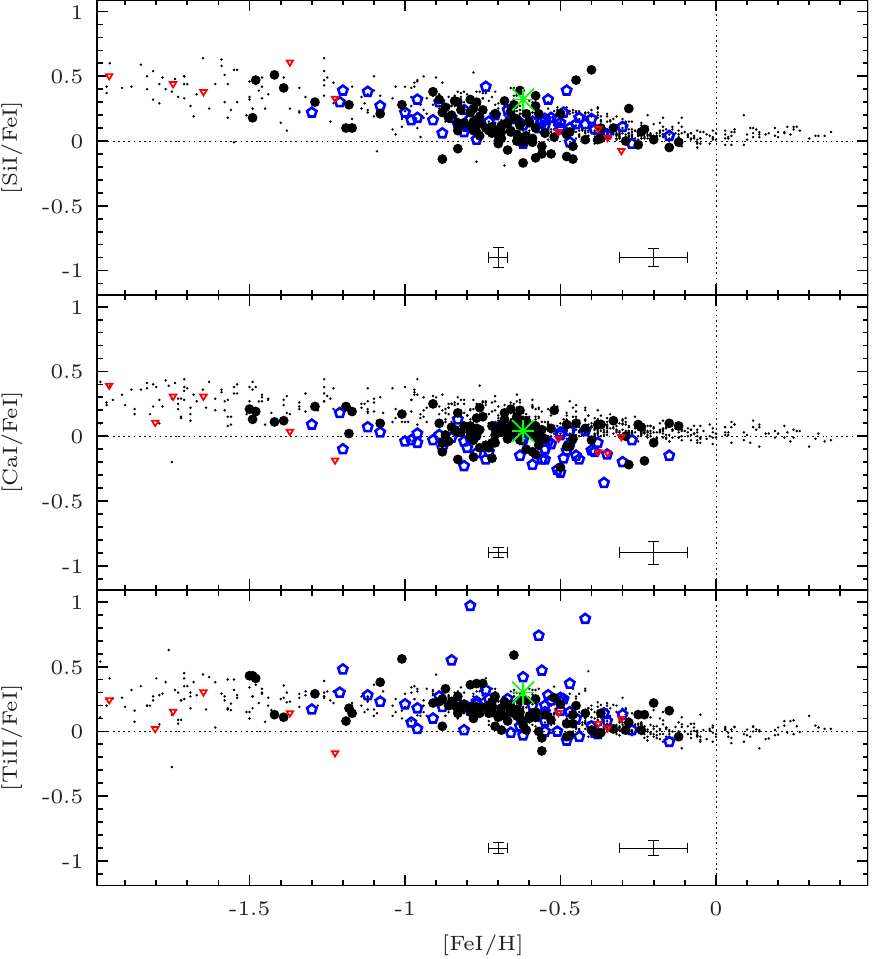}
    \caption{\label{vdsetal2012_SivsFe_CavsFe_TivsFe} First row: $\abratio{\ion{Si}{I}}{\ion{Fe}{I}}$ vs. $\abratio{\ion{Fe}{I}}{H}$. Second row: $\abratio{\ion{Ca}{I}}{\ion{Fe}{I}}$ vs. $\abratio{\ion{Fe}{I}}{H}$. Third row: $\abratio{\ion{Ti}{II}}{\ion{Fe}{I}}$ vs. $\abratio{\ion{Fe}{I}}{H}$. Legend: same as Figure~\ref{vdsetal2012_OvsFe_MgvsFe_OMgvsFe}.}
  \end{centering}
\end{figure}

Figures~\ref{vdsetal2012_OvsFe_MgvsFe_OMgvsFe} and \ref{vdsetal2012_SivsFe_CavsFe_TivsFe} show the abundance trends for $\abratio{O}{Fe}$, $\abratio{Mg}{Fe}$, $\abratio{Si}{Fe}$, $\abratio{Ca}{Fe}$ and $\abratio{Ti}{Fe}$. O, Mg, Si, Ca and Ti belong to the $\alpha$-elements and are used to track the epoch when \ac{SNeII} drove the chemical evolution of the galaxy. Indeed, $\alpha$-elements are formed by successive $\alpha$ captures occurring in the interiors of massive stars,  released to the \ac{ISM} by \ac{SNeII} explosions \citep{1957RvMP...29..547B}. As iron-peak elements are also processed in massive stars, it results in a constant $\abratio{\alpha}{Fe}$ ratio. When \ac{SNeIa} start to dominate the chemical enrichment and release huge amount of iron-peak elements \citep{2003ApJ...590L..83T}, $\abratio{\alpha}{Fe}$ decreases (\ac{SNeIa} efficiently produce iron-peak elements without producing $\alpha$-elements).

The bottom panel of Figure~\ref{vdsetal2012_OvsFe_MgvsFe_OMgvsFe} shows $\abratio{\alpha}{Fe} = \abratio{O+Mg}{2Fe}$ for the \ac{LMC} and the \ac{MW} (when O, Mg ratios were available, we computed $\abratio{\alpha}{Fe}$ the same way for \ac{MW}). We clearly see that compared to the \ac{MW}, the \ac{LMC} exhibits deficient $\abratio{\alpha}{Fe}$ for $\abratio{Fe}{H} \ge \SI{-1.3}{\dex}$. Those low $\abratio{\alpha}{Fe}$ ratios can be explained by a higher contribution of \ac{SNIa} to the chemical enrichment of the \ac{LMC}, compared to the \ac{MW} \citep[e.g.][]{1998MNRAS.299..535P}.

The comparison of the \ac{LMC} trends to those of the \ac{MW} shows that the $\alpha$-elements can be divided in two groups: on one hand, O and Mg, and on the other hand, Si, Ca and Ti. Indeed, the \ac{LMC} distributions of O and Mg (Figure~\ref{vdsetal2012_OvsFe_MgvsFe_OMgvsFe}) are below those of the \ac{MW} at all metallicities (except for the very most metal poor stars), while the \ac{LMC} distribution of Si, Ca and Ti (Fig.~\ref{vdsetal2012_SivsFe_CavsFe_TivsFe}) completely or partially overlap the \ac{MW} distributions. Although O, Mg, Si, Ca and Ti are all $\alpha$-elements, their production efficiency depends on the mass of the \ac{SNII} progenitor: while O and Mg are predicted to be mainly produced in very massive \ac{SNII} progenitors \citep{1995ApJS..101..181W}, Si, Ca and Ti are predicted to be produced in intermediate mass \ac{SNII} and, in smaller quantity, by \ac{SNIa} \citep{1995MNRAS.277..945T, 2002Ap&SS.281...25T}. The discrepancy between Galactic and \ac{LMC} trends for O and Mg is not an artifact of the abundance analysis: for instance, for a \SI{1}{\dex} metallicity bin centred around Arcturus, we have $\langle \abratio{O}{Fe}_{\mathrm{LMC Bar}} \rangle = \SI{0.1}{\dex}$, $\langle \abratio{O}{Fe}_{\mathrm{MW Discs}} \rangle = \SI{0.47}{\dex}$, $\abratio{O}{Fe}_{\mathrm{Arcturus}} = \SI{0.44}{\dex}$, hence $\Delta(\mathrm{MW} - \mathrm{LMC}) \approx \Delta(\mathrm{Arcturus} - \mathrm{LMC})$ (the same is true for Mg). Therefore, it suggests that the \ac{LMC} formed high mass stars less efficiently than the \ac{MW}.

Our two fields do not exhibit strong differences in their $\alpha$ trends: for O, Mg, Si and Ti, the trends of the bar and the disc overlap at all metallicities. On the other hand, we observe a larger scatter for the bar $\abratio{\alpha}{Fe}$ for $\SI{-0.8}{\dex} \le \abratio{Fe}{H} \le \SI{-0.4}{\dex}$: over this range of metallicities, $\rms (\abratio{\alpha}{Fe}) = \SI{0.08}{\dex}$ for the bar, $\rms (\abratio{\alpha}{Fe}) = \SI{0.05}{\dex}$ for the disc. According to the age-metallicity relation \citep{2005AJ....129.1465C}, this metallicity range corresponds to ages between $\SI{2}{\giga\Year}$ to $\SI{6}{\giga\Year}$ ago, thus the suspected epoch of the bar formation. The slight increase of the scatter between the two fields can be understood in the scenario where a new population is formed. Indeed, if the bar is the result of a new population formation, sustained by gas inflow, then the number of massive stars will increase and they will release significant amounts of freshly formed $\alpha$-elements into the \ac{ISM}. We should then expect an increase of $\abratio{\alpha}{Fe}$ a few \SI{}{\mega\Year} after the start of the star burst \citep{1991ApJ...367L..55G, 1995MNRAS.277..945T, 1998MNRAS.299..535P}. In fact, this increase would be too small to be clearly identifiable in our data because of uncertainties in both $\abratio{Fe}{H}$ and $\abratio{\alpha}{Fe}$, but a larger scatter should appear instead of a distinctive bump in the trend. So, if the observed scatter is true, then it supports the scenario of a new stellar population instead of a dynamically-driven bar.

Unlike the \ac{MW}, the plateau corresponding to the SNII-dominated regime is not clearly visible in the \ac{LMC} field star distribution. Two possibilities can lead to this: either there is a plateau but it appears at a much lower metallicity, which means that the chemical evolution has been very slow compared to the \ac{MW} (when the \ac{SNeIa} start to explode in the \ac{LMC}, the metallicity has reached a lower value than in the \ac{MW}). Or there is no plateau, which \citet{2012ApJ...761..180B} (their Fig.~2) explain by prompt \ac{SNeIa}, for which the onset of \ac{SNeIa} occurs as soon as \SI{100}{\mega\Year} after the formation of the progenitors. Since the lowest metallicity of our samples is only $\approx \SI{-1.6}{\dex}$, we cannot draw firm conclusions about the presence or absence of a plateau in the \acs{LMC}. However, \acs{LMC} globular clusters can also be used to infer the [$\alpha$/Fe] in the metal-poor regime. Except for O (for which metal-poor \acs{LMC} \acs{GC} stars exhibit chemical anomalies due to self-enrichment), there is a good agreement between \acs{LMC} \acs{GC} and \acs{LMC} field stars at both low ($ \SI{-1.5}{\dex} \le \abratio{Fe}{H} \le \SI{-1.2}{\dex} $) and high ($ \SI{-0.5}{\dex} \le \abratio{Fe}{H} \le \SI{-0.2}{\dex} $) metallicity. The metal-poor \acs{LMC} \acs{GC} populate the metallicity range $[-2, -1.2]$ and line up along a \ac{MW}-like plateau at low metallicity ($ \abratio{Fe}{H} \lessapprox \SI{-1.6}{\dex} $). Furthermore, \citet{2012AJ....144...88H} also find that extremely metal-poor RR-Lyrae stars in the \acs{LMC} populate a plateau similar to that of the \ac{MW}.

Let us note that \cite{2012arXiv1210.3968B} also used prompt \ac{SNeIa} in their models to reproduce the \ac{LMC} trends of Mg and Ca (based on abundances of \ac{LMC} field stars and \ac{GC} stars) and conclude that prompt \ac{SNeIa} have influenced the chemical evolution of the \ac{LMC}. We remark that to fully explain the low \ac{LMC} $\abratio{Ca}{Fe}$, \cite{2012arXiv1210.3968B} had to invoke galactic winds which remove Ca more efficiently than the others $\alpha$-elements, but this may not be necessary with our revised abundances (they used abundance measurements from \citealp{2008A&A...480..379P} and our revised abundances are $\approx \SI{0.1}{\dex}$ higher). 

Finally, we note that \cite{2012ApJ...761...33L} have measured O, Mg and Si for \num{89} stars in a filed in the \acs{LMC} disc (around the globular cluster NGC\,1786, some \SI{3}{degree} North-West of the bar centre) and found similar trends as ours.

\subsection{Heavy elements}
\label{Results_heavy}
\begin{figure}
  \begin{centering}
    \includegraphics{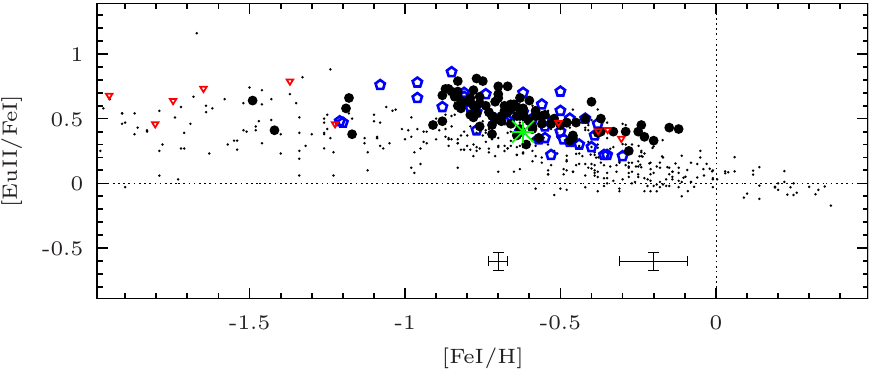}
    \caption{\label{vdsetal2012_EuvsFe} $\abratio{\ion{Eu}{II}}{\ion{Fe}{I}}$ vs. $\abratio{\ion{Fe}{I}}{H}$. Legend: same as Figure~\ref{vdsetal2012_OvsFe_MgvsFe_OMgvsFe}.}
  \end{centering}
\end{figure}

\begin{figure}
  \begin{centering}
    \includegraphics{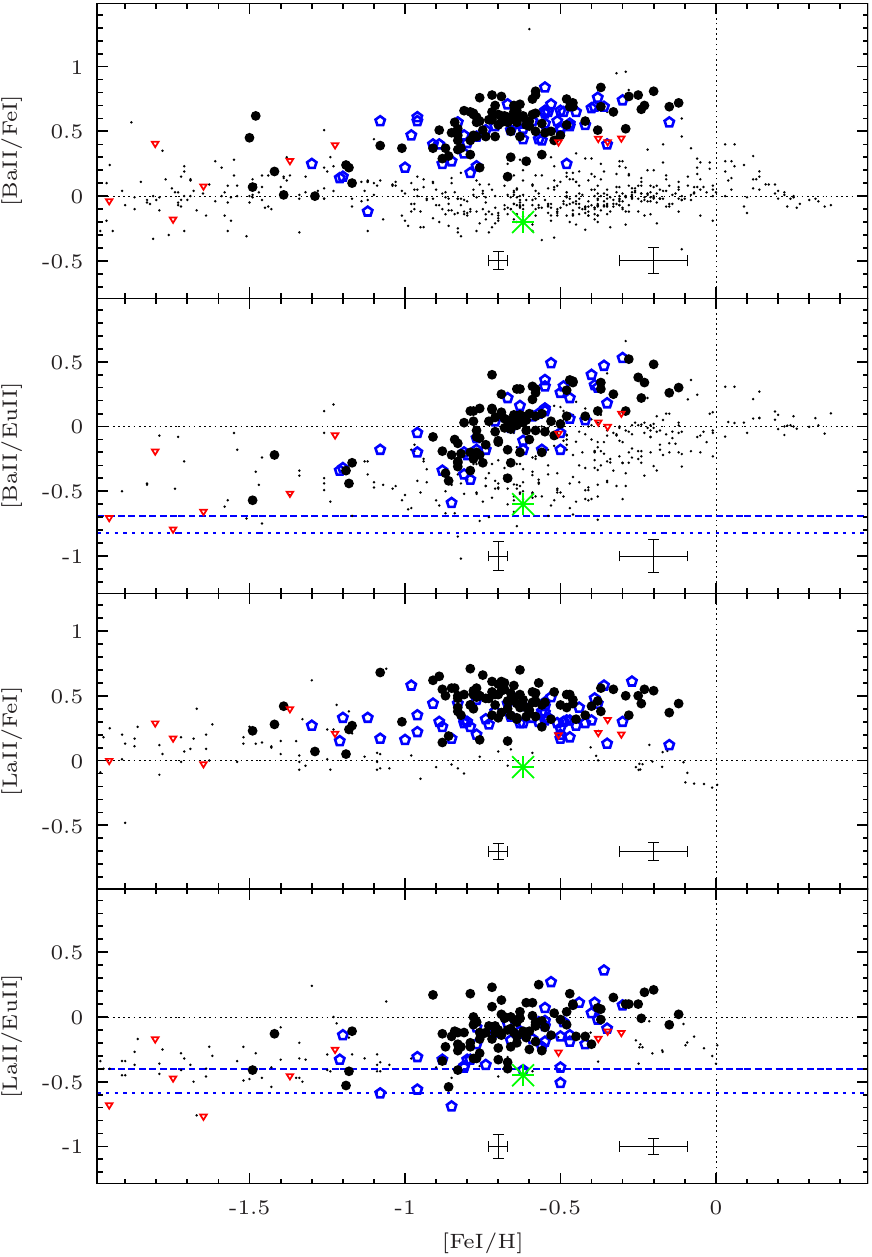}
    \caption{\label{vdsetal2012_BavsFe_BavsEu_LavsFe_LavsEu} First row: $\abratio{\ion{Ba}{II}}{\ion{Fe}{I}}$ vs. $\abratio{\ion{Fe}{I}}{H}$. Second row: $\abratio{\ion{Ba}{II}}{\ion{Eu}{II}}$ vs. $\abratio{\ion{Fe}{I}}{H}$. Third row: $\abratio{\ion{La}{II}}{\ion{Fe}{I}}$ vs. $\abratio{\ion{Fe}{I}}{H}$. Fourth row: $\abratio{\ion{La}{II}}{\ion{Eu}{II}}$ vs. $\abratio{\ion{Fe}{I}}{H}$. Legend: same legend as Figure~\ref{vdsetal2012_OvsFe_MgvsFe_OMgvsFe}; additional \acs{MW} data for Eu and La from \cite{2004ApJ...617.1091S, 2006AJ....131..431B}; horizontal blue dashed line: $\abratio{Ba_{r}}{Eu_{r}} = \SI{-0.69}{\dex}$ and $\abratio{La_{r}}{Eu_{r}} = \SI{-0.4}{\dex}$ \citep{1999ApJ...525..886A}; horizontal blue dotted line: $\abratio{Ba_{r}}{Eu_{r}} = \SI{-0.82}{\dex}$ and $\abratio{La_{r}}{Eu_{r}} = \SI{-0.59}{\dex}$ \citep{2008ARA&A..46..241S}.}
  \end{centering}
\end{figure}

\begin{figure}
  \begin{centering}
    \includegraphics{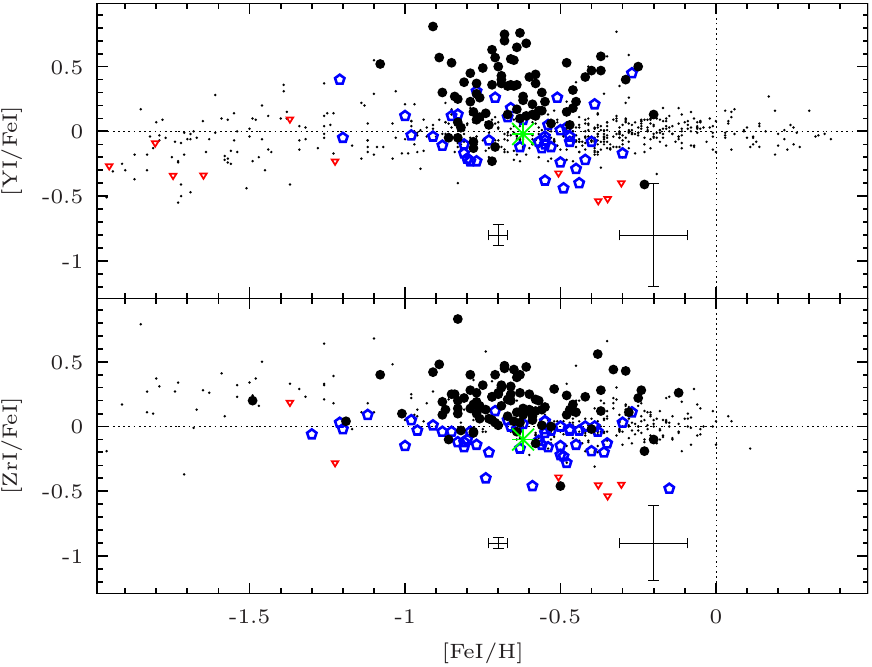}
    \caption{\label{vdsetal2012_YvsFe_ZrvsFe} First row: $\abratio{\ion{Y}{I}}{\ion{Fe}{I}}$ vs. $\abratio{\ion{Fe}{I}}{H}$. Second row: $\abratio{\ion{Zr}{I}}{\ion{Fe}{I}}$ vs. $\abratio{\ion{Fe}{I}}{H}$. Legend: same as Figure~\ref{vdsetal2012_OvsFe_MgvsFe_OMgvsFe}.}
  \end{centering}
\end{figure}

\begin{figure}
  \begin{centering}
    \includegraphics{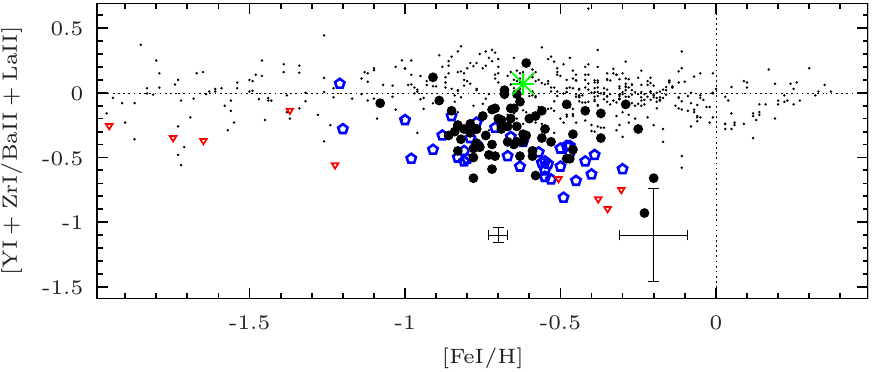}
    \caption{\label{vdsetal2012_YZrvsBaLa} $\abratio{\ion{Y}{I} + \ion{Zr}{I}}{\ion{Ba}{II} + \ion{La}{II}}$ vs. $\abratio{\ion{Fe}{I}}{H}$. Legend: same as Figure~\ref{vdsetal2012_OvsFe_MgvsFe_OMgvsFe}.}
  \end{centering}
\end{figure}

Figures~\ref{vdsetal2012_EuvsFe} to \ref{vdsetal2012_YZrvsBaLa} present the elemental distributions for the heavy elements Eu, Y, Zr, Ba, La. Unlike the elements lighter than iron, the heavy elements are produced by neutron captures through \emph{s}- and \emph{r}-processes. The naming of the two nucleosynthetic channels refers to the neutron flow: the \emph{slow}-process (\emph{s}-process) refers to neutron captures where only a few neutron captures happen before a radioactive $\beta$-decay ($\tau_{\mathrm{capture}} > \tau_{\mathrm{decay}}$ with $\tau_{\mathrm{capture}}$ the timescale of a neutron capture event and $\tau_{\mathrm{decay}}$ the timescale of $\beta$-decay) while the \emph{rapid}-process (\emph{r}-process) refers to processes where numerous neutron captures occur before a radioactive $\beta$-decay ($\tau_{\mathrm{capture}} < \tau_{\mathrm{decay}}$). While it is known that the envelopes of \ac{AGB} stars are the place of the \emph{s}-process \citep[\emph{e.g.},][]{1999ARA&A..37..239B}, there is no consensus as to where the \emph{r}-process is made, except that it should be linked to massive stars. The more promising candidates (providing the needed high neutron fluxes) are 
\ac{SNII} \citep{1996ApJ...466L.109W}, but neutron stars \citep{1999ApJ...525L.121F} also enter the lists \citep{2012arXiv1201.5112Q}. Unlike Y, Zr, Ba, and La which can be mainly produced by the \emph{s}-process (more than \SI{60}{\percent} in the solar system) with a minor contribution from the \emph{r}-process (\emph{e.g.}, $\approx \SI{85}{\percent}$ of the solar Ba was produced by the \emph{s}-process \citealp{2000ApJ...544..302B, 2008ARA&A..46..241S}), Eu is often considered as a pure \emph{r}-process element (the \emph{r}-process contribution to the solar Eu is of \SI{94}{\percent} according to \citealp{1999ApJ...525..886A} and \SI{97}{\percent} according to \citealp{2008ARA&A..46..241S}).

\paragraph{Europium}
In Figure~\ref{vdsetal2012_EuvsFe}, we see that the \ac{LMC} bar and disc Eu distributions agree very well: they both exhibit a constant $\abratio{Eu}{Fe} \approx \SI{0.5}{\dex}$ for $\abratio{Fe}{H} \le \SI{-0.8}{\dex}$, then a decreasing trend with increasing metallicity. 
While for the metal-poor stars the abundance ratios of the \ac{LMC} and the \ac{MW} halo overlap, it is clear that for $\abratio{Fe}{H} \ge \SI{-1}{\dex}$ the \ac{LMC} trend is above that of \ac{MW}. This enhancement for metal-rich stars is not an artifact of our analysis since Arcturus has the expected Eu abundance (\emph{i.e.} it falls on top of the \ac{MW} thick disc). This is in fact a chemical anomaly already noticed in \acs{LMC} supergiant stars \citep{1989ApJS...70..865R, 1995A&A...293..347H} and \acs{LMC} \acs{GC} stars \citep{2008AJ....136..375M, 2012ApJ...746...29C} and its origin still remains unclear. Different mechanisms can help in maintaining a high Eu ratio in late stages of the chemical evolution: (1) new star bursts will form a high number of massive stars which will, in turn, inject fresh Eu in the \ac{ISM}; (2) another source of \emph{r}-processed Eu; (3) a stronger contribution of \emph{s}-processed Eu. Explanation (1) is not supported by the \ac{SFH} of our two \ac{LMC} fields \citep{2002ApJ...566..239S}: although recent star bursts (about \SI{5}{\giga\Year} ago and less than \SI{1}{\giga\Year} ago) are expected in the bar, they are not expected in the inner disc; so it cannot explain the high ratios observed in both fields. Moreover, they would produce similar $\alpha$ enhancements, which are not observed. But it can explain the small difference between the \ac{LMC} bar and disc observed for the most metal-rich stars (above \SI{-0.5}{\dex}). Indeed, while the disc $\abratio{Eu}{Fe}$ still decreases and reaches lower values than the bar, the bar $\abratio{Eu}{Fe}$ seems to remain constant ($\abratio{Eu}{Fe} \approx \SI{0.4}{\dex}$). (2) or (3) are more likely to explain the differences between the \ac{LMC} and the \ac{MW}. The contribution of the \emph{s}-process to the solar system Eu is estimated to be a few percent \citep[\SI{3}{\percent} according to][]{2008ARA&A..46..241S} and therefore, it is dubious that the \emph{s}-process could be responsible for the Eu enhancement. However, in the \acs{MW}, a particular class of metal-poor stars --- carbon enhanced metal-poor (CEMP) stars --- exhibits carbon enhancements, understood as the result of a binary interaction with a (now deceased) \acs{AGB} companion. In addition, CEMP stars are called CEMP-\emph{s} (respectively CEMP-\emph{r}/\emph{s}) stars when they also have \emph{s} enhancements (respectively \emph{r} and \emph{s} enhancements). Several scenarios have been proposed to explain the origin of, \emph{e.g.}, \ce{Ba} and \ce{Eu} enhancements in CEMP-\emph{r}/\emph{s}) stars but they all invoke both \emph{s}- and \emph{r}-processes: three mass-transfers in a binary system (see \citealp{2003ApJ...588.1082C}); two successive mass accretions from a \num{8}-\num{10}~$\mathrm{M}_{\odot}$ companion (see \citealp{2006ApJ...636..842W}); pre-\emph{r} enrichment followed by \emph{s}-material accretion (see \citealp{2012MNRAS.422..849B}). But \cite{2010A&A...509A..93M} found a correlation between $\abratio{Pb}{Ba, La, Ce, }$ and $\abratio{N}{H}$, and interpreted it as evidence of the operation of the \ce{^{22}Ne({\alpha} {,} n)^{25}Mg} chain in the CEMP-\emph{r}/\emph{s} companion. As this chain is associated with high neutron density, \cite{2010A&A...509A..93M} claimed that the \ce{Eu} enhancement observed today in CEMP-\emph{r}/\emph{s} stars is due to the production of \ce{Eu} via the \emph{s}-process in the deceased \acs{AGB} companion, \emph{i.e.} they invoke a unique site to explain both \emph{s}- and \emph{r}-enhancements. In addition, in their study of CEMP-\emph{s} and CEMP-\emph{r}/\emph{s} stars, \cite{2012A&A...548A..34A} found a correlation between $\abratio{Ba}{Fe}$ and $\abratio{Eu}{Fe}$ and that $\abratio{Eu}{Fe}$ is coupled to the degree of C over-abundance, \emph{i.e.} \ce{Eu} was produced by the former \acs{AGB} companion. Consequently, \cite{2012A&A...548A..34A} claim that Ba and Eu have the same origin in CEMP-\emph{r}/\emph{s}, \emph{i.e.} they are produced by the \emph{s}-process occurring during the \acs{AGB} phase of the more massive star of a binary system. In essence, the CEMP-\emph{r}/\emph{s} stars would be stars polluted by metal-poor \acs{AGB} with significant Eu production. Thus, those new results and the dominant role played by \acs{AGB} stars in the chemical evolution of the \acs{LMC} (see below) support explanation (3).

\paragraph{Barium and lanthanum}
While the \ac{MW} has constant solar $\abratio{Ba}{Fe}$ and $\abratio{La}{Fe}$ ratios (with a weak increase towards high metallicities), both \ac{LMC} fields exhibit a dramatic increase of $\abratio{Ba}{Fe}$ and $\abratio{La}{Fe}$ with increasing metallicity (first and third panels of Figure~\ref{vdsetal2012_BavsFe_BavsEu_LavsFe_LavsEu}): the \ac{LMC} distributions agree with \ac{MW} halo trends, \emph{i.e.} for $\abratio{Fe}{H} \le \SI{-1.}{\dex}$, and are above the \ac{MW} elsewhere. The bar and disc field distributions overlap for both Ba and La. Ba has the strongest increase, starting from solar value at $\abratio{Fe}{H} = \SI{-1.5}{\dex}$ and reaching \SI{0.8}{\dex} for $\abratio{Fe}{H} \ge \SI{-0.3}{\dex}$. La, on the other hand, remains approximately constant from $\abratio{Fe}{H} \approx \SI{-0.7}{\dex}$ ($\abratio{La}{Fe} \approx \SI{0.5}{\dex}$ for the \ac{LMC} bar and $\approx \SI{0.4}{\dex}$ for the \ac{LMC} disc). Furthermore, there is an excellent match between the \acs{LMC} field and  \acs{GC} populations. This indicates that the production of Ba and La has been much more efficient in the \ac{LMC} than in the \ac{MW}. 

To identify the process responsible for this high production, we examine $\abratio{Ba}{Eu}$ and $\abratio{La}{Eu}$ in second and fourth panels of Figure~\ref{vdsetal2012_BavsFe_BavsEu_LavsFe_LavsEu}. We see that for \acs{LMC} \acs{GC} and field metal-poor stars (from \SI{-2.}{\dex} to \SI{-0.8}{\dex}), $\abratio{Ba, La}{Eu}$ is constant and compatible (within uncertainties) with a pure \emph{r}-process source (\citealp{1999ApJ...525..886A}: $\abratio{Ba_{r}}{Eu_{r}} = \SI{-0.69}{\dex}$ and $\abratio{La_{r}}{Eu_{r}} = \SI{-0.4}{\dex}$; \citealp{2008ARA&A..46..241S}: $\abratio{Ba_{r}}{Eu_{r}} = \SI{-0.82}{\dex}$ and $\abratio{La_{r}}{Eu_{r}} = \SI{-0.59}{\dex}$). On the other hand, for $\abratio{Fe}{H} \ge \SI{-0.8}{\dex}$, the increase of the \acs{LMC} $\abratio{Ba, La}{Eu}$ is interpreted as the rise of a new source of Ba and La, \emph{i.e.} the \emph{s}-process. The \ac{MW} exhibits similar patterns (constant ratio at low metallicity, then an increase) but two differences exist with the \acs{LMC}: first, the increase of $\abratio{Ba, La}{Eu}$ starts at lower metallicity in the \ac{LMC} ($\abratio{Fe}{H} \approx \SI{-0.8}{\dex}$) than in the \ac{MW} ($\abratio{Fe}{H} \approx \SI{-0.4}{\dex}$), reflecting the slower metal-enrichment in the \acs{LMC}; secondly, while for the \ac{MW}, $\abratio{Ba}{Eu}$ reaches a solar value, for the \ac{LMC}, $\abratio{Ba}{Eu}$ reaches a much higher value ($\abratio{Ba}{Eu} \approx \SI{0.4}{\dex}$). This suggests that the production of Ba and La by the \emph{s}-process has been much more efficient in the \ac{LMC} than in the \ac{MW}, and indicates that \acs{AGB} stars played a stronger role in the chemical enrichment of the \acs{LMC} compared to the \acs{MW}.

\paragraph{Yttrium and zirconium}
For $\abratio{Fe}{H} \le \SI{-1}{\dex}$, the \acs{LMC} bar and disc seem to have a solar Zr ratio (Figure~\ref{vdsetal2012_YvsFe_ZrvsFe}) and a solar Y ratio (at least for the disc field; we do not have enough data points for the bar field). For $\abratio{Fe}{H} > \SI{-1}{\dex}$, the two \acs{LMC} fields have a flat Y and Zr distribution with a large scatter but the \ac{LMC} disc and bar exhibit a different mean behaviour in their Y and Zr trends: in the mean, the bar has higher $\abratio{Y}{Fe}$ and $\abratio{Zr}{Fe}$ than the disc (bar: $\langle \abratio{Y}{Fe} \rangle = \SI{0.31}{\dex}$, $\langle \abratio{Zr}{Fe} \rangle = \SI{0.19}{\dex}$; disc: $\langle \abratio{Y}{Fe} \rangle = \SI{-0.04}{\dex}$, $\langle \abratio{Zr}{Fe} \rangle = \SI{-0.08}{\dex}$).   
We checked for possible systematic effects explaining the differences but found none (same kind of stars, same instrument and observing setups, similar data reduction procedure, same procedures to derive stellar parameters and abundances). In particular, we checked whether one of the three Zr lines used could be responsible for the difference. In the mean, each line gives higher Zr abundances for the bar than for the disc: for the bar, $\langle \abratio{Zr}{Fe} \rangle = \SI{0.22}{\dex}$, \SI{0.20}{\dex} and \SI{0.14}{\dex}; for the disc, $\langle \abratio{Zr}{Fe} \rangle = \SI{-0.04}{\dex}$, \SI{0.}{\dex}, and \SI{-0.14}{\dex} (respectively for the line at \SI{6127}{\angstrom}, \SI{6134}{\angstrom} and \SI{6143}{\angstrom}). We also looked for possible correlations between the derived abundances and the stellar parameters and found for both fields a clear correlation between $\abratio{Y, Zr}{Fe}$ and $\mathrm{T}_{\mathrm{eff}}$ or $\xi_{\mathrm{micro}}$, \emph{i.e} increasing abundance ratio with increasing temperature or microturbulence velocity. \acs{LMC} \acs{GC} stars of \cite{2008AJ....136..375M, 2010ApJ...717..277M} follow the same correlation as ours, and since they are lower temperature than our stars, they have Y and Zr abundances lower than ours. Furthermore, our two samples do not have the same temperature coverage (\SI{3900}{\kelvin} to \SI{5200}{\kelvin} for the bar, \SI{3800}{\kelvin} to \SI{4600}{\kelvin} for the inner disc). But if we select only stars in the temperature range $[4000, 4400]$, the dispersion slightly decreases in each field, but the two fields remain significantly different: $\langle \abratio{Y}{Fe} \rangle = \SI{0.25}{\dex}$, $\langle \abratio{Zr}{Fe} \rangle = \SI{0.15}{\dex}$ for the bar, and $\langle \abratio{Y}{Fe} \rangle = \SI{0.}{\dex}$, $\langle \abratio{Zr}{Fe} \rangle = \SI{-0.07}{\dex}$ for the disc.   
The typical random error (due to pixel noise) on the final Y or Zr abundances are respectively \SI{0.08}{\dex} and \SI{0.04}{\dex} while the typical systematic error (due to errors on effective temperature) are \SI{0.4}{\dex} and \SI{0.29}{\dex} respectively. Thus, errors can explain the observed scatter but cannot explain the offset between our two \acs{LMC} populations.

Such a discrepancy is not seen for Ba and La for which the distributions of our two \acs{LMC} fields agree rather well. Interestingly, Y and Zr belong to the first peak of the \emph{s}-process, while Ba and La on the other hand belong to the second peak (the positions of the peaks correspond to magic numbers for nuclear stability). The observed differences may be an effect of the metallicity of the \acs{AGB} stars producing the \emph{s}-elements because the second peak is favoured when metal-poor \ac{AGB} stars dominate the chemical enrichment \citep[\emph{e.g.},][]{1997ApJ...478..332S, 1998ApJ...497..388G, 2011ApJS..197...17C}. This suggests that \ac{AGB} stars were more metal-poor in the disc than in the bar of the \acs{LMC}. We note that the metal-rich \acs{LMC} \acs{GC} have $\abratio{Y}{Fe}$ and $\abratio{Zr}{Fe}$ clearly lower than those of the \acs{LMC} bar, and probably similar to the \acs{LMC} inner disc, which is understandable since their projected locations lie in the \ac{LMC} disc. Thus, the differences observed between the \acs{LMC} bar and disc for Y and Zr for $\abratio{Fe}{H} \gtrapprox \SI{-1}{\dex}$ speak in favour of a different chemical evolution path: unlike the disc, the bar experienced a recent episode of stellar formation (a few Gyrs ago) which generated metal-rich \acs{AGB} that explain the present Y an Zr ratios.

Figure~\ref{vdsetal2012_YZrvsBaLa} shows $\abratio{Y + Zr}{Ba + La}$. We recall that Y and Zr are light-\emph{s} (ls) elements, while Ba and La are heavy-\emph{s} elements. Thus the ratio $\abratio{ls}{hs}$ tracks the relative importance of metal-poor and metal-rich \ac{AGB}. For $\abratio{Fe}{H} \gtrapprox \SI{-0.8}{\dex}$, we know from above that the \emph{s}-process dominates the chemical enrichment. We remark that the \ac{LMC} trend is below that of the \ac{MW}, which suggests that the \ac{AGB} stars which dominated the \ac{LMC} enrichment were more metal-poor than those of the \ac{MW}. While the bar distribution is flat, the disc distribution seems to slightly decrease with increasing metallicity. \cite{2012ApJ...746...29C} also found a decrease of $\abratio{Y}{Ba}$ in intermediate-age clusters. This could be a mass effect (in the \ac{LMC}, metal-poor low mass \ac{AGB} stars still contribute significantly to the enrichment) but only a consistent chemical evolution modelling can confirm this explanation. We remark that the match between \acs{LMC} \acs{GC} and our \acs{LMC} fields is again excellent.

\subsection{Sodium and iron-peak elements}
\begin{figure}
  \begin{centering}
    \includegraphics{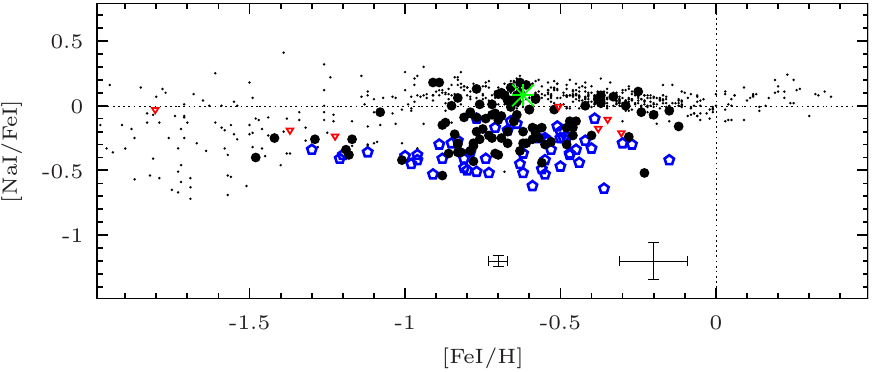}
    \caption{\label{vdsetal2012_NavsFe} $\abratio{\ion{Na}{I}}{\ion{Fe}{I}}$ vs. $\abratio{\ion{Fe}{I}}{H}$. Legend: same as Figure~\ref{vdsetal2012_OvsFe_MgvsFe_OMgvsFe}.}
  \end{centering}
\end{figure}

\begin{figure}
  \begin{centering}
    \includegraphics{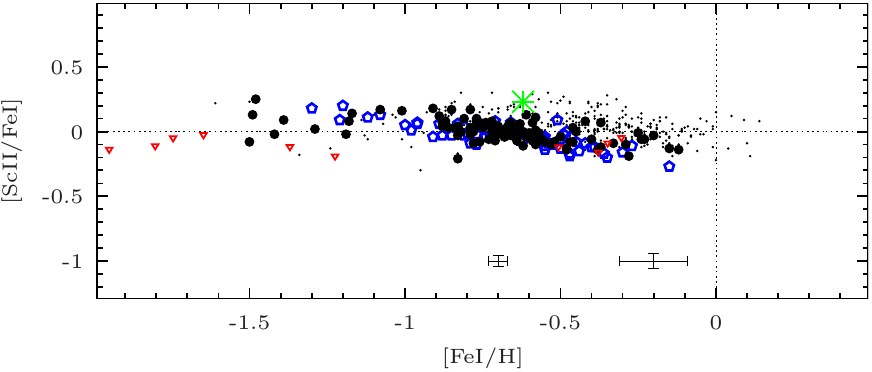}
    \caption{\label{vdsetal2012_ScvsFe} $\abratio{\ion{Sc}{II}}{\ion{Fe}{I}}$ vs. $\abratio{\ion{Fe}{I}}{H}$. Legend: same as Figure~\ref{vdsetal2012_OvsFe_MgvsFe_OMgvsFe}.}
  \end{centering}
\end{figure}

\begin{figure}
  \begin{centering}
    \includegraphics{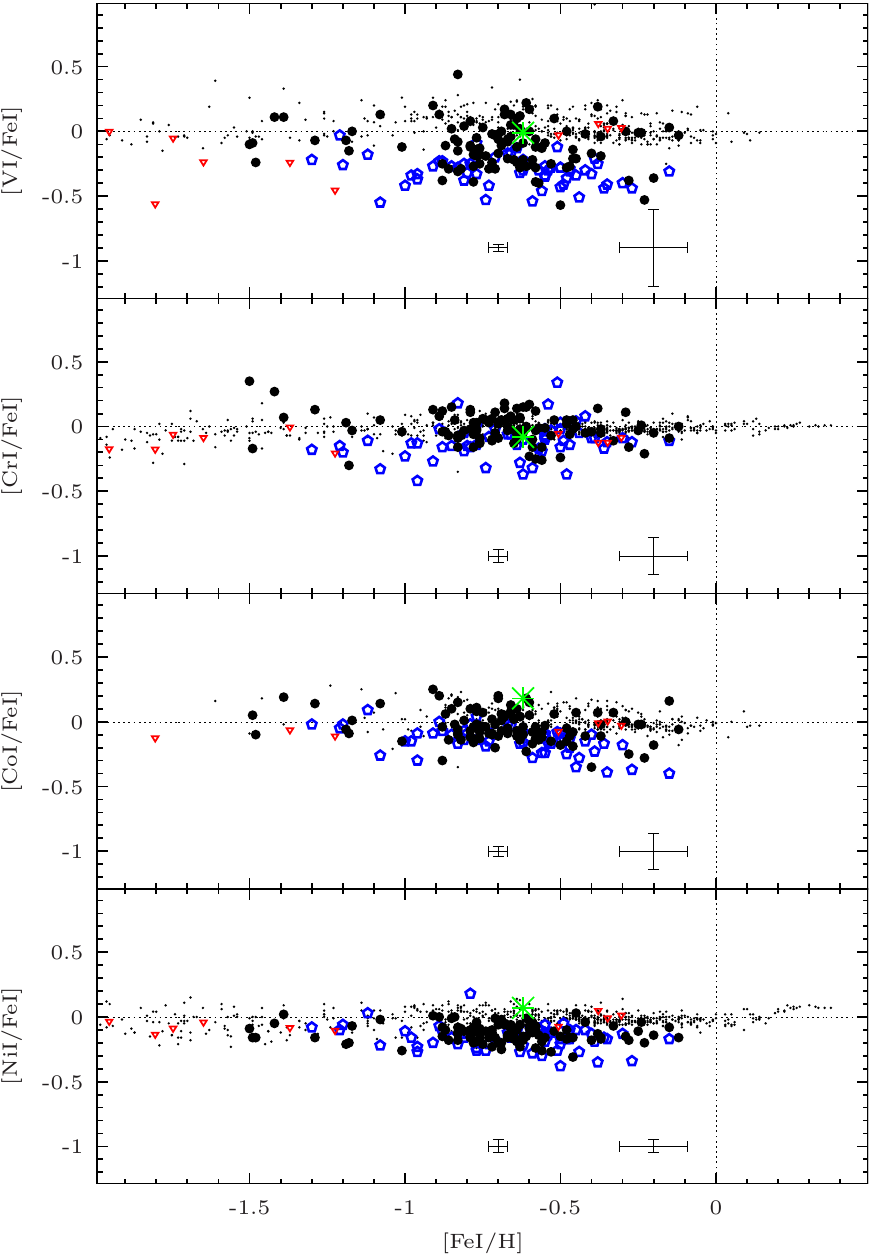}
    \caption{\label{vdsetal2012_VvsFe_CrvsFe_CovsFe_NivsFe} First row: $\abratio{\ion{V}{I}}{\ion{Fe}{I}}$ vs. $\abratio{\ion{Fe}{I}}{H}$. Second row: $\abratio{\ion{Cr}{I}}{\ion{Fe}{I}}$ vs. $\abratio{\ion{Fe}{I}}{H}$. Third row: $\abratio{\ion{Co}{I}}{\ion{Fe}{I}}$ vs. $\abratio{\ion{Fe}{I}}{H}$. Fourth row: $\abratio{\ion{Ni}{I}}{\ion{Fe}{I}}$ vs. $\abratio{\ion{Fe}{I}}{H}$. Legend: same as Figure~\ref{vdsetal2012_OvsFe_MgvsFe_OMgvsFe}.}
  \end{centering}
\end{figure}

Figures~\ref{vdsetal2012_NavsFe} to \ref{vdsetal2012_VvsFe_CrvsFe_CovsFe_NivsFe} show the elemental distributions of Na and iron-peak elements Sc, V, Cr, Co and Ni. 

\paragraph{Sodium}
In Figure~\ref{vdsetal2012_NavsFe}, for stars whose metallicity is below \SI{-1.1}{\dex}, we see that the \ac{LMC} bar and disc $\abratio{Na}{Fe}$ agree rather well within uncertainties (bar: $\langle \abratio{Na}{Fe} \rangle = \SI{-0.28}{\dex}$; disc: $\langle \abratio{Na}{Fe} \rangle = \SI{-0.37}{\dex}$) and they overlap the \ac{MW} halo distribution. On the other hand, for $\abratio{Fe}{H} \ge \SI{-1.1}{\dex}$, the two \ac{LMC} distributions become different. The bar $\abratio{Na}{Fe}$ seems to increase with increasing metallicity and reaches solar values ($\langle \abratio{Na}{Fe} \rangle = \SI{-0.13}{\dex}$, $\rms = \SI{0.17}{\dex}$), thus overlapping the Galactic trends. On the other hand, the disc $\abratio{Na}{Fe}$ remains sub-solar with a flat distribution ($\langle \abratio{Na}{Fe} \rangle = \SI{-0.35}{\dex}$, $\rms = \SI{0.13}{\dex}$). Both fields exhibit a large scatter in this metallicity regime: although only five bar stars and four disc stars are observed for $\abratio{Fe}{H} \le \SI{-1.1}{\dex}$, we can guess that the scatter is smaller than for $\abratio{Fe}{H} \ge \SI{-1.1}{\dex}$ (if the scatter were the same, it would be unlikely to have five or four measures concentrated within \SI{0.2}{\dex}). None of the three Na lines used is responsible for the difference; in the mean, each line gives higher Na abundances for the bar than for the disc: for the bar, $\langle \abratio{Na}{Fe} \rangle = \SI{-0.11}{\dex}$, \SI{-0.15}{\dex} and \SI{0.12}{\dex}; for the disc, $\langle \abratio{Na}{Fe} \rangle = \SI{-0.31}{\dex}$, \SI{-0.35}{\dex}, and \SI{-0.36}{\dex} (respectively for the line at \SI{5688}{\angstrom}, \SI{6154}{\angstrom} and \SI{6160}{\angstrom}).   
As for Y and Zr, there is a correlation between $\abratio{Na}{Fe}$ and $\mathrm{T}_{\mathrm{eff}}$ and again, if we select only the stars in the temperature range $[4000, 4400]$, the dispersion slightly decreases but we still see the different mean behaviour. We note that there is excellent agreement at both low and high metallicity between the \acs{LMC} \acs{GC} and our \acs{LMC} fields. As for Y and Zr, the typical random and systematic error on the final Na abundance are \SI{0.04}{\dex} and \SI{0.14}{\dex} respectively, and can explain the scatter but not the offset between the two fields.

The production of Na is still uncertain and is thought to occur in high-mass \ac{SNII} \citep{1995ApJS..101..181W} and \ac{AGB} stars \citep{2000A&A...362..599G,2006MmSAI..77..774C,2010MNRAS.404.1529B}. Issues on the abundance measurement have been reported, \emph{e.g.} in \cite{2004A&A...424..951P} where the authors find a disagreement between Na abundances of giant and dwarf stars belonging to the same cluster. Different explanations are quoted to explain these issues: departure from local thermodynamic equilibrium, surface Na abundances modified by the first dredge-up or uncertainties on atomic data \citep{2012MNRAS.422.1562S}. It is therefore difficult to understand the \ac{LMC} trends relative to those of the \ac{MW} (most of the \ac{MW} abundances were measured in dwarf stars) but comparing the two \ac{LMC} fields is still valid. The discrepancy between the \ac{LMC} bar and disc fields tells us that the production of Na has been more efficient in the bar than in the disc: it could be the result of the star burst that gave birth to the new population in the central parts of the \ac{LMC}.

\paragraph{Scandium}
Figure~\ref{vdsetal2012_ScvsFe} shows the \ac{LMC} bar and disc Sc distributions. The bar and disc have similar $\abratio{Sc}{Fe}$. They overlap the \ac{MW} halo but are below the \ac{MW} discs. As noticed for the \ac{MW} \citep{2000A&A...353..722N,2003MNRAS.340..304R,2006MNRAS.367.1329R}, the Sc in the \ac{LMC} behaves approximately like Ca or Ti: small scatter at all metallicities; $\abratio{Sc}{Fe}$ decreases with increasing metallicity; the distribution for the most metal-poor stars ($\abratio{Fe}{H} \le \SI{-1}{\dex}$) is compatible with a plateau (especially for the bar); the amplitude of the decrease between the metal-poor edge and the metal-rich edge is of \SI{0.2}{\dex}. \cite{2000ApJ...537L..57P} claimed that the $\alpha$-like pattern of Sc could be due to poor \ac{hfs} data but \cite{2003MNRAS.340..304R,2006MNRAS.367.1329R} used weak \ion{Sc}{II} lines in dwarf stars for which the \ac{hfs} has little effect on the derived abundances. For our giant stars, the \acs{hfs} must be taken into account since \ion{Sc}{II} lines are strengthened; and we see that Arcturus $\abratio{Sc}{Fe}$ lies on the top of the thick disc distribution, as expected.

\paragraph{Other iron-peak elements}
Figure~\ref{vdsetal2012_VvsFe_CrvsFe_CovsFe_NivsFe} presents the abundance distributions of V, Cr, Co and Ni. Although they all belong to the iron-peak and are mainly produced by \ac{SNIa}, these elements exhibit different patterns. The abundance distributions of V, Cr and Co are characterised by a rather large scatter, V being the most dramatic case. Cr and Co have flat distributions overlapping those of the \ac{MW} for both \ac{LMC} fields while Ni is sub-solar at all metallicities for both \ac{LMC} fields. 

On the other hand, for V, the bar and disc distributions agree only for $\abratio{Fe}{H} \le \SI{-1.1}{\dex}$. For $\abratio{Fe}{H} \ge \SI{-1.1}{\dex}$, in the mean, the bar has higher $\abratio{V}{Fe}$ than the disc (bar: $\langle \abratio{V}{Fe} \rangle = \SI{-0.11}{\dex}$; disc: $\langle \abratio{V}{Fe} \rangle = \SI{-0.30}{\dex}$).  
Among the iron-peak elements, V is the one with the highest number of measured lines (\num{7} lines, most of the time) but it exhibits the largest scatter, which should in principle be a sign that the scatter is astrophysical. As for Na, the disagreement between the two \ac{LMC} fields appears to reach \SI{-1}{\dex}. We performed similar sanity checks for V as we did for Y, Zr and Na, and found a correlation between the derived abundance and the temperature; but a selection on temperature leaves the discrepancy unchanged. The typical random and systematic error respectively \SI{0.04}{\dex} and \SI{0.29}{\dex} and can explain the scatter but not the offset between the two fields.

\subsection{Copper}
\begin{figure}
  \begin{centering}
    \includegraphics{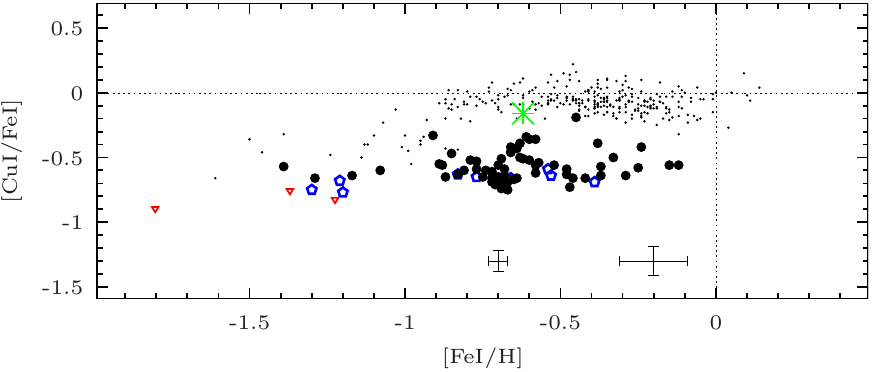}
    \caption{\label{vdsetal2012_CuvsFe} $\abratio{\ion{Cu}{I}}{\ion{Fe}{I}}$ vs. $\abratio{\ion{Fe}{I}}{H}$. Legend: same as Figure~\ref{vdsetal2012_OvsFe_MgvsFe_OMgvsFe}.}
  \end{centering}
\end{figure}

Figure~\ref{vdsetal2012_CuvsFe} shows the abundance trends of Cu. While the \ac{LMC} bar and disc ratios match those of the \ac{MW} for $\abratio{Fe}{H} \le \SI{-1.1}{\dex}$, the \ac{LMC} ratios are significantly lower than those of the \ac{MW} for higher metallicities: $\langle \abratio{Cu}{Fe} \rangle = \SI{-0.56}{\dex}$ for the bar and $\langle \abratio{Cu}{Fe} \rangle = \SI{-0.64}{\dex}$ for the disc. Since we found the expected value for Arcturus, the observed deficiency for Cu is not an artifact of our abundance analysis. The origin of Cu is still heavily debated since Cu is thought to have both primary and secondary production: \cite{2007MNRAS.378L..59R} see the origin of Cu in neutron captures occurring in massive stars dying as \ac{SNII} (primary production), \cite{2004ApJ...601..864T} invoke a minor contribution from \emph{s}-process in \ac{AGB} (secondary production) and \cite{2002A&A...396..189M} consider the thermonuclear nucleosynthesis in \ac{SNIa} as the main source (secondary production). We saw in Section~\ref{Results_alpha} that a stronger contribution of \ac{SNIa} is needed to explain the low $\alpha$ ratios and in Section~\ref{Results_heavy}, that a stronger contribution of \ac{AGB} is needed to explain the high Ba and La ratios. In addition, \cite{2010ApJ...710.1557P} found that more than half of the solar copper is produced through weak \emph{s}-process occurring in massive stars ($25\mathrm{ M}_{\odot}$). Therefore, \ac{SNIa} and \ac{AGB} stars cannot be the main site of Cu production in the \ac{LMC} and the hypothesis of massive stars being the main source of Cu seems to be more plausible: Cu in the \acs{LMC} has mainly a primary origin. As with the $\alpha$ elements O and Mg, our results for Cu support the scenario of a chemical enrichment dominated  by intermediate mass \ac{AGB} stars and \ac{SNIa}, with a smaller contribution from very massive stars (compared to the \ac{MW}).

\section{Summary and conclusions}
\label{Conclusion}
To compare the chemical history of the \acs{LMC} to that of the \acs{MW} and disentangle the chemical evolution of the \acs{LMC} bar and disc, we performed a detailed chemical analysis of \acs{LMC} giant stars using high-resolution spectra obtained at ESOVLT. Our sample was made up of \num{106} newly observed stars in a field centred on the LMC bar, plus 58 stars observed by \citet{2008A&A...480..379P} in an inner disc field $\sim$\SI{2}{deg} South of the bar.  We took great care to insure the homogeneity of the two samples, and furthermore used the local thick disc giant Arcturus to insure a proper comparison to Milky Way samples. Our main findings can be summarised as follows:

\begin{itemize}

\item The two samples cover the metallicity range $\abratio{Fe}{H}$ from $-1.5$ to \SI{-0.1}{\dex}, covering the full \acs{LMC} disc metallicity distribution \citep{2005AJ....129.1465C,2008AJ....135..836C}, albeit leaving out the most metal-poor (and less numerous) tail of the distribution despite our deliberate overpopulation of this tail in the target selection. $80\%$ of the sample is contained between $-1.1$ and \SI{-0.4}{\dex}. As this sample is metallicity-biased, it cannot be used for metallicity distribution studies.

\item In the metallicity range covered by both types of objects, the \ac{LMC} field and \ac{GC} elemental abundances exhibit an excellent agreement for all elements (except for O and Na at low metallicity; the low-metallicity clusters have experienced self-enrichment, creating anti-correlated O-Na star to star variations).

\item The $\alpha$-element ratios $\abratio{Mg}{Fe}$ and $\abratio{O}{Fe}$ are lower in the \acs{LMC} than in the \acs{MW} suggesting a smaller contribution of massive stars (with respect to \acs{SNIa}) in the \ac{LMC}, or a slower enrichment. The presence of a plateau for $\abratio{\alpha}{Fe}$ is not convincingly probed by our sample due to a lack of metal-poor members, although the most metal-poor \ac{GC} do seem to lie on a plateau. $\abratio{Ba, La}{Eu}$ exhibit a strong increase from $\abratio{Fe}{H} \approx \SI{-0.8}{\dex}$ with increasing metallicity, showing that the chemical enrichment of the \acs{LMC} has been slower than that of the \acs{MW}, and that the neutron-capture elements were dominated by the \emph{s}-process, mainly occurring in \acs{AGB} stars. The \acs{LMC} has lower $\abratio{Y+Zr}{Ba+La}$ ratios than the \acs{MW} indicating that these \acs{AGB}s were more metal-poor in the \acs{LMC}. 

\item Eu does not follow the expected trend which could be an indication of an efficient \emph{s}-production of this element, despite the usually assumed almost pure \emph{r}-process origin of this element. This finding is supported by the recent work by \citet{2012A&A...548A..34A} who advocate a strong \emph{s}-process contribution to Eu in a certain category of extremely metal-poor carbon and \emph{s}-process enhanced stars (the so-called \emph{sr}-stars).

\item Cu is almost constant over the metallicity range and about \SI{0.5}{\dex} lower in the \acs{LMC} than in the \acs{MW} showing that in the \acs{LMC} Cu has mainly a primary origin (through weak \emph{s}-process in massive stars).

\item The \acs{LMC} bar and disc exhibit subtle differences in their $\abratio{\alpha}{Fe}$ (slightly larger scatter for the bar in the metallicity range $[-1, -0.5]$), their $\abratio{Eu}{Fe}$ (the bar trend is above the disc trend for $\abratio{Fe}{H} \ge \SI{-0.5}{\dex}$, their Y and Zr, their Na and their V (offset between the bar and the disc distributions). These differences are possibly related to the formation of a new stellar population in the central part of the \acs{LMC}: the resulting new generation of massive stars will inject freshly synthesised $\alpha$-elements (hence the increased scatter observed in the bar) and Eu (hence the higher $\abratio{Eu}{Fe}$ ratios in the bar) and the new generation of metal-rich \acs{AGB} stars will produce Na, Y and Zr (hence the offset). These findings strengthen a scenario where the \acs{LMC} bar is not a mere dynamically driven (or interaction driven) overdensity, but implied a fresh episode of star formation. This scenario also supported by the star formation history derived in the bar, that highlights an increased star formation \SI{2}{}-\SI{5}{\giga\Year} ago, with no clear counterparts in other locations in the \acs{LMC} disc \citep{2002ApJ...566..239S}. More globally, \citet{2009AJ....138.1243H} have established a map of star formation histories across the whole \acs{LMC} and find again that the dominant star formation episode that occurred some \SI{5}{\giga\Year} is more pronounced in the bar than anywhere else in the \acs{LMC}. Even younger bursts of star formation seem to follow the bar morphology, around \SI{500}{} and \SI{100}{\mega\Year} ago. \citet{2008ApJ...682L..89G}, although their sample does not include the bar per se, also highlight that the younger populations in the \ac{LMC} are found closer to the centre together with a positive age-gradient of the youngest star formation episode towards the outskirts. All these findings regarding the star formation history of the \acs{LMC} bar and disc strengthen a scenario where the bar is the strongest manifestation of the higher recent star formation activity in the central parts of the \acs{LMC}.
\end{itemize}

\begin{acknowledgements}
  M. Van der Swaelmen thanks ESO and the CNRS for funding his PhD work. We thank Betrand Plez for his program to compute hyperfine structure. \daospec has been written by P. B. Stetson for the Dominion Astrophysical Observatory of the Herzberg Institute of Astrophysics, National Research Council, Canada. This research has made use of NASA's Astrophysics Data System Bibliographic Services, of the VALD and Kurucz databases. M. Van der Swaelmen and V. Hill thank the support of the French Programme National Galaxies (INSU) which allowed this research to be carried out. We would like to pay a special tribute to our collaborator and friend Luciana Pompeia who passed away much too soon. She was a key person early in this LMC project and we have special thoughts towards her familly, friend and collegues.
\end{acknowledgements}

\bibliographystyle{aa}
\bibliography{VanderSwaelmenEtAl2013}

\end{document}